\documentclass[12pt]{article}
\pdfoutput=1

\usepackage{cite}
\usepackage{booktabs}
\usepackage[english]{babel}
\usepackage{amsmath,amssymb,amsbsy,amstext, amsthm, simplewick}
\usepackage{hyperref}
\usepackage{graphicx}
\usepackage{amsfonts}
\usepackage{amssymb}
\usepackage[small]{caption}
\usepackage{upgreek}
\usepackage[svgnames,dvipsnames,x11names,table]{xcolor}
\usepackage{multirow}
\usepackage{geometry}
\usepackage[hang,flushmargin]{footmisc}
\usepackage{bm}
\usepackage{braket}
\usepackage{subcaption}
\usepackage{mathtools}
\usepackage{setspace}
\usepackage{cleveref}
\usepackage{comment}
\usepackage{scalerel}
\usepackage[normalem]{ulem}
\usepackage{slashed}
\usepackage{enumitem}
\usepackage{dsfont}
\usepackage{tikz}
\usepackage{colortbl}
\usepackage[table]{xcolor}
\usetikzlibrary{decorations.markings}
\usetikzlibrary{shapes.misc}

\definecolor{colorC0}{HTML}{1f77b4}
\definecolor{colorC1}{HTML}{ff7f0e}
\definecolor{colorC2}{HTML}{2ca02c}

\newcolumntype{d}{>{\columncolor{colorC0!10}}c}
\newcolumntype{e}{>{\columncolor{colorC1!10}}c}
\newcolumntype{f}{>{\columncolor{colorC2!10}}c}

\makeatletter
\g@addto@macro\bfseries{\boldmath}
\makeatother

\hypersetup{
    colorlinks=true,
    linkcolor={red!50!black},
    citecolor={blue!50!black},
    urlcolor={blue!80!black}
}

\usepackage{colortbl}

\makeatletter
\newlength{\apb@width}
\newcommand{\autoparbox}[2][c]{\settowidth{\apb@width}{#2}\parbox[#1]{\apb@width}{#2}}

\makeatother

\definecolor{lightgray}{gray}{0.9}

\usepackage[framemethod=default]{mdframed}
\newmdenv[skipabove=7pt,
skipbelow=7pt,
rightline=false,
leftline=false,
topline=false,
bottomline=false,
backgroundcolor=gray!10,
linecolor=gray,
innerleftmargin=5pt,
innerrightmargin=5pt,
innertopmargin=5pt,
innerbottommargin=5pt,
leftmargin=0cm,
rightmargin=0cm,
linewidth=4pt]{eBox}

\usepackage[most]{tcolorbox}
\tcbset{colback=white, colframe=black,
        highlight math style= {enhanced, %<-- needed for the ?remember? options
            colframe=red,colback=red!10!white,boxsep=0pt}
        }
\definecolor{light-gray}{gray}{0.95}

\crefname{table}{Table}{Tables}
\crefname{equation}{Eq.}{Eqs.}
\crefname{appendix}{App.}{Apps.}
\crefname{section}{Sec.}{Secs.}
\crefname{figure}{Fig.}{Figs.}

\numberwithin{equation}{section}

\def\beq{\begin{equation}}
\def\eeq{\end{equation}}

\def\bea{\begin{eqnarray}}
\def\eea{\end{eqnarray}}

\def\Neff{N_{\rm eff}}
\def\beq{\begin{equation}}
\def\eeq{\end{equation}}
\def\bea{\begin{eqnarray}}
\def\eea{\end{eqnarray}}

\def\fnl{f_{\rm NL}}

\def\taunl{\tau_{\rm NL}}

\def\l{{\ell}}

\def\H{{\cal H}}

\def\k{{{\bm k}}}

\def\q{{\bm q}}

\def\v{{\bm v}}
\def\x{{\bm x}}

\def\vl{{\bm \ell}}
\def\n{{\hat{\bm n}}}
\def\r{{\hat{\bm r}}}
\def\vd{{\bm d}}

\def\ddm{\delta_{\rm dm}}
\def\fdm{f_{\rm dm}}

\def\db{\delta_{\rm b}}
\def\fb{f_{\rm b}}

\def\vn{{\nabla}}
\def\L{{\bm L}}

\def\he{{\hat{\bm e}}}
\def\hx{{\hat{\bm x}}}
\def\hy{{\hat{\bm y}}}
\def\hz{{\hat{\bm z}}}

\def\vA{{\bm A}}

\def\mnubao{\sum m_\nu^\mathrm{BAO}}
\def\mnuclustering{\sum m_\nu^\mathrm{clustering}}
\def\mnu{\sum m_\nu}
\def\mnutilde{\sum \tilde{m}_\nu}

\DeclareRobustCommand{\SkipTocEntry}[4]{}

\definecolor{colorTC}{rgb}{.2,.7,.2}

\definecolor{acolor}{rgb}{0.4, 0.2, 0.4}

\begin{document}

\begin{titlepage}
\setcounter{page}{1} \baselineskip=15.5pt
\thispagestyle{empty}
$\quad$
\vskip 70 pt

\begin{center}
%{\fontsize{20.74}{24} \bf  The Dark Force Awakens}
%{\fontsize{20.74}{24} \bf  Neutrinos After Dark}
%{\fontsize{20.74}{24} \bf Neutrinos in the Dark}
%{\fontsize{20.74}{24} \bf Dark Forces Gathering}
{\fontsize{20.74}{24} \bf New Interpretations of the Cosmological Preference for a Negative Neutrino Mass}
\end{center}

\vskip 20pt
\begin{center}
\noindent
{\fontsize{12}{18}\selectfont Peter W.~Graham$^{1,2}$, Daniel Green$^3$, and Joel Meyers$^4$} 
\end{center}

\begin{center}
\vskip 4pt
\textit{$^1$ {\small Leinweber Institute for Theoretical Physics at Stanford, Department of Physics,
Stanford University, Stanford, California 94305, USA}\\
$^2${\small Kavli Institute for Particle Astrophysics and Cosmology,
Department of Physics, Stanford University, Stanford, California 94305, USA}\\
$^3${\small Department of Physics, University of California, San Diego,  La Jolla, CA 92093, USA}\\
$^4${\small Department of Physics, Southern Methodist University, Dallas, TX 75275, USA}}
\end{center}

\vspace{0.4cm}
 \begin{center}{\bf Abstract}
  \end{center}

Recent observations of the cosmic microwave background (CMB) and baryon acoustic oscillations (BAO) show some tension with a $\Lambda$CDM cosmology. For one, the cosmological parameters determined by the CMB are at odds with the expansion history determined by latest BAO measurements. In addition, the combined data has placed uncomfortably strong constraints on neutrino mass. Both effects can be interpreted as negative neutrino mass, one describing the change to the expansion history and the other one describing enhanced lensing.  In this paper, we show the current tensions can be solved with a single change either to the lensing of the CMB or the expansion of the universe. We show additional lensing could arise from a variety of models with new light fields. However, these models rarely give the same signal in temperature and polarization, giving a concrete test of the scenario. Alternatively, dark sector models can explain the changes to the expansion by changing the evolution of the matter density. These models introduce new forces, giving rise to long range signals in the three-point statistics of galaxies. We discuss a range of other examples which all illustrate the pattern that additional signals should appear if these tensions are explained by beyond the Standard Model physics.

\noindent

\end{titlepage}
\setcounter{page}{2}

\restoregeometry

\begin{spacing}{1.2}
\newpage
\setcounter{tocdepth}{2}
\tableofcontents
\end{spacing}

\setstretch{1.1}
\newpage

\section{Introduction}

The current generation of cosmic microwave background (CMB) and galaxy surveys have reached the sensitivity where evidence for the cosmological effects of neutrino masses was expected to emerge~\cite{Font-Ribera:2013rwa}. Surprisingly, the current constraints from the CMB with the measurements of the expansion history from the Dark Energy Spectroscopic Instrument~\cite{DESI:2016fyo} (DESI), via the baryon acoustic oscillations (BAO), show no evidence for non-zero neutrino mass~\cite{DESI:2024hhd}. In fact, when including most recent results from the South Pole Telescope (SPT), the range of masses consistent with neutrino oscillations, $\sum m_\nu > 58$ meV~\cite{ParticleDataGroup:2024cfk}, is excluded at 98\% confidence\footnote{This constraint on $\mnu$ from the SPT analysis is tighter than that derived from a similar analysis with ACT data~\cite{ACT:2025tim}. The tighter constraint is at least partially driven by the choice of prior on the optical depth employed by SPT, as we discuss further in Sec.~\ref{subsec:AlteringCMB}.}~\cite{SPT-3G:2025bzu}. Even more surprisingly, observations seem to prefer large negative values of neutrino mass when the definition is extended beyond the positive (physical) regime~\cite{Craig:2024tky,Green:2024xbb,DESI:2025ejh}.

One of the central questions about this unexpected situation is whether the failure to measure a positive neutrino mass reflects a property of the expansion history, the clustering of matter, or some other effect altogether. As DESI has reported evidence for non-standard dark energy~\cite{DESI:2025ejh}, it is natural to think that these are all symptoms of the expansion history. DESI does measure a smaller value of $\Omega_m$ from its BAO measurement than Planck does from the primary CMB~\cite{Planck:2018vyg}. This might suggest that the preference for negative neutrino mass is just capturing this reduced amount of matter as determined by expansion~\cite{Loverde:2024nfi,Lynch:2025ine}.

Nevertheless, it is also the case that the cosmological signal that will best enable the detection of neutrino mass is the neutrino-induced suppression of the clustering of matter~\cite{Dolgov:2002wy,Lesgourgues:2006nd,Hannestad:2010kz,Green:2021xzn}. Neutrinos are too light to fall into the gravitational wells created by the dark matter and baryons, therefore slowing the rate of clustering as compared to cold dark matter~\cite{Bond:1980ha,Hu:1997mj}. This effect can be observed via the gravitational lensing of the CMB~\cite{Kaplinghat:2003bh} as a smaller amplitude for the lensing power spectrum in the presence of massive neutrinos. For CMB lensing, an apparent ``negative neutrino mass" should be understood as an increase to the lensing amplitude, rather than the expected suppression. This could arise physically from changes to the way structure is formed, or from a change to the statistics of the CMB in a way that is similar to mode coupling introduced by CMB lensing~\cite{Craig:2024tky}.

The challenge of measuring neutrino mass in cosmology, and the confusion over the source of the current tension, arises from the degeneracies between the lensing amplitude and other cosmological parameters~\cite{Allison:2015qca,CMB-S4:2016ple} (see also~\cite{Reboucas:2024smm,Allali:2024aiv,Herold:2024enb,Noriega:2024lzo,Bertolez-Martinez:2024wez,DESI:2025gwf}). Specifically, the amplitude of lensing is controlled by neutrino mass, the primordial amplitude of fluctuations (the measurement of which is degenerate with the optical depth to reionization), and the amount of matter in the universe. Changing the expansion history could change the inferred value of $\Omega_m$ and therefore change the expected amount of lensing without altering the physics of clustering~\cite{Pan:2015bgi}. The converse is also true: the inference of a tension in the expansion history is sensitive to the determination of $\Omega_m$ in the CMB, which depends on information gleaned from CMB lensing\footnote{Here it is important that CMB lensing affects temperature and polarization power spectra (e.g.~from peak smearing), in addition to the lensing reconstruction (four-point function). As a result, the measurement of $\Omega_m$ from $C_\ell^{TT}$, $C_\ell^{TE}$, and $C_\ell^{EE}$ still contains (two-point) lensing information, even without including the reconstructed lensing power $C_\ell^{\phi\phi}$ (four-point) likelihood.}. 

The goal of this paper is to better understand the interplay between expansion, clustering, and statistics in the tension between the CMB and DESI BAO measurements. We will introduce two distinct definitions of neutrino mass, one that alters the distance-redshift relation observed via BAO (similar to~\cite{Lynch:2025ine}), and another that changes the amplitude of CMB lensing (as in~\cite{Craig:2024tky,Green:2024xbb}). We will find that measurements of the CMB combined with DESI BAO suggest that allowing either one of these parameters to be negative can be sufficient to resolve the tension and allow for positive neutrino mass as inferred by the other. Yet, despite this fact, it is not a given that a model that alters clustering or expansion will lead to a good fit for all data. For example, we find that allowing for dynamical dark energy does not resolve the need for negative neutrino mass in clustering. 

We focus on trying to identify additional tests that might point to the origin of either the negative neutrino mass signal or a possible resolution of the tension. We begin with a focus on the clustering measurement. The signal of clustering is a change to the statistics of the CMB and this can be biased by other modulations of the CMB. We explore the space of models that can introduce lensing-like modifications to the CMB statistics and find that a key signal of non-lensing modulation is the difference between the amplitude measured by the $TT$ and $EB$ lensing estimators. It is straightforward to introduce isocurvature modes that would produce a bias in the lensing amplitude derived from the $TT$ estimator but these modifications would not generically generate $B$-mode polarization.

Next, we explore the possibility of dark forces that introduce changes to the evolution of the low-redshift universe. At face value, an attractive long-range dark force would increase matter clustering and therefore lead to excess CMB lensing. However, as shown in~\cite{Bottaro:2024pcb}, the back-reaction of the dark matter on the force mediating particle changes the time-evolution of dark matter density and has a larger impact on the distances inferred through BAO measurements. In this regard, concrete models of dark forces appear as a lower value of $\Omega_m$ at low redshift compared to the measurements from the CMB. Nevertheless, the long-range force leads to apparent violations of the equivalence principle and gives a unique signal in a multi-tracer galaxy bispectrum that is near the level that could, in principle, be measured by DESI.

A general conclusion of these models is that explanations of the current signal do not require both the clustering and BAO definitions of neutrino mass to be negative. Other analyses of negative neutrino mass have focused on the regime where they are equal to match the behavior of physical neutrinos (see eg ~\cite{Naredo-Tuero:2024sgf,Elbers:2024sha,DESI:2025ejh}), but we see that neither the data nor models compel us to consider this possibility. Moreover, since modifications to clustering and to expansion give rise to different observational signals, an explanation in terms of systematic effects is also unlikely to bias these two neutrino mass parameters in the same way. For these reasons, requiring neutrino mass to have the same effect on all observables is only well motivated for masses associated with physical neutrinos\footnote{New physics in the neutrino sector can lower the cosmological inference $0<\sum m_\nu < 58$~meV in a way that leaves all the signals of conventional neutrinos with  smaller masses~\cite{Craig:2024tky}. These models do not extend to negative masses. } where $\sum m_\nu \geq 0$.

This paper is organized as follows: in Section~\ref{sec:twomnus}, we introduce the two different definitions of neutrino mass and discuss their current constraints with CMB and BAO data. In Section~\ref{sec:mod}, we discuss various models that modulate the CMB temperature and polarization in a way that could bias the lensing measurement and give the appearance of negative neutrino mass in clustering. In Section~\ref{sec:darkforce}, we discuss the potential for long range forces in the dark sector to generate a negative neutrino mass signal. We show the phenomenology is strongly affected by backreaction and that, due to the change to the expansion history, the model ends up behaving largely as a negative neutrino mass signal in the BAO. In Section~\ref{sec:expanding}, we discuss other mechanisms that could yield the phenomenology needed to resolve the neutrino mass tension in this two-dimensional space. We conclude in Section~\ref{sec:conclusions}.

We include two appendices to expand on the results in the main text.  Appendix~\ref{app:fisher} gives additional details on how our Fisher matrix calculations were performed for both the CMB modulation and the large scale structure bispectrum. In Appendix~\ref{app:lensing} we review the flat sky limit of the CMB and CMB lensing.

\subsection{Executive Summary}
\label{sec:Brief_Summary}

Here we provide a brief summary of what will be discussed in more detail throughout the paper, highlighting some key definitions and results.

In Section~\ref{sec:twomnus}, we separate the effects of neutrino mass into two quantities, $\mnuclustering$, which alters the amplitude of CMB lensing as defined in Eq.~\eqref{eq:mnuclustering_def}, and $\mnubao$, which impacts cosmological distances as defined in Eq.~\eqref{eq:mnubao_def}.  The effects of physical neutrino mass on CMB and BAO observables are recovered for $\mnuclustering=\mnubao>0$.  Figure~\ref{fig:mnu_BAO_clustering} shows observational constraints on these parameters and the degeneracy between them, demonstrating how current data favors excess clustering and/or a modified expansion history compared to $\Lambda$CDM with the minimal sum of neutrino masses $\mnu>58$~meV.  Figure~\ref{fig:DX_mnubao} shows how $\mnubao$ affects the distances measured by BAO.  We show how allowing for dynamical dark energy significantly weakens constraints on $\mnubao$ in Figure~\ref{fig:mnu_BAO_clustering_w0wa} and Table~\ref{tab:Constraints}, though the preference for excess clustering ($\mnuclustering<0$) remains, particularly when the dark energy is restricted to satisfy the null energy condition.  In Section~\ref{subsec:Other_Tensions} we discuss the relation of the preference for negative neutrino mass to other cosmological tensions and show how the inferred value of $H_0$ varies with $\mnuclustering$ and $\mnubao$ in Figure~\ref{fig:mnu_H0}. In Section~\ref{subsec:Forecasts} we show forecasts for how future data is expected to improve constraints on neutrino mass parameters, with results presented in Figure~\ref{fig:fisher_mnu} and Table~\ref{tab:Forecasts}.

In Section~\ref{sec:mod}, we consider how new physics might bias our inference of the lensing amplitude, which could then show up in observational constraints as a preference for $\mnuclustering<0$.  In Section~\ref{subsec:Lensing_Bias}, we show how a modulation of the temperature fluctuations by a field $\chi$ defined in Eq.~\eqref{eq:Tchi}, modulation of the temperature gradient by a field $\sigma$ defined in Eq.~\eqref{eq:Tsigma}, or mode-coupling induced during inflation by a field $\varphi$ defined in Eq.~\eqref{eq:time_ind_modes} could lead to a bias in the temperature-based reconstruction of the CMB lensing potential $\phi$, shown in Eq.~\eqref{eq:phi_bias_result}.  However, we further show that this new physics will generally not lead to the same effects on CMB polarization as gravitational lensing and could thus be discriminated from lensing by, for example, comparing the reconstruction of the lensing potential from the $TT$ estimator to that from the $EB$ estimator, the sensitivity of which we forecast in Figure~\ref{fig:lensing_SN}.  In particular, we show that generating $B$-mode polarization requires non-trivial coupling at late times, since it arises from the curl of the baryon velocity (see Eq.~\eqref{eq:B_curl_vb}).  One could arrange for this by imposing a new force on the baryon-photon fluid, as in Eq.~\eqref{eq:sigma_vb_coupling}, though such new forces directly impacting the visible sector are typically quite constrained.

In Section~\ref{subsec:Isocurvature_Models}, we consider a set of specific models that could mimic the effects of lensing and describe the additional constraints that must be evaded in these scenarios.  We show how a field modulating the time at which inflation ends, coupled via a term like that shown in Eq.~\eqref{eq:S_int_sigma} is capable of producing the gradient modulation shown in Eq.~\eqref{eq:Tsigma}, and is therefore capable of biasing a temperature-based (though not polarization-based) reconstruction of the lensing potential.  We then consider models impacting physics around or after recombination that could lead to modifications to both temperature and polarization, which we treat with an effective modulation of the speed of light, as in Eq.~\eqref{eq:ceff}.  All such models are subject to a tight constraint shown in Eq.~\eqref{eq:speed_bound} from the observed agreement between the speed of gravitational waves and speed of light, which requires that modulations to the speed of light are strongly suppressed in the recent cosmological past.  Realizing a change to the effective speed of light with couplings to an axion-like field $a$, as shown in Eq.~\eqref{eq:axion_couplings} is subject to tight constraints from axion searches.  A two-index tensor $B_{\alpha\beta}$ coupling to photons like in Eq.~\eqref{eq:B_coupling} provides the correct structure, and could be realized by the quadratic coupling of a hidden photon $C^\mu$ as in Eq.~\eqref{eq:C_coupling} or by an effective cosmological solid described by three scalar field $\sigma^I$ with non-vanishing vacuum expectation values (VEVs) as in Eq.~\eqref{eq:sigma_vevs}.  Each of these scenarios is subject to further constraints described in the main text.

In Section~\ref{sec:darkforce}, we explore the possibility of new physics that could physically enhance the amplitude of cosmological clustering, rather than biasing our inference of clustering as measured through CMB lensing.  We show how a new long range force acting on dark matter would affect cosmological evolution, showing that it necessarily impacts both clustering and expansion.  We do this in two steps. First, in Section~\ref{subsec:Clustering_By_Hand} we work through the impact of a long range force on cosmological clustering ignoring the effect of the mediator on the background evolution.  We see in Eq.~\eqref{eq:Enhanced_Growth} that during matter domination, the matter growth function $D_m(\tau)=\tau^{2\gamma}$ is enhanced compared to $\Lambda$CDM cosmology $\gamma=1+\frac{6}{5}f_\mathrm{DM}^2\alpha^2$, where $f_\mathrm{dm}$ is the fraction of non-relativistic matter in the form of dark matter and $\alpha$ is the strength of the new long range force acting on dark matter relative to that of gravitation (see Eq.~\eqref{eq:Poisson}).  We further show that the apparent violation of the equivalence principle due to the long range force on dark matter produces a non-vanishing contribution to the tree-level bispectrum, Eq.~\eqref{eq:mmr_bispectrum}, that provides a compelling signature of this scenario.

However, since a new long range force must be mediated by a new light field, we must also include the effects of the new degrees of freedom on cosmological evolution.  We turn to this more complete treatment in Section~\ref{subsec:Backreaction}, where we consider coupling of scalar dark matter $\chi$ to a light scalar mediator $s$, described by Eq.~\eqref{eq:DM_action}.  The backreaction of the light mediator causes the density of the $\chi$ particles to redshift differently than ordinary non-relativistic matter as shown in Eq.~\eqref{eq:DM_scaling}, thus leading to a modified expansion history during matter domination; see Eq.~\eqref{eq:Modified_Expansion}.

A long range force acting on dark matter therefore impacts both cosmological clustering  and the cosmological expansion history.  In Section~\ref{subsec:Dark_Force_Observations}, we estimate how the impacts of a long range force can be mapped onto $\mnuclustering$ and $\mnubao$ in Eqs.~\eqref{eq:Dark_Force_mnuclustering} and \eqref{eq:Dark_Force_mnubao}, respectively, and compare to the observational constraints in Figure~\ref{fig:mnu_dark_force}.  The strength of the long range force, characterized by  $\beta\approx2\alpha^2$, picks out a particular line in the $\mnubao$-$\mnuclustering$ plane, and we show how our constraints on these parameters agree with previous observational constraints on $\beta$.  As discussed in Section~\ref{subsec:Clustering_By_Hand}, the long range force predicts signatures that can be seen in the galaxy bispectrum (Eq.~\eqref{eq:AAB_bispectrum}), for which we forecast the observational prospects in Figure~\ref{fig:bispectrum}.  We then go on to discuss complementary astrophysical constraints on new long range forces, showing that cosmology provides the best limits in the small coupling regime.

In Section~\ref{sec:expanding}, we consider other models of new physics that may impact our measurements of $\mnubao$ and $\mnuclustering$.  In Section~\ref{subsec:AlteringBAO}, we discuss how new ingredients like spatial curvature or negative energy density could modify the expansion history and thus impact $\mnubao$ constraints, and we show constraints on a scenario with 
spatial curvature in Figure~\ref{fig:mnu_omk}.  We also briefly discuss decaying dark matter, which impacts both the expansion history and clustering.  In Section~\ref{subsec:AlteringCMB}, we discuss how a biased measurement of the optical depth to reionization, either due to systematic error or new physics, would impact the measurement of $\mnuclustering$ and present one relevant example in Figure~\ref{fig:mnu_tauprior}.  We also discuss how a modification to the initial conditions through the presence of primordial isocurvature fluctuations could lead to enhanced clustering compared to the standard scenario, and we show the impact of isocurvature on the primary CMB spectra and on lensing in Figure~\ref{fig:iso}.  These scenarios provide examples of how we can gain useful insights about various new physics models by considering constraints on $\mnubao$ and $\mnuclustering$.

\section{A Neutrino Mass for Expansion and Clustering }\label{sec:twomnus}

The gravitational influence of massive neutrinos affects cosmological observables in multiple ways. In this Section, we will break down the physical impacts of neutrino mass on cosmological observables and analyze what current data suggests about the possible origins of a negative neutrino mass signal.

\subsection{Physical Signals of Massive Neutrinos}

Cosmic neutrinos decoupled from the thermal plasma at a temperature of around 1~MeV. At lower temperatures, the weak interactions that allowed neutrinos to exchange energy with electrons and positrons are inefficient compared to the expansion rate of the universe. Assuming $m_\nu\lesssim 250$~meV, cosmic neutrinos remained relativistic from decoupling through the epoch of recombination. This condition is consistent with existing constraints on neutrino mass from conservative CMB-only analyses~\cite{Planck:2018vyg} and laboratory measurements of beta decay endpoints~\cite{KATRIN:2024cdt}. For these reasons, we will assume throughout this work that all neutrino mass eigenstates are sufficiently light so as to have been relativistic at recombination. Given the mass-squared splittings of the neutrino mass eigenstates, as inferred from neutrino flavor oscillation experiments~\cite{ParticleDataGroup:2024cfk}, and the temperature of the cosmic neutrino background\footnote{The cosmic neutrino background is well-described up to small corrections~\cite{Akita:2020szl,Froustey:2020mcq,Bennett:2020zkv,Bond:2024ivb} by a relativistic Fermi-Dirac distribution with $T_\nu = (4/11)^{1/3}T_\gamma$ where $T_\gamma$ is the temperature of the CMB.}, at least two mass eigenstates of cosmic neutrinos should be non-relativistic today.  As a result, massive neutrinos contributed to the radiation density in the early universe and to the non-relativistic matter density during more recent cosmological history.  This transition from radiation-like behavior to matter-like behavior impacts both the cosmological expansion history and the clustering of matter.

Acoustic peak heights in the CMB power spectrum are sensitive to the baryon density $\Omega_\mathrm{b} h^2$ and cold dark matter density $\Omega_\mathrm{c} h^2$, while the angular scale of the peaks, $\theta_\star$, is set primarily by the distance to the surface of last scattering. The latter depends, in turn, upon the post-recombination expansion history and thus the total non-relativistic matter density at late times~\cite{Planck:2016tof}.  Measurements of the BAO scale as a function of redshift, as achieved in spectroscopic galaxy surveys like DESI, directly probe the late-time expansion history allowing a determination of the energy budget of the universe and an inference of $\Omega_m$, including the contribution from non-relativistic neutrinos.  Combined observations of the CMB and BAO thereby allow for a comparison of the baryon and cold dark matter densities to the total non-relativistic matter density.  This comparison allows for constraints on the neutrino mass, through the impact that non-relativistic neutrinos have on the expansion history~\cite{Pan:2015bgi,Loverde:2024nfi,Lynch:2025ine}.

The large thermal velocities of cosmic neutrinos cause a suppressed growth of structure on scales smaller than the free-streaming length of neutrinos, when compared to a universe containing only massless neutrinos.  This suppressed clustering results from both the fact that neutrinos themselves do not cluster on these scales, and that the baryons and cold dark matter exhibit reduced clustering in the presence of non-relativistic neutrinos. The latter arises because the neutrino density increases the expansion rate without a corresponding increase to the depth of the potential wells in which the matter overdensities reside~\cite{Lesgourgues:2006nd,Wong:2011ip,Lesgourgues:2012uu,Green:2021xzn,Green:2022bre}.  In the presence of massive neutrinos, the matter power spectrum takes the form
\begin{equation}
    P_{f_\nu}(k\gg k_\mathrm{fs},z) \approx \left(1-2 f_\nu - \frac{6}{5}f_\nu \log \frac{1+z_\nu}{1+z} \right) P_{f_\nu = 0}(k\gg k_\mathrm{fs},z) \, ,
    \label{eq:Pk_suppression}
\end{equation}
where $f_\nu$ is the fraction of non-relativistic matter in the form of neutrinos, $z_\nu$ is the redshift at which neutrinos become non-relativistic, and $k_{\rm fs}$ is the neutrino free-streaming scale. These quantities are given by
\begin{align}
 f_\nu = \Omega_\nu/\Omega_m &\approx 4\times 10^{-3} \left(\frac{\sum m_\nu}{58~\mathrm{meV}} \right) \\
z_\nu \approx 100  \left(\frac{m_\nu}{50~\mathrm{meV}}\right) \qquad & k_{\rm fs} = 0.04 \, h \, {\rm Mpc}^{-1} \times \frac{1}{1+z} \, \left(\frac{\sum m_\nu}{58 \, {\rm meV}}\right) \ .
\end{align}
The suppression of the power spectrum is calculated relative to $P_{f_\nu=0}(k)$, the matter power spectrum in a universe with vanishing neutrino mass. This would give a present-day matter power spectrum that is suppressed by about 3\% on scales smaller than the neutrino free-streaming length for the minimal sum of neutrino masses consistent with flavor oscillation measurements, $\sum m_\nu=58$~meV.

Measurement of the CMB lensing power spectrum provides a particularly valuable means to measure the amplitude of the matter power spectrum.  Weak gravitational lensing of the CMB is caused by cosmic structure that intervenes between our telescopes and the surface of last scattering, and measurements of CMB lensing thereby provide us with a map of the line-of-sight integrated matter density throughout the universe~\cite{Lewis:2006fu}.  The CMB lensing power spectrum on the scales to which current and near-future observations are most sensitive can be reliably computed with linear perturbation theory and is mostly insensitive to the details of baryonic effects that can be important for the formation of structure on smaller scales~\cite{McCarthy:2020dgq,McCarthy:2021lfp}.  Furthermore, CMB lensing is a purely gravitational effect, which allows for a direct measurement of the matter overdensities responsible for the deflection. This circumvents the uncertainties that can arise when estimating the matter power spectrum with the use of biased tracers such as the density of galaxies or counts of galaxy clusters~\cite{Green:2021xzn}.

In order to translate a measurement of the CMB lensing power spectrum into a constraint on the sum of neutrino masses, we compare the observed lensing power spectrum to what it would have been in a universe containing only massless neutrinos.  This requires that we tightly constrain the other parameters that contribute to the lensing amplitude.  Within $\Lambda$CDM+$\sum m_\nu$ cosmology, two especially important parameters are the primordial amplitude of scalar fluctuations as parameterized by $A_s$ and the matter fraction $\Omega_m$.  Measurements of overall amplitude of the CMB power spectra for $\ell\gtrsim 20$ tightly constrain the combination $A_s e^{-2\tau}$ where $\tau$ is the optical depth to reionization.  The optical depth $\tau$ can be measured via observation of the reionization bump in the CMB polarization power spectrum on large angular scales $\ell<20$, thereby partially breaking the degeneracy between $A_s$ and $\tau$.  It is unfortunately quite challenging to measure the large angular scale CMB polarization necessary to extract the optical depth with traditional ground-based observations, but there exist proposals to improve optical depth measurement using novel ground-based observations~\cite{Essinger-Hileman:2014pja}, from a balloon~\cite{Errard:2022fcm}, or with an upcoming satellite~\cite{NASAPICO:2019thw,LiteBIRD:2022cnt}.  The matter fraction $\Omega_m$ is best measured through its impact on the expansion history, and can be constrained with BAO observations as discussed above.

Whether we rely on their impact on the expansion history or on the growth of structure, cosmological constraints on neutrino masses are derived by combining multiple data sets (the CMB and BAO in the present case).  The tightest constraints are derived by using the available data to constrain all of the observable impacts of neutrino mass simultaneously.  However, we can gain some additional insight into what aspects of the data are responsible for the  constraints by treating the physical effects separately.  This is particularly valuable in the present situation where current data seems to be in tension with standard expectations and in fact favor a negative value of the neutrino mass.

We introduce two separate parameters designed to capture the observational effects of neutrino mass on cosmology.  One parameter, $\mnubao$, affects the distance-redshift relation as measured through observations of BAO.  The other parameter, $\mnuclustering$, affects the amplitude of the matter power spectrum on scales smaller than the neutrino free-streaming length.  Each of these parameters is constructed to match the effect of physical neutrino mass on the appropriate observables in the regime $\mnu>0$, but each parameter is allowed to take values independent of the other and is allowed to take negative values. This is implemented in practice by treating the response of each observable to changes in neutrino mass as a Taylor expansion about $\mnu=0$, such that 
\begin{align}
    C_L^{\phi\phi} =  \left.C_L^{\phi\phi}\right|_{\mnu=0} + \left.\frac{\partial  C_L^{\phi\phi}}{\partial \mnu}\right|_{\mnu=0} \mnuclustering \, , 
    \label{eq:mnuclustering_def} \\
    (D_X(z)/r_d) =  \left.(D_X(z)/r_d)\right|_{\mnu=0} + \left.\frac{\partial  (D_X(z)/r_d)}{\partial \mnu}\right|_{\mnu=0} \mnubao \, ,  
    \label{eq:mnubao_def}
\end{align} 
where in the second line, $r_d$ is the size of the sound horizon at the baryon drag epoch and $D_X\in[D_M,D_H,D_V]$ refer to the transverse comoving distance, the Hubble distance ($D_H(z) = c/H(z)$), and the angle-averaged distance ($D_V(z)=\left[zD_M(z)^2D_H(z)\right]^{1/3}$) to any redshift as measured by BAO.\footnote{Numerical derivatives of the observables are computed holding fixed $\theta_\star$, $\Omega_\mathrm{b}h^2$, and $\Omega_\mathrm{c}h^2$.  Note that this is subtly different than the definition of $\mnutilde$ used in Refs.~\cite{Craig:2024tky,Green:2024xbb} which was defined with derivatives that were calculated with fixed $H_0$ and $\Omega_mh^2$.  The choice of fixed parameters in calculating derivatives here is made to ensure that there is a non-vanishing impact of $\mnubao$ on the BAO observables, then $\mnuclustering$ is defined with the same fixed parameters for consistency.  We checked that the distinction between $\mnuclustering$ and $\mnutilde$ has little impact when deriving observational constraints via a Markov chain Monte Carlo analysis.}  By construction, predictions of physical neutrino mass are recovered for $\mnubao = \mnuclustering > 0$.  By defining separate parameters, we can disentangle the roles of expansion history and clustering on the constraints, and we can understand how parameter degeneracies affect our interpretation of the data.

%%%%%%%%%%%%%%%%%
\subsection{Observational Constraints}
\label{subsec:constraints}
%%%%%%%%%%%%%%%%%

Our two neutrino mass parameters are implemented in a modified version of \texttt{CAMB}~\cite{Lewis:1999bs,Howlett:2012mh}, used for our Boltzmann calculations.  We use the likelihood for CMB temperature and polarization from Planck's 2018 data release~\cite{Aghanim:2019ame} and ACT DR6~\cite{ACT:2025fju} with the \texttt{SRoll2} analysis of large-scale polarization~\cite{Pagano:2019tci}, along with the combination of ACT DR6 CMB lensing~\cite{ACT:2023dou,ACT:2023kun} and Planck CMB lensing~\cite{Carron:2022eyg}. We utilize also DESI DR2 BAO~\cite{DESI:2025zgx}.  We refer to this data combination as `Planck+ACT+DESI' throughout the paper.  We adopt the precision settings for \texttt{CAMB} that match those used in the ACT DR6 cosmological analysis~\cite{ACT:2025tim}, including the use of \texttt{CosmoRec}~\cite{Chluba:2010ca,Chluba:2010fy} for recombination calculations.\footnote{Due to the significantly higher computational resource requirements in non-flat models, we reduced the \texttt{lens\_potential\_accuracy} parameter in \texttt{CAMB} to 2 in the analysis involving spatial curvature parameter $\Omega_K$, while we set this parameter to 8 for all other analyses.}   Our parameter constraints were obtained with \texttt{cobaya}~\cite{Torrado:2020dgo}, using the Markov chain Monte Carlo sampler adapted from \texttt{CosmoMC}~\cite{Lewis:2002ah,Lewis:2013hha} with the fast-dragging procedure~\cite{Neal:2005uqf}.  Results were analyzed and plotted using \texttt{GetDist}~\cite{Lewis:2019xzd}.

%%%%%%%%%%%%
\begin{figure}[t!]
    \centering
    \includegraphics[width=0.95\textwidth]{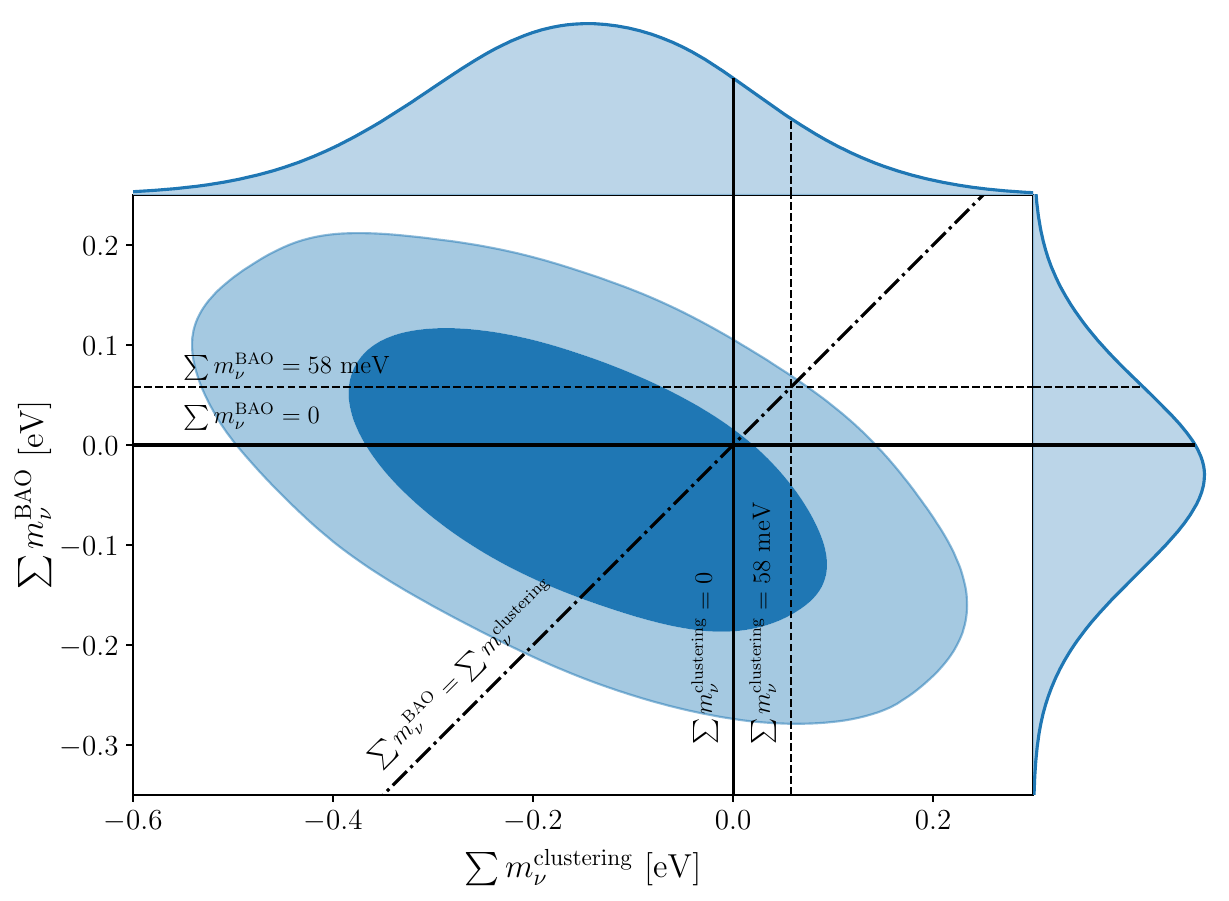}
    \caption{Constraints on $\mnubao$ and $\mnuclustering$ in $\Lambda$CDM+$\mnubao$+$\mnuclustering$ cosmology using data from Planck+ACT+DESI.  The posterior peaks at negative values for both $\mnubao$ and $\mnuclustering$, indicating that data prefers both modified distances and enhanced clustering compared to a universe with only massless neutrinos, opposite from the expected impact of positive neutrino mass for both parameters.  The direction of degeneracy between the two neutrino mass parameters shows that physics altering either the expansion history or the clustering could bring the behavior of the other in line with the expectation from positive neutrino mass.  For example, new physics causing enhanced clustering ($\mnuclustering<0$) would shift parameter inferences such that the observed expansion history is consistent with expectations of a model containing the minimal positive neutrino mass inferred from flavor oscillation experiments ($\mnubao>58$~meV) at less that $1\sigma$.}
    \label{fig:mnu_BAO_clustering}
\end{figure}
%%%%%%%%%%%%%

\begin{figure}[t!]
    \centering
    \includegraphics[width=0.65\textwidth]{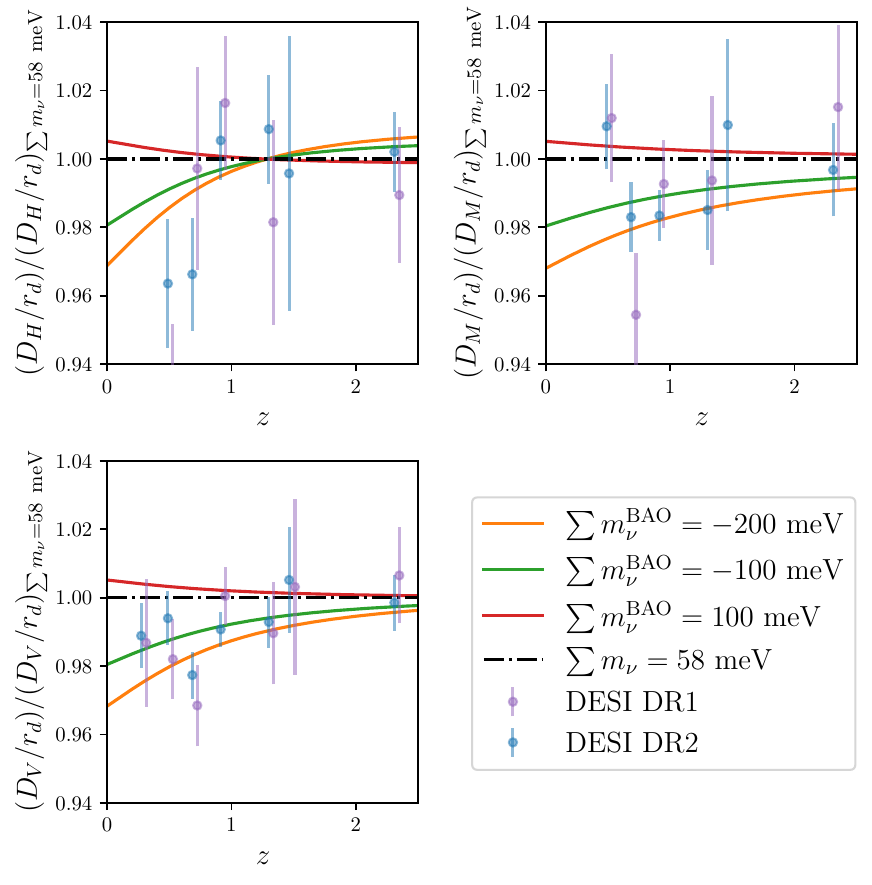}
    \caption{Effect of the parameter $\mnubao$ on the distances measured by BAO surveys compared to the observational errors of DESI DR1 and DESI DR2.  Distances are shown relative to those predicted from the Planck best fit $\Lambda$CDM cosmology.}
    \label{fig:DX_mnubao}
\end{figure}

In Figure~\ref{fig:mnu_BAO_clustering} we show constraints derived from Planck+ACT+DESI on the neutrino mass parameters in the $\Lambda$CDM+$\mnubao$+$\mnuclustering$ cosmology.  We see that each of $\mnubao$ and $\mnuclustering$ have best fit values below zero, indicating that the data favors modified cosmological distances and enhanced clustering compared to a universe with only massless neutrinos, contrary to the expected effects of massive neutrinos for both the expansion history and the clustering of matter.  There are further lessons to be learned from the degeneracy direction between the two parameters.  The shape of the constraints implies that enhanced clustering ($\mnuclustering<0$) would allow for an expansion history consistent with the minimal neutrino mass ($\mnubao>58$~meV).  Conversely, a modified expansion history that causes altered cosmological distances as measured by BAO ($\mnubao<0$, see Fig.~\ref{fig:DX_mnubao}) would allow for the suppression of clustering expected from positive neutrino mass ($\mnuclustering>$58~meV). 

%%%%%%%%%%%%
%%%%%%%%%%%%%

%%%%%%%%%%%%
\begin{figure}[t!]
    \centering
    \includegraphics[width=0.95\textwidth]{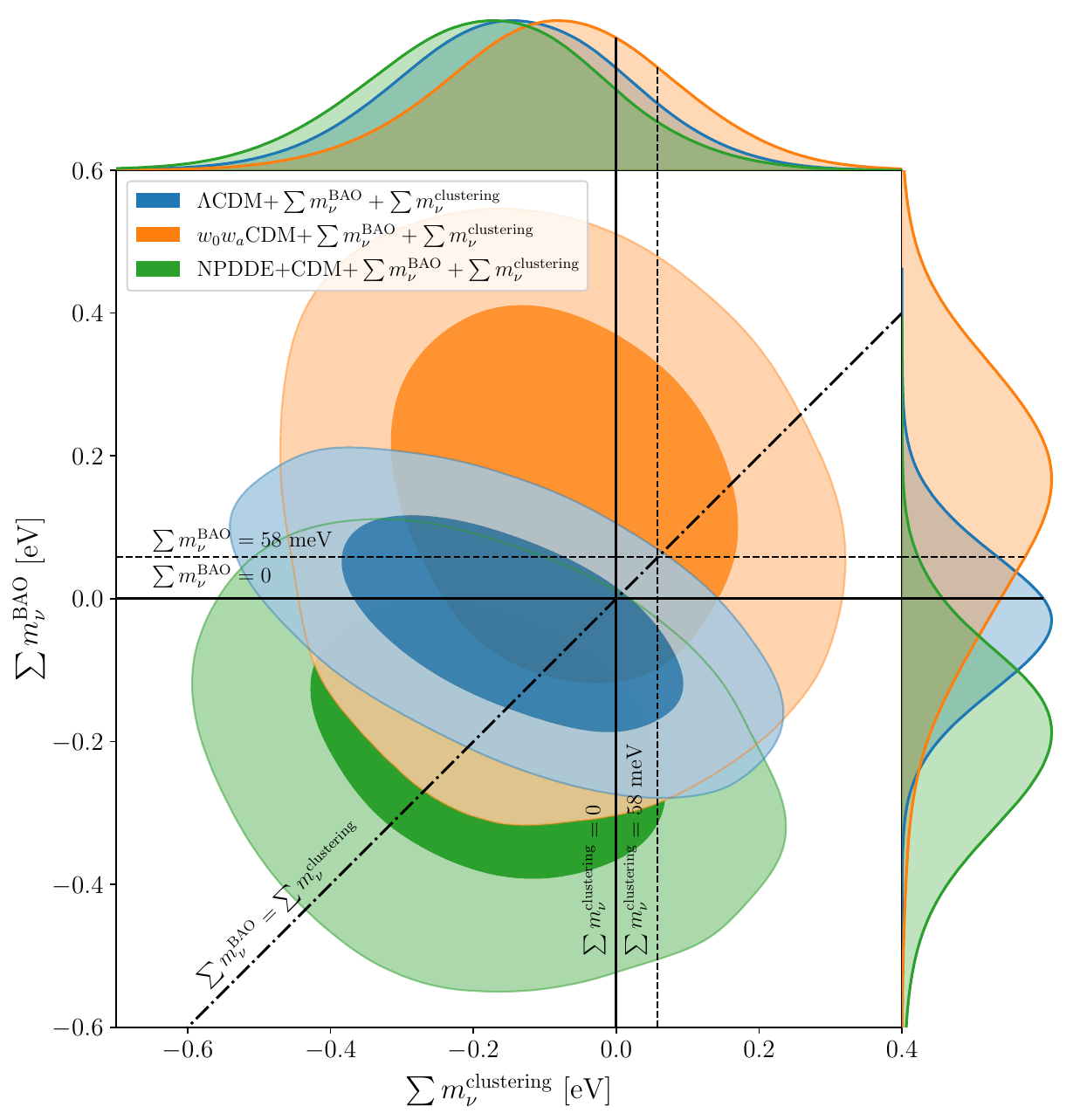}
    \caption{Same as Figure~\ref{fig:mnu_BAO_clustering}, including also dynamical dark energy models.  One can see that allowing for dynamical dark energy primarily affects constraints on the parameter $\mnubao$ with a relatively small change to $\mnuclustering$.  The case of non-phantom dynamical dark energy (NPDDE) shows a stronger preference for negative neutrino mass, since the behavior of both clustering and expansion due to dark energy obeying the null energy condition is opposite that expected from positive neutrino mass. }
    \label{fig:mnu_BAO_clustering_w0wa}
\end{figure}
%%%%%%%%%%%%%

However, not all models that modify the expansion history will lead to a preference for positive neutrino mass as measured through clustering.  As emphasized by the DESI collaboration~\cite{DESI:2024mwx,DESI:2025zgx}, models involving evolving dark energy provide a better fit to various data combinations than does $\Lambda$CDM cosmology.  In Fig.~\ref{fig:mnu_BAO_clustering_w0wa}, we show constraints on the parameters $\mnubao$ and $\mnuclustering$ in cosmological models including dynamical dark energy, one in which $w_0$ and $w_a$ are free, and another in which  $w_0$ and $w_a$ are allowed to vary subject to the constraint that the dark energy always satisfies the null energy condition, i.e.~that $w(z)\geq -1$ throughout cosmic history.  This latter case, called here non-phantom dynamical dark energy (NPDDE), is implemented by requiring $w_0\geq -1$ and $w_0+w_a \geq -1$, ensuring that $w(z) = w_0 + w_a \frac{z}{1+z} \geq 1$ for all redshifts $z$~\cite{Vagnozzi:2018jhn}.  One can see that compared to the posteriors in $\Lambda$CDM+$\mnubao$+$\mnuclustering$ cosmology, the constraints on the neutrino mass parameters are weakened in models containing dynamical dark energy, as would be expected when marginalizing over additional parameters.  Additionally, in the $w_0w_a$ model, the peak of the $\mnubao$ posterior shifts to positive values, while the NPDDE model prefers more negative values for $\mnubao$.  On the other hand, the $\mnuclustering$ posterior remains peaked at negative values in all three models.\footnote{The analysis presented in Ref.~\cite{Green:2024xbb} corresponds to finding the posterior for $\mnuclustering$ while fixing $\mnubao=0$, which then more strongly favors $\mnuclustering<0$ in the NPDDE model.}

In Table~\ref{tab:Constraints}, we show constraints derived from Planck+ACT+DESI on neutrino mass and dark energy parameters in various cosmological models.  The first row shows the traditional analysis described by $\Lambda$CDM cosmology with a physical neutrino mass restricted to be positive $\mnu \geq 0$.  We show several other models including $\mnuclustering$ and $\mnubao$.  The case with $\mnuclustering=\mnubao>0$ most closely mimics the standard analysis, and the constraints in those cases match very well, as would be expected.  The results show a consistent preference for $\mnuclustering$ to be below 58~meV (and less than 0 in all models except one with dynamical dark energy where $\mnuclustering=\mnubao$), indicating that the data exhibits excess clustering as compared to the expectation for a universe with the minimal sum of neutrino masses.  
\vskip 10pt

The summary of the current data is that either $\mnuclustering < 0$ with $\mnubao= 58$~meV or $\mnubao < 0$ with $\mnuclustering= 58$~meV could be sufficient to explain both the tensions in lensing amplitude and BAO distances. This, for example, is consistent with the observation that a higher value of the optical depth could resolve all these tensions~\cite{Sailer:2025lxj,Jhaveri:2025neg}. However, as seen with the case of dynamical dark energy, this does not imply that any model that modifies either the BAO or the lensing amplitude will actually push the central values of $\mnu$ towards the physical value. Adding additional parameters generally reduces the significance of the tension by increasing uncertainties, but may leave the preference of negative neutrino mass in the posteriors. As we explore these and other tensions and models to address them, we will see the pattern that $\mnuclustering <  0$ tends to remain the best fit to the data.

%%%%%%%%%%%%%%%%%%
% Cosntraint table
%%%%%%%%%%%%%%%%%%
\begin{table}
    \centering
    \scriptsize
    \begin{tabular}{l||ccc|cc}
        & \cellcolor{gray!15} $\sum m_\nu$ & \cellcolor{gray!15} $\mnuclustering$ &  \cellcolor{gray!15}$\mnubao$ & \cellcolor{gray!15} $w_0$ & \cellcolor{gray!15} $w_a$  \\
       \hline \hline
        \cellcolor{gray!15} $\Lambda$CDM+$\sum m_\nu$ & $<32$ & - & - & - & -  \\
        \hline  \cellcolor{gray!15} $\Lambda$CDM+$(\mnuclustering>0)$ & - & $<74$ & - & - & -  \\
        \cellcolor{gray!15} $\Lambda$CDM+$\mnuclustering$ & - & $-180\pm130$ & - & - & -  \\
        \hline  \cellcolor{gray!15} $\Lambda$CDM+$(\mnuclustering=\mnubao>0)$ & - & \multicolumn{2}{c|}{$<33$} & - & -  \\
        \cellcolor{gray!15} $\Lambda$CDM+$(\mnuclustering=\mnubao)$ & - & \multicolumn{2}{c|}{$-71\pm57$} & - & -  \\
        \hline \cellcolor{gray!15} $\Lambda$CDM+$\mnuclustering+\mnubao$ & - & $-150\pm160$ & $-32\pm99$ & - & -  \\
        \hline \cellcolor{gray!15} $w_0w_a$CDM+$(\mnuclustering=\mnubao)$ & - &  \multicolumn{2}{c|}{$24^{+120}_{-95}$} & $-0.54^{+0.24}_{-0.29}$ & $-1.30^{+0.91}_{-0.72}$  \\
        \cellcolor{gray!15} NPDDE+CDM+$(\mnuclustering=\mnubao)$ & - &  \multicolumn{2}{c|}{$-196^{+93}_{-71}$} & $< -0.905$ & $0.036^{+0.089}_{-0.14}$  \\
        \hline \cellcolor{gray!15} $w_0w_a$CDM+$\mnuclustering+\mnubao$ & - &   $-90^{+170}_{-150}$ & $130^{+190}_{-160}$ & $-0.50^{+0.26}_{-0.30}$ & $-1.46^{+0.96}_{-0.82}$  \\
        \cellcolor{gray!15} NPDDE+CDM+$\mnuclustering+\mnubao$ & - &  $-190\pm 160$ & $-210^{+140}_{-120}$ & $< -0.906$ & $0.046^{+0.091}_{-0.15}$  \\
    \end{tabular}
    \caption{Constraints on neutrino mass and dark energy parameters in various cosmological models using current Planck+ACT+DESI data.  We show here $1\sigma$ bounds with neutrino mass parameters expressed in units of meV.}
    \label{tab:Constraints}
\end{table}
%%%%%%%%%%%%%%

%%%%%%%%%%%%%%
%% Other Tensions
%%%%%%%%%%%%%%

\subsection{Relation to Other Tensions}
\label{subsec:Other_Tensions}

%%%%%%%%%%%%%%
%% Hubble Tension
%%%%%%%%%%%%%%

\subsubsection{Hubble Tension}
\label{subsec:Hubble_Tension}

\begin{figure}[t!]
    \centering
    \includegraphics[width=0.95\textwidth]{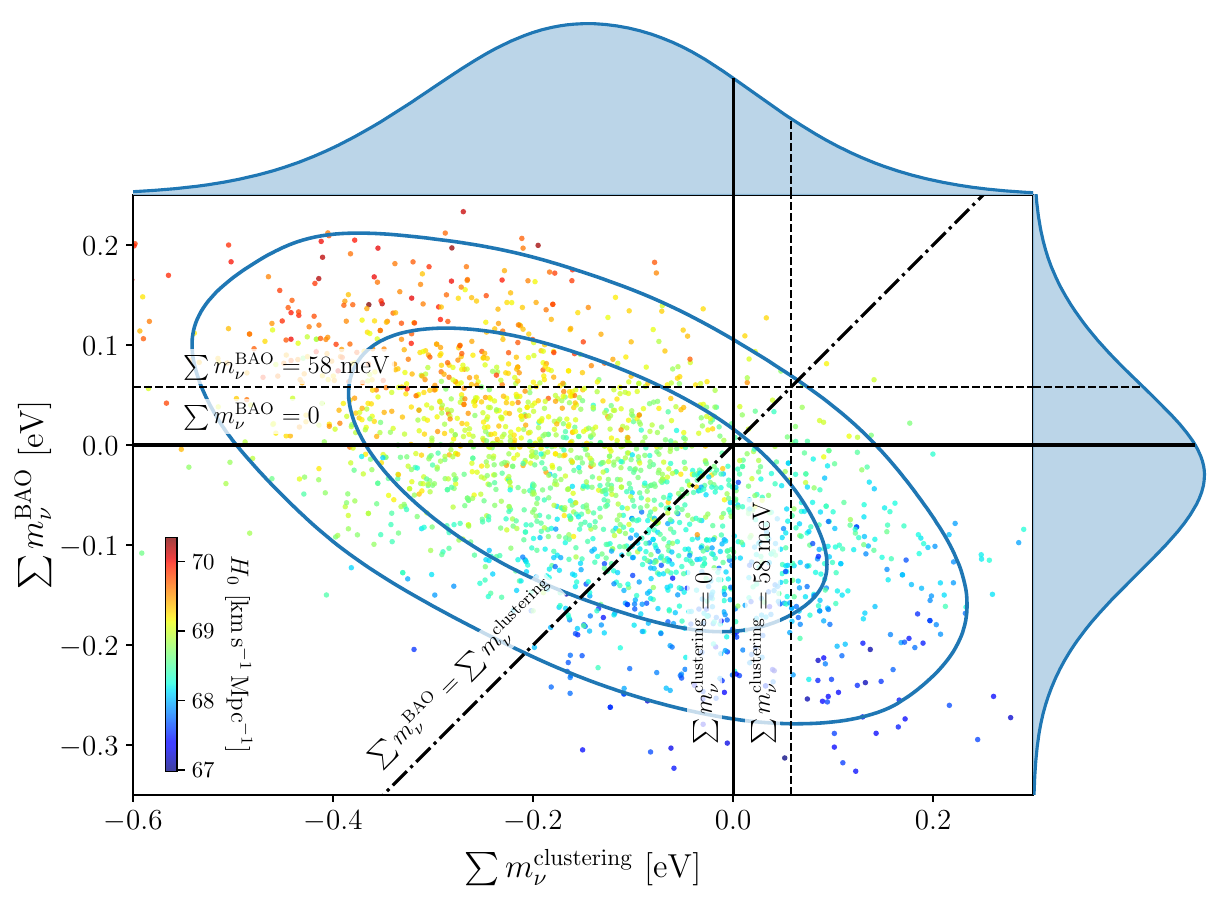}
    \caption{Constraints on neutrino mass parameters in $\Lambda$CDM+$\mnubao$+$\mnuclustering$ cosmology overlaid with the value of the Hubble constant for various samples in the parameter space.  One can see that samples with positive $\mnubao$ and negative $\mnuclustering$ satisfy the constraints and favor larger values of $H_0$. }
    \label{fig:mnu_H0}
\end{figure}

Local measurements of the expansion rate~\cite{Riess:2021jrx,Freedman:2024eph} tend to favor values of $H_0$ larger than those inferred from observations of the CMB~\cite{Planck:2018vyg,SPT-3G:2022hvq,SPT-3G:2024atg,ACT:2025fju} and BAO~\cite{Cuceu:2019for,DESI:2025zgx} within $\Lambda$CDM cosmology.  It is natural to ask how the modifications to clustering (via $\mnuclustering$) and distances measured by BAO (parametrized by $\mnubao$) impact the inference of $H_0$.  In Fig.~\ref{fig:mnu_H0} we show constraints on these parameters along with the sampled values of $H_0$.  We find that $\mnubao$ tends to negatively correlate with the inferred value of $H_0$.  Given the partial degeneracy between $\mnubao$ and $\mnuclustering$, we therefore find that models that provide a good fit to the data by exhibiting excess clustering ($\mnuclustering<0$) will favor larger values of $H_0$, while those that address the preference for negative neutrino mass through a change to the distances inferred from BAO ($\mnubao<0$) will favor smaller values of $H_0$.  Note, however, that the parameter $\mnubao$ changes only the BAO-inferred distances; a physical, self-consistent modification to the expansion history may exhibit different behavior for $H_0$ than what is shown in Fig.~\ref{fig:mnu_H0}.

%%%%%%%%%%%%%%
%% H_0 samples figure
%%%%%%%%%%%%%%
%%%%%%%%%%%%
%%%%%%%%%%%%%

%%%%%%%%%%%%%%
%% Phase Shift
%%%%%%%%%%%%%%

\subsubsection{Phase Shift of Acoustic Peaks}
\label{subsec:Phase_Shift}

Free-streaming radiation in the early universe causes a shift in the phase of acoustic peaks in the CMB power spectra~\cite{Bashinsky:2003tk,Follin:2015hya,Baumann:2015rya,Montefalcone:2025unv} and in the BAO spectrum~\cite{Bashinsky:2003tk,Baumann:2017lmt,Baumann:2019keh,Green:2020fjb}. Analysis of DESI BAO data led to the surprising conclusion that the measured BAO phase shift is larger than would be expected due to the free-streaming of Standard Model neutrinos by $2.6\sigma$ in $\Lambda$CDM using Planck CMB + DESI DR1 BAO data~\cite{Whitford:2024ecj} (the parameter characterizing the phase shift $\beta_\phi$ is measured to be $\beta_\phi=2.69^{+0.59}_{-0.66}$ while the Standard Model expectation is $\beta_\phi=1$).  However, it is shown in Ref.~\cite{Whitford:2024ecj} that the preference for excess phase shift weakens to $1.9\sigma$ when the lensing amplitude $A_\mathrm{lens}$ is also allowed to vary in the analysis ($\beta_\phi=2.05\pm0.55$).  Since excess lensing is favored by CMB-only analyses when $A_\mathrm{lens}$ is allowed to vary~\cite{Planck:2018lbu}, this indicates that models that give $\mnuclustering<0$ may also help to mitigate the preference for a larger than expected phase shift.  Conversely, in $w_0w_a$CDM cosmology, the best fit value of the phase shift is even larger than in $\Lambda$CDM though with wider error bars giving a $2.5\sigma$ tension with the expectation from standard neutrinos ($\beta_\phi=3.7^{+1.2}_{-1.1}$).

These results suggest that models that allow $\mnuclustering < 0$ are likely to improve and potentially resolve this tension in the phase shift. Although these would seem to be unrelated, the BAO phase shift measurement uses a prior on the $\Lambda$CDM parameters from the primary CMB (marginalized over $\Neff$) to constrain the BAO frequency in order to give a more precise measurement of the phase. Because the frequency and phase are degenerate, shifts in the cosmological parameters (such as $H_0$) also shift the inferred values of the phase. However, since the phase $\beta_\phi$ is determined from the combined measurement of the BAO in all the redshift bins, a reanalysis is required to determine the precise impact that $\mnuclustering$ and $\mnubao$ have on the measurement of $\beta_\phi$.

%%%%%%%%%%%%%%
%% Forecasts
%%%%%%%%%%%%%%

\subsection{Constraints with Future Data}
\label{subsec:Forecasts}

%%%%%%%%%%%%%%%%%%
% Forecast table w/physical mnu
%%%%%%%%%%%%%%%%%%%%%
\begin{table}
    \centering
    \scriptsize
    \begin{tabular}{c|c||>{\cellcolor{gray!15}}ccc|>{\cellcolor{gray!15}}ccc}
         \multicolumn{2}{c||}{ }   & $\mnu$ & $\mnuclustering$ & $\mnubao$ & $\mnu$ & $\mnuclustering$ & $\mnubao$ \\
         \hline
         $\Delta_T$~[$\mu$K-arcmin] & $\sigma(\tau)$  & \multicolumn{3}{c|}{+DESI 5yr} & \multicolumn{3}{c}{+DESI 5yr+Spec-S5} \\
        \hline \hline
        10 & 0.006  & 37 & 85 & 62 & 36 & 85 & 60 \\
        6 & 0.006  & 37 & 81 & 59 & 35 & 81 & 57 \\
        1 & 0.006  & 36 & 65 & 51 & 34 & 65 & 49 \\
        \hline
        10 & 0.002  & 27 & 69 & 62 & 26 & 69 & 60 \\
        6 & 0.002  & 26 & 65 & 59 & 25 & 65 & 57 \\
        1 & 0.002  & 25 & 53 & 51 & 24 & 53 & 49 \\
    \end{tabular}
    \caption{Forecasts for $1\sigma$ errors on $\mnu$ in units of meV for $\Lambda$CDM+$\mnu$ cosmology shown in gray and on $\mnuclustering$ and $\mnubao$ in units of meV for $\Lambda$CDM+$\mnubao$+$\mnuclustering$ cosmology.  For the forecasts shown here, we assume that CMB surveys are modeled by white noise spectra at $\ell\geq30$ with a $1.4$~arcmin beam covering $f_\mathrm{sky}=0.5$. }
    \label{tab:Forecasts}
\end{table}
%%%%%%%%%%%%%%%%%%%%%

We can also address how future cosmological data will improve our constraints on the observable imprints of neutrino mass.  We carried out a set of Fisher forecasts showing how constraints on neutrino mass parameters will be improved by future small-scale CMB observations, large-scale CMB polarization observations (affecting the inference of the optical depth $\tau$), and BAO observations.  We treated CMB experiments with simple white noise characterized by $\Delta_P=\sqrt{2}\Delta_T$, assuming a beam size of 1.4~arcmin, taking the sky coverage to be $f_\mathrm{sky}=0.5$, and restricting $30<\ell<5000$ for $TE$ and $EE$ spectra, while for the $TT$ power spectrum we use $\ell_\mathrm{max}^{TT}=3000$.  We show results for $\Delta_T=10$~$\mu$K-arcmin (comparable to the white-noise level of ACT DR6~\cite{ACT:2025xdm,ACT:2025fju}), $\Delta_T=6$~$\mu$K-arcmin (comparable to the white-noise level anticipated from Simons Observatory~\cite{SimonsObservatory:2018koc,SimonsObservatory:2025wwn}), and $\Delta_T=1$~$\mu$K-arcmin (comparable to the white-noise level proposed for CMB-S4~\cite{CMB-S4:2016ple,Abazajian:2019eic}).  We show forecasts for the current uncertainty on optical depth from Planck $\sigma(\tau)=0.006$~\cite{Planck:2018vyg,Pagano:2019tci} and also for a future cosmic variance limited measurement of the optical depth $\sigma(\tau)=0.002$ as can be obtained by a future CMB satellite like LiteBIRD~\cite{LiteBIRD:2022cnt} or PICO~\cite{NASAPICO:2019thw} or balloon-borne experiment like Taurus~\cite{Taurus:2024dyi}.  We consider the full 5-year DESI BAO results with anticipated errors on $D_M(z)/r_d$ and $D_H(z)/r_d$ given by Ref.~\cite{Font-Ribera:2013rwa}.  We also include BAO from a future Stage-5 Spectroscopic survey providing 0.1\% precision on BAO parameters in the redshift range $2.1<z<4.5$~\cite{Spec-S5:2025uom}.
Boltzmann calculations were carried out with the same modified version of \texttt{CAMB} used for parameter constraints above, lensing reconstruction noise was calculated using \texttt{CLASS\_delens}\footnote{\url{https://github.com/selimhotinli/class_delens}}, and Fisher calculations were implemented with \texttt{FisherLens}\footnote{\url{https://github.com/ctrendafilova/FisherLens}}~\cite{Hotinli:2021umk}.

\begin{figure}[t!]
    \centering
    \includegraphics[width=0.95\textwidth]{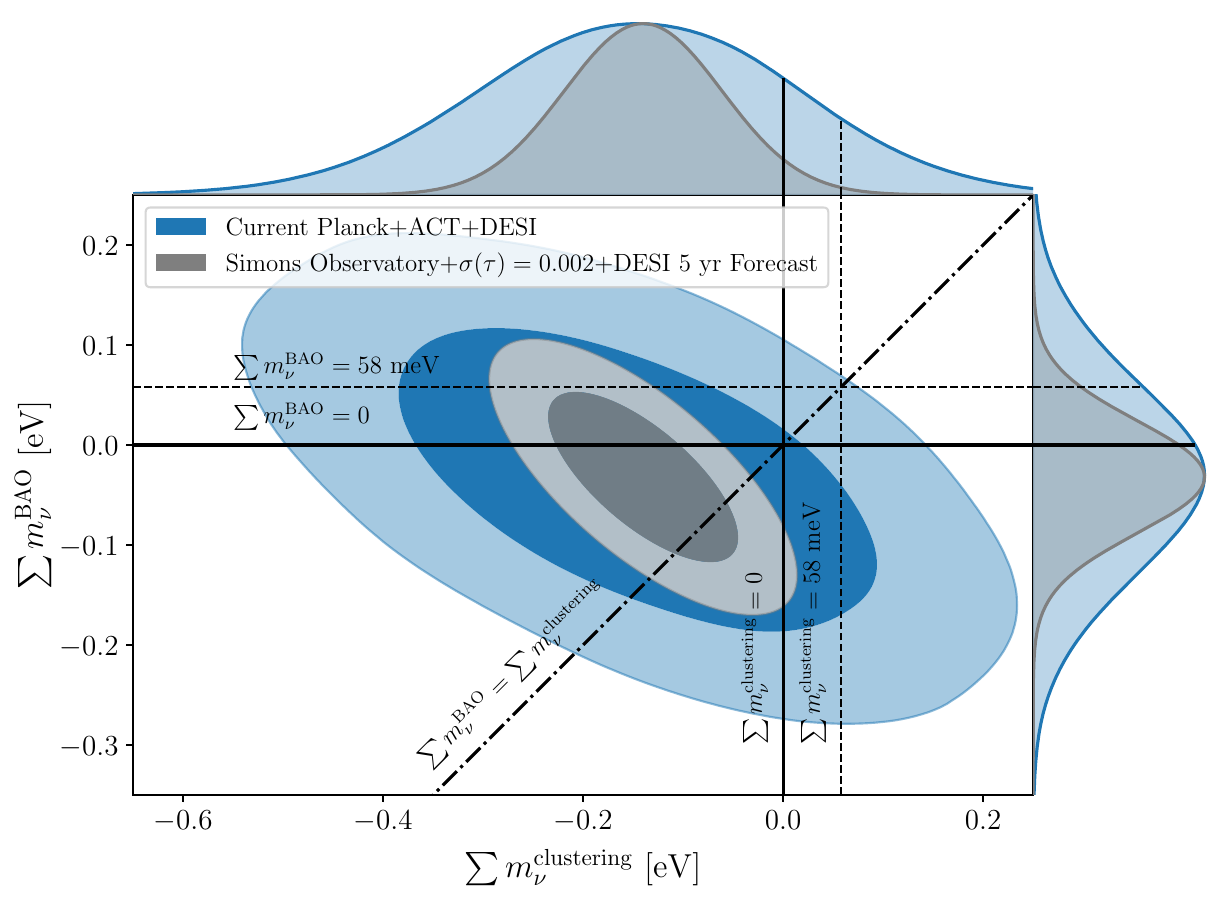}
    \caption{Forecast for future constraints on $\mnuclustering$ and $\mnubao$ expected from a CMB survey like Simons Observatory~\cite{SimonsObservatory:2018koc} (modeled here with white noise level $\Delta_T=6~\mu$K-arcmin at $\ell\geq30$, a 1.4~arcmin beam, covering $f_\mathrm{sky}=0.5$), BAO from the full DESI survey~\cite{Font-Ribera:2013rwa}, and a cosmic variance limited measurement of the optical depth $\sigma(\tau)=0.002$, compared to current constraints from Planck+ACT+DESI.}
    \label{fig:fisher_mnu}
\end{figure}

We show the results of these forecasts in Table~\ref{tab:Forecasts} for two different models.  First we show forecasts for $1\sigma$ constraints on the physical neutrino mass in $\Lambda$CDM+$\mnu$ cosmology.\footnote{It may be noted that the forecasts for $\sigma\left(\mnu\right)$ presented here are less optimistic than those shown elsewhere, such as in Refs.~\cite{Allison:2015qca,CMB-S4:2016ple}.  This difference can be traced to the treatment of the BAO uncertainties.  The forecasts employed here use uncertainties on $D_M(z)/r_d$ and $D_H(z)/r_d$ at each redshift rather than a single combined `volume distance' parameter $D_V(z)/r_d= [zD_M(z)^2D_H(z)]^{1/3}/r_d$ at each redshift, and employing the latter treatment of BAO recovers the more optimistic expectations for neutrino mass constraints.  See also Ref.~\cite{Font-Ribera:2013rwa} for discussion of this distinction.}  One can see that with DESI 5-year data included, these constraints are mostly limited by the current uncertainty on the optical depth $\tau$, and the constraints only marginally improve with lower-noise CMB observations and/or high redshift BAO information.  Next we show $1\sigma$ constraints on $\mnuclustering$ and $\mnubao$ in $\Lambda$CDM+$\mnubao$+$\mnuclustering$ cosmology. These parameters exhibit tighter constraints with lower noise CMB observations, and $\mnuclustering$ benefits from a better determination of the optical depth.  See Figure~\ref{fig:fisher_mnu} for a comparison of current constraints with forecasted constraints on $\mnubao$ and $\mnuclustering$.  High-redshift BAO observations do not significantly improve the constraints on either parameter (and we confirmed that this conclusion is robust to rather wide variations in the uncertainty on BAO parameters in the redshift range $2.1<z<4.5$).  However, high-redshift BAO information does improve constraints on $\mnubao$ significantly in models where the dark energy is dynamical (see Appendix~\ref{app:DE_forecasts}).

%%%%%%%%%%%%
%%%%%%%%%%%%%

It is noteworthy that the forecasted constraints on the physical neutrino mass are significantly tighter than the constraints on either of $\mnubao$ or $\mnuclustering$. This can be understood from the partial degeneracy of these parameters exhibited in Figure~\ref{fig:fisher_mnu}.  We can see there that the line that describes physical neutrino mass ($\mnubao=\mnuclustering$) lies in the direction nearly orthogonal to the direction of degeneracy between the parameters, meaning that the uncertainty on $\mnu$ is reasonably approximated by the width of the ellipse on its minor axis, rather than being dominated by the uncertainty in one parameter or the other.

\section{Lensing and Statistics}\label{sec:mod}

From the point of view of a map of the CMB, lensing is a local change to the statistics of the temperature and polarization fluctuations. It correlates different angular scales, allowing one to reconstruct the lensing potential from this apparent deviation from statistical isotropy. Similar deviations can be achieved via primordial non-Gaussianity or isocurvature perturbations. However, it is also important that measurement of the lensing amplitude relies on multiple distinct statistical effects associated with each of the CMB two-point statistics. Other sources of statistical anisotropy need not reproduce all of the effects of CMB lensing. Our goal in this section is to determine what types of changes to the statistics could appear as a lensing excess and how we can distinguish them from true lenses.

In this section, we will work in the flat sky limit so that angular direction on the sky $\n = (x,y)$ is a two-dimensional vector and $T(\n) = T(x,y)$. Similarly, we will define the Fourier transform
\beq
T(\vl) = \int d^2 n \, e^{-i \n \cdot \vl} T(\n) \ .
\eeq
We will define the length of vectors without the bold, e.g.~$|\vl| \equiv \ell$. 

An essential difference between lensing and other changes to the statistics will be the impact on polarization. CMB polarization transforms as a spin-2 field. One defines the $E$ and $B$ modes of this field so that gradients of a scalar generate only the $E$ modes~\cite{Zaldarriaga:1996xe,Kamionkowski:1996ks}. In contrast, gravitational lensing of the CMB and other fields that directly modulate the polarization do generate $B$-modes. We review relevant aspects of flat sky limit of CMB polarization in Appendix~\ref{app:lensing}.

 We will demonstrate in this section that changing the statistics of the scalar metric fluctuations can produce a lensing-like signal in temperature but cannot generate a $B$-mode lensing signal. As a result, comparing the lensing amplitude from $TT$- and $EB$-based lensing reconstruction will be a sensitive test of a wide range of models. 

\subsection{Lensing and Non-Gaussianity}
\label{subsec:Lensing_Bias}

\subsubsection*{Temperature}

The dominant source of information about lensing in Planck~\cite{Planck:2018lbu} and ACT~\cite{ACT:2023dou,ACT:2023kun} is the lensing-induced trispectrum in temperature (the $TTTT$ trispectrum). This can be understood as a consequence of lensing deflection of the CMB temperature fluctuations:
\beq
\tilde T(\n) = T(\n + \vd) = T(\n + \vn \phi) \ ,
\eeq
where ($T$) $\tilde T$ is the (un-) lensed temperature, $\vd =\vn \phi$ is the lensing deflection, and $\phi$ is the lensing potential~\cite{Lewis:2006fu}. At linear order in $\phi$, CMB lensing leads to off-diagonal correlations in the temperature fluctuations
\bea
\langle \tilde T(\vl)\tilde T(\vl') \rangle_{\rm CMB} &=& \int \frac{d^2 \ell''}{(2\pi)^2}  \langle T (\vl)  T (\vl' - \vl'')\rangle_{\rm CMB} \vl\cdot \vl'' \phi(\vl'') +\{ \vl \leftrightarrow \vl'\} \\
&=&  \phi(\L )  \left(\L\cdot \vl \bar C_\ell^{TT} +\L\cdot \vl' \bar C_{\ell'}^{TT} \right)
\eea
where $\L=\vl+\vl'$ and $\bar{C}_\ell^{TT}$ denotes the unlensed temperature power spectrum without noise or foregrounds. As we will review below, it is from these off-diagonal correlations that we can reconstruct $\phi(\L)$ and, from it,  the lensing power spectrum.

Non-Gaussianity associated with multifield inflation, or isocurvature perturbations, can take a very similar form. Suppose we posit the existence of an isocurvature field $\chi(\n)$ that modulates the temperature fluctuations locally via
\beq\label{eq:Tchi}
T_\chi(\n) = T(\n) + \chi(\n) T(\n) \ .
\eeq
We could also imagine a modulating field, $\sigma(\n)$, that couples to the gradient of temperature
\beq\label{eq:Tsigma}
T_\sigma(\n) = T(\n) + \vn \sigma(\n) \cdot \vn T(\n) \ .
\eeq
This type of coupling typically only arises after inflation. During inflation, it is common to find models with an additional field~\cite{Chen:2009zp}, $\varphi(\x,t)$, that is derivatively coupled to the inflaton. However, the coupling between different Fourier modes depends on the time of freeze-out of the shortest mode. As a consequence, typical inflationary models will give
\beq\label{eq:time_ind_modes}
T_\varphi(\n) =  T(\n) + \vn \varphi(\n) \cdot \frac{\vn}{\nabla^2} T(\n)\ ,
\eeq
where $\varphi(\n) =\varphi(\x\to \n,t_\star)$ the local value of the field at the time of freeze-out of the short modes, $t_\star$. The condition for freeze-out, $a^2(t_\star) H^2 = k^2 \to -\nabla^2$, converts the time when the shorter mode crosses the horizon into a suppression by the ratio of the wavenumbers. Ultimately, this is just the statement that gradients redshift away~\cite{Baumann:2011nk,Assassi:2012zq}, and therefore the coupling between modes at different scales is suppressed.

Each of these different mechanisms gives rise to a mode coupling similar to lensing
\bea
\langle \tilde T(\vl)\tilde T(\vl') \rangle_{\rm CMB} &=&   \Phi(\L )  f^{(\Phi)}(\vl, \L-\vl)
\eea
for $\Phi \in [ \phi, \chi, \sigma, \varphi]$. The specific modulating functions take the form 
\bea
f^\phi &=& f^\sigma  = \L\cdot \vl \bar C_\ell^{TT} +\L\cdot \vl' \bar C_{\ell'}^{TT} 
\\
f^\chi &=& \bar C_\ell^{TT}  +\bar C_{\ell'}^{TT}  \qquad
f^\varphi = \frac{\L\cdot \vl}{\ell^2} \bar C_\ell^{TT} +\frac{\L\cdot \vl'}{\ell'{}^2} \bar C_{\ell'}^{TT} \ .
\eea
Although there are qualitative similarities between these models, our goal is to quantify to what degree a modulating field, such as $\chi$, $\sigma$, or $\varphi$, could be confused with lensing, giving rise to a larger lensing amplitude. To do so, direct constraints on these models must be sufficiently weak that evidence for these interactions would first appear as an apparent excess of lensing.

For any of these models, we can reconstruct the field $\Phi \in [\phi, \chi, \sigma, \varphi]$ from the off-diagonal correlations using
\beq
\hat \Phi(\L) = \int d^2 \ell F^{(\Phi)}(\vl, \L-\vl) T(\vl) T(\L-\vl) \ .
\eeq
We choose $F^{(\Phi)}(\vl, \L-\vl)$ to minimize the variance of the reconstruction of $\Phi$ while requiring that it is unbiased, 
\beq
\langle \hat \Phi(\L) \rangle_{\rm CMB} = \Phi(\L) \to \int d^2 \ell F^{(\Phi)}(\vl, \L-\vl) f^{(\Phi)}(\vl, \L-\vl) = 1 \ .
\eeq 
We can impose that it is unbiased with a Lagrange multiplier $\lambda$ so that we are minimizing
\beq
 \langle |\hat \Phi -\Phi|^2 \rangle_{\Phi, {\rm CMB}}  - \lambda\left(\int  F f -1\right) 
\eeq
with respect to both $F$ and $\lambda$. Extremizing with respect of $F$ and $\lambda$ gives
\beq
%  F^{(\Phi)}(\vl, \L-\vl)  = \frac{\lambda_{\Phi} f^{(\Phi)}(\vl,\L-\vl)}{ 2 C^{TT}(\ell) C^{TT}(\L-\vl) }  \ , \quad 
% \lambda_{\Phi}=\left( \int \frac{d^2\ell}{(2\pi)^2} \frac{|f^{(\Phi)}(\vl,\L-\vl)|^2}{2 C^{TT}(\ell) C^{TT}(\L-\ell)  }\right)^{-1} \ .
 F^{(\Phi)}(\vl, \L-\vl)  = \frac{\lambda_{\Phi} f^{(\Phi)}(\vl,\L-\vl)}{ 2 C^{TT}_\ell C^{TT}_{|\L-\vl|} }  \ , \quad 
\lambda_{\Phi}(L)=\left( \int \frac{d^2\ell}{(2\pi)^2} \frac{|f^{(\Phi)}(\vl,\L-\vl)|^2}{2 C^{TT}_\ell C^{TT}_{|\L-\vl|}  }\right)^{-1} \ ,
\eeq
where the unbarred $C_\ell^{TT}$ refer to the total observed temperature power spectra including noise, foregrounds, etc. The reconstruction noise is 
\beq\label{eq:noisecurve}
% N_\Phi = \langle |\hat \Phi -\Phi|^2 \rangle_{\Phi, {\rm CMB}} = \lambda(L)^2 \int \frac{d^2 \ell_1 }{(2\pi)^2} \frac{|f(\vl_1,\L-\vl_1)|^2  }{2 C^{TT}_{\ell_1} C^{TT}_{|\L-\vl_1|} }  = \lambda(L) \ .
N_\Phi = \langle |\hat \Phi -\Phi|^2 \rangle_{\Phi, {\rm CMB}} = \lambda_\Phi(L)^2 \int \frac{d^2 \ell }{(2\pi)^2} \frac{|f^{(\Phi)}(\vl,\L-\vl)|^2  }{2 C^{TT}_{\ell} C^{TT}_{|\L-\vl|} }  = \lambda_\Phi(L) \ .
\eeq
This is the noise that will appear in the power spectrum reconstruction of each $\Phi$. More details on these noise curves and how they appear in measurements of the modulation amplitude can be found in Appendix~\ref{app:CMB_for}.

We are now ready to address the key question: can the modulating fields $\chi$, $\sigma$ or $\varphi$ mimic the lensing potential $\phi$ via $TT$ lensing reconstruction and thus yield a biased lensing power spectrum? %This bias would also increase the amplitude of the lensing power spectrum. 
Of course, such a bias would also have to be consistent with bounds on other kinds of primordial non-Gaussianity~\cite{Planck:2019kim}, particularly from the CMB trispectrum~\cite{Smith:2015uia,Philcox:2025wts}. We can answer this question for any field $\Phi \neq \phi$ as follows. If such a field is present, when we reconstruct $\hat \phi$ it will be biased by the field $\Phi$ via
\bea
% \hat \phi_{\rm bias}(\L) &=& \Phi(\L) \int \frac{d^2 \ell}{(2\pi)^2} F^\phi(\ell,\L-\vl) f^{(\Phi)}(\ell,\L-\vl) \\
% &=& \Phi(\L) \lambda_\phi(\L) \int \frac{d^2 \ell}{(2\pi)^2} \frac{ f^\phi(\ell,\L-\vl) f^{(\Phi)}(\ell,\L-\vl) }{2 C^{TT}(\ell) C^{TT}(\L-\ell)} \ ,
\hat \phi_{\rm bias}(\L) &=& \Phi(\L) \int \frac{d^2 \ell}{(2\pi)^2} F^\phi(\vl,\L-\vl) f^{(\Phi)}(\vl,\L-\vl) \\
&=& \Phi(\L) \lambda_\phi(\L) \int \frac{d^2 \ell}{(2\pi)^2} \frac{ f^\phi(\vl,\L-\vl) f^{(\Phi)}(\vl,\L-\vl) }{2 C^{TT}_\ell C^{TT}_{|\L-\vl|}} \ .
\eea
We will focus on the modes in the regime $L \ll \ell$. In addition, we will define $\ell_{\rm max}$ as the largest $\ell$ with high signal to noise, namely those modes for which $\bar C_\ell^{TT} \approx C_{\ell}^{TT}$. Using these approximations, it is easy to see that $\lambda_\phi(L) \propto (L^2 \ell_{\rm max}^4)^{-1}$ and the biases will be given by
\beq\label{eq:phi_bias_result}
\hat \phi_{\rm bias}(\L)  \approx \sigma(\L) + \frac{1}{\rm \ell_{\rm max}^2} \left(\chi(\L) + \varphi(\L)\log L/\ell_{\rm max} \right) \ .
\eeq
For Planck, we already have $\ell_{\rm max} > 10^3$ and therefore the biases from all but $\sigma(\L)$ are highly suppressed. These integrals can be calculated precisely, but given a suppression by a factor of $10^6$ there is effectively no way for $\chi$ or $\varphi$ to impact lensing reconstruction within current constraints.

The main result here is that scalar couplings to isocurvature modes ($\chi$) or vector couplings during inflation ($\varphi$) are not viable mechanisms for biasing the lensing potential. In temperature, lensing is a derivative coupling but also is purely local and therefore couples all Fourier models equally. In contrast, inflation dilutes the gradients of the long wavelength modes, reducing the correlations between distantly separated scales.

\subsubsection*{Lensing and Polarization}

Gravitational lensing of the CMB impacts polarization fluctuations as well as temperature fluctuations. Most significantly, lensing converts $E$ modes to $B$ modes, which suggests that generic modulation signals that correlate modes in temperature may impact polarization data in a way that distinguishes them from lensing.   Future CMB data will very sensitive to this $E$-to-$B$ signal and therefore the different impact on temperature and polarization offers a valuable window into any mechanism that could explain an enhanced lensing amplitude.

CMB lensing is a particular kind of modulation of the CMB. As we will review now, generating $B$ modes is a non-trivial feature of lensing which is more difficult to mimic than changes to the temperature statistics in models involving only mode coupling.

We can understand the origin of CMB polarization in the tight coupling limit, such that the Stokes parameters of the polarization field, $Q$ and $U$, are given by
\begin{equation}
% (Q+i U)(\n)= -\frac{4}{3\sigma_T}  \int \frac{d^2\Omega'}{4\pi} (\n \cdot \hO')  T_2(\hO') \qquad  T_2=\lambda_p n^i n^j \partial_i v_{j} \ , 
%(Q+i U)(\n)= -\frac{3}{4}\sigma_T  \int \frac{d^2n'}{4\pi} (\he_+ \cdot \n')  T_2(\n') \qquad  T_2(\n')=\lambda_p n'^i n'^j \partial_i v_{bj} \ , 
(Q+i U)(\n)= -\frac{3}{4}\sigma_T  \int \frac{d^2\Omega'}{4\pi} (\he_+ \cdot \r')  T_2(\r') \qquad  T_2(\r')=\lambda_p \hat{r}^{\prime \, i} \hat{r}^{\prime \, j} \partial_i v_{bj} \ , 
\end{equation}
where $T_2$ is the temperature quadrupole, 
%$\he_\pm = \hx + i \hy$,
$\he_{\pm} = \he_x \pm i \he_y$ for $\n$ in the $\hz$-direction,
$\sigma_T$ is the Thomson cross-section, $\lambda_p$ is the mean-free path of the photons, ${\bm v}_b$ is the velocity of baryons, and the integral is performed over the sphere of directions of incident photons denoted by $\Omega'$ (see Appendix~\ref{app:lensing} for additional details of polarization in the flat sky limit). 
Integrating over incident angles, the polarization field can be written as~\cite{Zaldarriaga2000b}
\begin{equation}\label{eq:thomp_P}
% \left.(Q+i U)(\x) \approx \epsilon \Delta \tau_R e_+^a e_+^b \nabla_a v_b\right|_{\tau_R}
\left.(Q+i U)(\n) \approx \epsilon \Delta \tau_\star e_+^i e_+^j \nabla_i v_{bj}\right|_{\tau_\star}
\end{equation}
where $\tau_\star$ is the conformal time at recombination and $\Delta \tau_\star$ is the width of recombination.
%$e_\pm = \hat x + i \hat y$ 
(see e.g.~\cite{Zaldarriaga:1995gi} for a more detailed derivation). Using 
\beq
%\nabla^2 (E+i B) = e_-^a e_-^b \nabla_a \nabla_b (Q+iU)
\nabla^2 (E+i B) = e_-^i e_-^j \nabla_i \nabla_j (Q+iU)
\eeq
and $ e_-^i e_+^j \nabla_i \nabla_j = \nabla^2$,
%$ e_-^a e_+^b \nabla_a \nabla_b = \nabla^2$,
the primordial $B$ mode along the line of sight in the $\hz$-direction is given by
\beq\label{eq:B_curl_vb}
% B \propto {\rm Im} \left[ e_-^a e_+^b \nabla_a v_b \right] = \partial_x v_y - \partial_y v_x = \hat z \cdot \vec \nabla \times \vec v \ .
B \propto {\rm Im} \left[ e_-^i e_+^j \nabla_i v_{bj} \right] = \partial_x v_{by} - \partial_y v_{bx} = \hz \cdot \vn \times \v_b \ .
\eeq
We see that the $B$ modes will vanish as long as the baryon velocity is described by a scalar velocity potential $\v_b = \vn \theta_b$. Of course this is just the familiar statement that scalar fluctuations cannot generate $B$ modes. However, it shows that even a vector modulation of the scalar metric perturbation, for example,
\beq
% \Phi(\x) = \Phi_g(\x) + \vec A(\x) \cdot \vec \nabla \Phi_g(\x) 
\Phi(\x) = \Phi_g(\x) + \vA(\x) \cdot \vn \Phi_g(\x) 
\eeq
will also not generate $B$-modes even though its statistics are influenced by a vector field 
$\vA(\x)$.
%$\vec A(\x)$.
Concretely, we can have 
$\vA = \vn \sigma$
%$\vec A = \vec \nabla \sigma$
to arrive at Equation~(\ref{eq:Tsigma}), which will bias the lensing reconstruction from temperature but does not generate $B$-mode polarization.

Lensing generates $B$ modes because it directly modulates the polarization field itself,
\beq
% \tilde P_{ab}(\x) = P_{ab}(\x+ \vn \phi(\x)) \approx P_{ab}(\x) + \vec \nabla \phi \cdot \vn P_{ab}(\x) + \ldots \ .
\tilde P_{ab}(\n) = P_{ab}(\n+ \vn \phi(\n)) \approx P_{ab}(\n) + \vn \phi \cdot \vn P_{ab}(\n) + \ldots \ .
\eeq
Using $Q+i U = P = e^a_+ e^b_+ P_{ab}$, the lensed Stokes parameters, $\tilde Q$ and $\tilde U$, are given by
\beq
% \tilde P  = \tilde Q + i \tilde U  = Q+i U + \vn^c \phi  \vn_c (Q+i U)(\x) + \ldots  \ .
\tilde P  = \tilde Q + i \tilde U  = Q+i U + \vn^c \phi  \vn_c (Q+i U)(\n) + \ldots  \ .
\eeq
Now using the Thomson scattering result in Equation~(\ref{eq:thomp_P}) we get
\beq
% \nabla^2 \tilde B \propto {\rm Im} \left[ e_-^a e_+^b \nabla_a \nabla^c \phi \nabla_c \nabla^2 v_b \right] + \ldots \ ,
\nabla^2 \tilde B \propto {\rm Im} \left[ e_-^i e_+^j \nabla_i \nabla^k \phi \nabla_k \nabla^2 v_{bj} \right] + \ldots \ ,
\eeq
where we used 
${\rm Im} \left[ e_-^i e_+^j \nabla_i v_{bj} \right] = 0$
%${\rm Im} \left[ e_-^a e_+^b \nabla_a v_b \right] = 0$
for a curl-free velocity. This expression does not vanish in the presence of a non-zero $E$ mode and lensing field. The induced $B$ mode can also be written in terms of the individual Fourier modes, $E(\vl)$ and $\phi(\L)$, as
\beq
\tilde B(\vl)= - {\rm Im} \int \frac{d^2 \ell'}{2 \pi} \vl^{\prime} \cdot(\vl-\vl') e^{2 i(\phi_{\vl'}-\phi_\vl)} \phi(\vl-\vl^{\prime}) E(\vl') \ .
\eeq
In this sense, it is possible to generate a $B$ mode from (the gradient of) a scalar field that directly modulates the polarization field.

From these two observations, we can see that a model that introduces bias to the $TT$ lensing estimator does not necessarily produce a bias in the $EB$ estimator. In order to generate $B$ modes, we could directly modulate the velocity perturbation, presumably by coupling to baryons. For example, we could have a force on the baryon-photon fluid proportional to the gradient of our scalar $\sigma$ so that
\beq\label{eq:sigma_vb_coupling}
% v^a(\x) = v^a_g(\x) +  \nabla^b \sigma(\x) \nabla_b v^a(\x) + \ldots
% v^i(\x) = v^i_g(\x) +  \nabla^j \sigma(\x) \nabla_j v^i(\x) + \ldots
v^i_b(\x) = v^i_{bg}(\x) +  \nabla^j \sigma(\x) \nabla_j v^i_b(\x) + \ldots
\eeq
which would give 
\beq
% B(\x) \propto \epsilon^{ca} \nabla_c \nabla^b \sigma(\x) \nabla_b \nabla_a \theta(\x) \, .
% B(\x) \propto \epsilon^{ki} \nabla_k \nabla^j \sigma(\x) \nabla_j \nabla_i \theta(\x) \, .
B(\n) \propto \epsilon^{ki} \nabla_k \nabla^j \sigma(\n,r_\star) \nabla_j \nabla_i \theta_b(\n,r_\star) \, ,
\eeq
where $r_\star$ is the distance to last scattering.
As long as $\sigma$ and $\theta_b$ are statistically independent fluctuations, we will generate  non-zero $B$ modes. Alternatively, a vector coupling of an isocurvature mode directly to the polarization field will generate a similar type of modulation.

%%%%%%%%%%%%%%
%% Lensing SN Plot
%%%%%%%%%%%%%%
%%%%%%%%%%%%
\begin{figure}[t!]
    \centering
    \includegraphics[width=0.7\textwidth]{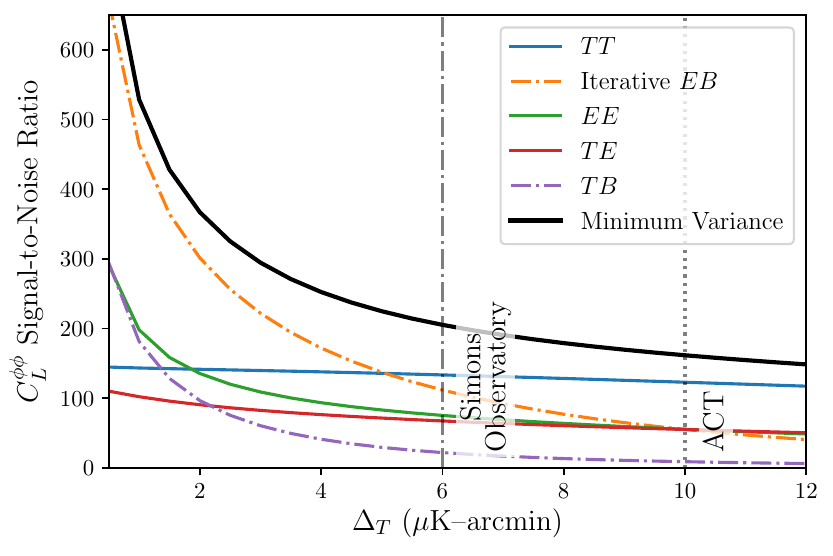}
    \caption{Signal-to-noise ratio of CMB lensing reconstruction as a function of survey depth $\Delta_T$ assuming a CMB survey with beam size of $1.4$~arcmin and a sky coverage of $f_\mathrm{sky}=0.5$.  In order to account for the effects of astrophysical foregrounds, we use $\ell_\mathrm{max}^T=3000$ and $\ell_\mathrm{max}^P=5000$ with $\ell_\mathrm{min}=30$.  We show the signal-to-noise ratio for each estimator as well as the minimum variance combination of all estimators.  We distinguish the $EB$ and $TB$ estimators which rely on measurement of lensing-induced $B$-mode polarization.  The results for the $EB$ estimator shown here include the improvements from iterative delensing~\cite{Hirata:2003ka,Smith:2010gu} as implemented in \texttt{CLASS\_delens}~\cite{Hotinli:2021umk}. }
    \label{fig:lensing_SN}
\end{figure}
%%%%%%%%%%%%%

We will see in the next section that there is no particular reason for the polarization and temperature lensing biases to have the same amplitudes and that they are generally model dependent. This provides an opportunity to understand these scenarios with new data from current and future CMB observations. From Figure~\ref{fig:lensing_SN} one can see that Planck and ACT are currently dominated by $TT$ lensing reconstruction, while future surveys like Simons Observatory and lower noise experiments will be dominated by the $EB$ estimator. Interestingly, SPT-3G recently analyzed only polarization data on 1500 degrees of sky~\cite{SPT-3G:2024atg} and reached a map noise amplitude of 4.9 $\mu$K-arcmin. At these noise levels, the $EB$ estimator is at least comparable to $EE$ and finds a larger-than-expected lensing amplitude similar to Planck and ACT. We cannot immediately determine that this is driven by the generation of $B$ modes, due to the $EE$ estimator contribution, but it suggests that the current data is at the level where some of the modulation models could be excluded.

Overall, we find one of the most straightforward and promising ways to evaluate non-lensing sources of CMB modulation is to compare the lensing reconstruction amplitude among different lensing estimators, similar to what was done for two- and four-point lensing reconstruction in~\cite{SPT-3G:2024atg}.

\subsection{Models of Isocurvature Modulation}
\label{subsec:Isocurvature_Models}

There are two distinct strategies for generating an apparent excess lensing signal from modulation. The first is to introduce an isocurvature mode around the time of reheating in order to change the statistics of the adiabatic modes. The second is for this isocurvature mode to be present around the time of recombination, where it can directly alter the evolution of the density fluctuations or propagation of photons. The former faces fewer observational constraints but does not generically give rise to a lensing-like signal.

\subsubsection*{Generating Non-Gaussianity at Reheating}

We have already seen that for temperature-only lensing reconstruction, we can bias the lensing amplitude through the statistical model in Equation~(\ref{eq:Tsigma}). 
\beq
T_\sigma(\n) = T(\n) + \vn \sigma(\n) \cdot \vn T(\n) \ .
\eeq
We can achieve such a statistical coupling of the CMB temperature modes by an interaction between the field responsible for ending inflation, $\pi$, and a scalar field $\sigma$, coupled via an interaction 
\beq\label{eq:S_int_sigma}
S_{\rm int} = -\int dt \, d^3 x \, a(t) c(t) \dot \pi \partial^i \pi \partial_i \sigma \ ,
\eeq
where $c(t)$ is a time-dependent coupling function. This interaction is not Lorentz invariant but is allowed because time translations are spontaneously broken by the background of the inflaton~\cite{Senatore:2010wk}. The Lorentz-violating nature of this interaction is important so that the derivatives are non-removable via integration by parts. We also need a strongly time-dependent coupling $c(t) \approx \delta(t-t_R)$, where $t_R$ is the time of reheating, in order to break scale invariance (time-translations) so that all the Fourier modes are coupled equally. A time-independent coupling would lead instead to Equation~(\ref{eq:time_ind_modes}) and would not bias the lensing amplitude.

The resulting correlation between $\pi$ and $\sigma$ can be calculated via the in-in formalism~\cite{Maldacena:2002vr,Weinberg:2005vy} to find
\beq\label{eq:pps}
\langle \pi(\k_1) \pi(\k_2) \sigma(\k_3) \rangle = i \int^{t_R}_{-\infty} dt \, \langle [H_{\rm int}(t), \pi(\k_1,t_R) \pi(\k_2,t_R) \sigma(\k_3,t_R) ] \rangle \ .
\eeq
Taking $H_{\rm int} = \int d^3 x \, a(t)c(t) \dot \pi \partial^i \pi \partial_i \sigma$ and using the canonical commutator $[\pi(\x,t_R), \dot \pi(\x',t_R)] =i a^{-3}\delta(\x-\x')$, one finds the correlation
\beq
\langle \pi(\k_1) \pi(\k_2) \sigma(\k_3) \rangle'  \propto a^{-2}(t_R) (\k_1 \cdot \k_3 P_\pi(k_1) + \k_2 \cdot \k_3 P_\pi(k_2) ) P_\sigma(k_3) \ ,
\eeq
where we defined $\langle .. \rangle  = \langle .. \rangle' (2\pi)^3 \delta(\sum \k_i)$. In deriving this result, we used the fact that the interaction was localized at $t_R$ so that additional terms vanish as $t \to t_R$. 

From this point of view, we see that we can straightforwardly generate a lensing-like signal in temperature without running into serious constraints. This works because origin of the lensing-like signal arises without ever directly coupling to Standard Model fields (non-gravitationally). However, the downside is that we do not generate a polarization signal. Next we will move to models that can bias all of the lensing statistics and their more serious existing observational constraints.

\subsubsection*{Extra Lensing at Recombination}

At first sight, it would seem quite reasonable that a  coupling between photons and some new field could exactly mimic a CMB lensing signal.  The new field, existing between us and the surface of last scattering, could act as an actual, physical lens for the CMB. For example, lensing can be formulated as a change to the effective propagation speed of light. In conformal Newtonian gauge, the metric takes the form
\beq
ds^2 = a^2(\tau)( - (1+2 \Phi) (c d\tau)^2 +(1- 2 \Phi) d \x^2 ) \ ,
\eeq
where we have restored $c=1$ for emphasis. Since light travels along null directions, we have the null rays obey
\beq
(1+2 \Phi)(c d\tau)^2 -(1- 2 \Phi) d \x^2 =0 \quad \to \quad \frac{dx}{d\tau} \approx c (1+2 \Phi) \ ,
\eeq
so that the effective speed is
\beq
c_{\rm eff}(\x,\tau) = c (1+2 \Phi(\x,\tau)) \ .
\eeq
Following from Fermat's principle, one can re-derive weak gravitational lensing as being a consequence of this apparent change to the speed of light (in these coordinates). 

Light traveling through a medium that couples to the photon would similarly be expected to generate a change to the speed of propagation depending on some new background field, $\Theta(\x,t)$ so that
\beq\label{eq:ceff}
c_{\rm eff}(\x,\tau) = c (1 - \Theta(\x,\tau) ) \ .
\eeq
In this sense, we can treat $2 \Phi(\x,t)\to \Theta(\x,\tau)$ as a realization of how this new field would appear in the CMB. In terms of the lensing potential, we would then have~\cite{Lewis:2006fu},

\beq
\nabla \phi(\hat n) = - \int_0^{\chi_*} d \chi \frac{\left(\chi_*-\chi\right)}{\chi_* \chi} \nabla \Theta\left(\chi \hat{\mathbf{n}} ; \tau_0-\chi\right) ,
\eeq
where $\chi_\star$ is the comoving distance to recombination and $\tau_0$ is the conformal age of the universe. The goal of the rest of this section is to offer realizations of $\Theta(\x,\tau)$ using different microscopic descriptions.

The most stringent, and model independent, constraint on this idea comes from the observations of the multi-messenger event GW170817 and GRB 170817A. The difference in arrival times of the gravitational and electromagnetic signals bounds the difference between the speeds to be~\cite{LIGOScientific:2017zic}
\beq\label{eq:speed_bound}
-3 \times 10^{-15} < c_{\rm grav} - c_\gamma < 7\times 10^{-16} \ ,
\eeq 
averaged over a distance of approximately $40 \, \text{Mpc}$. Given that we need an effective lensing amplitude of $\phi \approx 10^{-3}$, we will have to dramatically reduce $\Theta(\x,\tau)$ between the peak of the effective lensing and more recent cosmic history. We can estimate how quickly this effective lens must redshift away by noting that we will need $\Theta(\x,\tau) \sim 10^{-3}$ at $z \approx 10^3$ but $\Theta \lesssim 10^{-15}$ today. 
Assuming $\Theta \propto a^{-n}$, one finds that for $n\geq 5$, we can certainly achieve sufficient suppression of the signal by the current epoch while maintaining a large lensing amplitude. Interestingly, even redshifting as radiation, $n = 4$, may just avoid the constraint.  As we will see below this is a natural, well-motivated possibility.  Nevertheless, given the large powers involved, order one factors are important to the viability of these models and warrant a more careful analysis.  Nevertheless, models with $n=4$ do make an interesting prediction that even a relatively small improvement in the test of the equality of the speed of gravity and speed of light would yield important results.
Regardless of the specific value of $n$, the fact the signal redshifts away, rather than growing with time like the case of gravitational lensing, means that the apparent lensing arises from higher redshift than the we expect from the lensing kernel.

Naturally, one might hope that a weakly coupled background field, like an axion, could accomplish this goal and potentially make up (some of) the dark matter in the process. Unfortunately, mimicking lensing in simple, well-motivated models is challenging for several reasons. The most basic is that a change to the speed generally appears to break Lorentz invariance so that the photon's effective Lagrangian (in Coulomb gauge) includes
\beq
{\cal L} \supset \frac{1}{g^2}  \left( \dot A_i \dot A^i - c_{\rm eff}^2 \partial_i A_j \partial^i A^j \right) \approx \frac{1}{g^2} \left(F_{0i}F^{0i} - c_{\rm eff}^2 F_{ij} F^{ij} \right) \ .
\eeq
In contrast, the typical axion-like couplings (or simplest scalar field couplings) we might consider would be
\beq\label{eq:axion_couplings}
\mathcal{L} \supset -\frac{1}{4} a(\x,t) \left(g_{a\gamma \gamma} \tilde F_{\mu \nu} F^{\mu \nu}  + \bar g F_{\mu \nu} F^{\mu\nu} \right) \ ,
\eeq
where $a(\x,t)$ is our axion/scalar field (not to be confused with the scale factor). At leading order, a background value of either $g_{a\gamma \gamma} \, a(\x,t)$ or $\bar g \, a(\x,t)$ will break Lorentz invariance of the theory. However, it does not break the boosts of the photon at leading order. In principle, in the strong coupling limit, this could generate a speed of sound through multiple scatterings off the background, like a real world material. This is not applicable in our universe due to existing bounds on the axion-photon coupling, for example from the helioscope CAST which requires $g_{a\gamma \gamma} < 6\times 10^{-11}$ GeV$^{-1}$~\cite{CAST:2017uph} (or even more stringent bounds on $\bar g$). Using the number density of axion dark matter (see e.g.~\cite{OHare:2024nmr} for review), the mean free path of photons due to scattering with individual axions is much larger than the observable universe. 

A Lorentz-violating background in the form of a symmetric, two-index tensor field $B$ coupled to the photon would have the appropriate structure to induce an index of refraction, 
\beq\label{eq:B_coupling}
{\cal L} \supset  \frac{1}{g^2} \left( F_{\mu \nu} F^{\mu \nu} + \frac{B^{\alpha}_{\phantom{\alpha} \beta} }{\Lambda^2}  F_{\mu \alpha} F^{\mu \beta}  \right) \, ,
\eeq
where $\Lambda$ is some high scale suppressing this higher-dimension operator.
 If for example we take a VEV  with an $\mathcal{O}(1)$  difference between the temporal and spatial components and it has rough size $\langle B^{\mu \nu} \rangle \sim b(x)$ then we get a change to the effective index of refraction $\delta n \sim \mathcal{O} \left( \frac{b(x)}{\Lambda^2} \right)$.
 This can vary in space even without changing the direction of the VEV if $b(x)$ varies spatially, thus producing the effect of extra lensing of the CMB that we are looking for.
Of course this structure in principle does not need to come from a dynamical two-index tensor field.  It could come from another field (a scalar for example) so long as it happens to couple to just the relevant terms of $F_{\mu \alpha} F^{\mu \beta}$.  
If $B$ is actually a dynamical field with a kinetic term (or contains a dynamical field, e.g.~a scalar field)  then one must be sure that either the scale suppressing the coupling, $\Lambda$, or the mass of the field is large enough to avoid bounds on new particles coupled to the photon today. This could also be the effective operator for a more complicated material giving an effective index of refraction.

For example, one possible implementation could be a new, massive vector field (a hidden photon) $C^\mu$ with a coupling of the form:
\beq\label{eq:C_coupling}
{\cal L} \supset 
   \frac{1}{g^2} \left( F_{\mu \nu} F^{\mu \nu} + \frac{C^
\alpha C_\beta}{\Lambda^2}  F_{\mu \alpha} F^{\mu \beta}  \right) .
\eeq
A background for this field, $\langle C^\alpha  \rangle \neq 0$, could generate the structure above.
We will assume that the coupling quadratic in $C_\alpha$ is relevant here while the coupling linear in $C$ (the usual kinetic mixing) is not. The effective speed of light is
\beq
\langle c_{\rm eff} \rangle \sim c\left (1-   \frac{\langle C  \rangle^2}{\Lambda^2} \right) \, .
\eeq
In order to mimic a several percent correction to the lensing potential, $\phi \sim 10^{-3}$, we need $\langle C  \rangle_\text{CMB} \sim 10^{-2} \, \Lambda$ around the time of recombination\footnote{The lensing kernel peaks at $z\approx 2$, but due to the large amount of redshifting needed to avoid the low-$z$ constraint on $c_{\rm eff}$ the signal arises from much earlier times. Combining the rapid decay of $C_\alpha$ with the lensing kernel introduces an additional suppression of ${\cal O}(10^{-1})$.}. For simplicity let us imagine that $C_\alpha$ is a light field (mass below Hubble) and that the dominant modes of $C_\alpha$ that have been excited are in a narrow range of $k$ which enter the horizon roughly around the time of  CMB formation.  This should give extra lensing in roughly the right modes of the CMB.  Many other scenarios giving similar phenomenology are possible but this will serve as an example. Then the energy density contained in these modes at the time of CMB formation is naturally small $\rho_C \sim \langle C  \rangle_\text{CMB}^2 H_\text{CMB}^2 \sim \rho_\text{CMB} \frac{\langle C  \rangle_\text{CMB}^2}{M_\text{pl}^2}$. In order to avoid the multi-messenger  bound on the speed of light, Equation~(\ref{eq:speed_bound}), it is certainly safe if this energy density redshifts as $a^{-5}$.  However the natural, default redshifting of the energy density in these vector field modes is $\propto a^{-4}$.  This appears right on the edge and may be just safe from the constraint.  Interestingly, despite the impressive power of the current constraint, this would provide motivation for even a relatively small improvement as the current test appears to be just reaching an important regime.

There are many bounds on the existence of such a new light particle, but $\Lambda$ can be taken large enough to evade them.  For example $C_\alpha$ particles can be produced in supernovae.  A very rough estimate of this bound comes from rescaling the well-known bound on an axion-like coupling to this coupling.  In the axion case the rate of emission is proportional to the axion coupling squared $g_{a\gamma\gamma}^2$ and the limit is $g_{a\gamma\gamma} \gtrsim 10^9 \, \text{GeV}$.  In our case the rate is $\propto \Lambda^{-4}$ and the extra powers must be made up by powers of the supernova temperature $\sim 50 \, \text{MeV}$.  So the rough limit is $\Lambda \gtrsim \sqrt{(10^9 \, \text{GeV}) \times (50 \, \text{MeV})} \sim 10^4 \, \text{GeV}$ which is easily satisfied and still allows a small $\langle C  \rangle$ which is easily a very small fraction of the energy of the universe.  $C_\alpha$ also mediates a new, fifth force on Standard Model particles, but since this requires (one-loop) exchange of two $C_\alpha$'s, this constraint is similarly easy to satisfy.

One could also implement this idea within the context of an effective solid~\cite{Leutwyler:1996er,Son:2005ak,Endlich:2010hf}, or more accurately its generalization to a cosmological solid~\cite{Endlich:2012pz}. Here one would need three scalar fields, $\sigma^I$ whose background VEVs 
\beq\label{eq:sigma_vevs}
\langle \sigma^I \rangle = f^2 x^i \delta_{i I} \ , 
\eeq 
break the internal and spacetime rotations to the diagonal subgroup, where $f$ is a constant with units of energy. Coupling these fields to the photon, one can generate a change to the speed of light via
\beq
\frac{1}{\Lambda^4}\partial_\lambda \sigma^I \partial_\rho \sigma_I F^{\lambda \mu} F_{\mu}^\rho \to c_{\rm eff} =c \left(1- \frac{f^4}{\Lambda^4} \right) \ .
\eeq
In principle, direct constraints on these couplings are easy to avoid since it is a high dimension operator and $\Lambda$ (and $f$) can be large (in principle). Indirect constraints then depend on having a more precise model of the background evolution. Following the construction of solid inflation, if the equation of state allows the solid to exist only for $z> 1$, then it could cause the same lensing effects in the CMB while avoiding constraints on photon propagation nearby.

There are certainly other models to cause extra lensing that are worth considering including for example a scalar with coupling $\partial_\lambda \partial_\rho \sigma \, F^{\lambda \mu} F_{\mu}^\rho$ or in general a  field that for some reason has a Lorentz violating coupling to the photon $ {\cal L} \supset \frac{1}{g^2}  \left( \dot A_i \dot A^i - f(\sigma) \partial_i A_j \partial^i A^j \right)$.  It would be interesting to explore whether any of these models have a mechanism which naturally produces the desired state of the lens field from a simple cosmological history.

\section{The Dark Force Awakens}\label{sec:darkforce}

The most natural way to explain the excess lensing of the CMB is to directly increase the amplitude of clustering. The formation of structure is driven by the force of gravity acting on dark matter. While long range interactions beyond gravity are strongly excluded when it comes to how new forces act the Standard Model content of the universe~\cite{Will:2001mx,Tino:2020nla}, dark matter is not subject to the same kinds of constraints from tests within the solar system~\cite{Kesden:2006zb,Keselman:2009oaz}. To the degree that it is possible to explain current observations with new physics in the dark sector alone, it is a non-trivial question how else we might test these models observationally or experimentally.

Introducing a long range force by hand into the fluid equations provides a physical mechanism by which to realize the negative neutrino mass parameter $\mnuclustering$ in all CMB observables and in any other measurement of late time clustering, including from weak lensing. It additionally predicts equivalence principle violation in galaxy clustering in a way that can be the target of future observations~\cite{Creminelli:2013nua}. However, physical models of such a force lead to additional changes to the evolution of the universe~\cite{Archidiacono:2022iuu,Bottaro:2023wkd}. In addition to the growth of structure, the interactions between dark matter and a mediator field change the evolution of the matter density, and therefore the expansion of the universe, in the form of $\mnubao$. In this precise sense, dark forces have non-trivial impacts on both lensing and the BAO.

In this section, we will isolate both effects and discuss their physical significance to observations. We will then discuss the recent evidence for such a force~\cite{Bottaro:2024pcb}, which combines both effects, and how it depends on each contribution individually.

\subsection{Increasing Clustering by Hand}
\label{subsec:Clustering_By_Hand}

From the point of view of the large scale structure of the universe, the effect of a new long range force is to change the appearance of the fluid equations describing the late universe. Let us first imagine that we can change these equations directly to add a new force, without including a microscopic description. This parameterization directly changes the linear growth and also the non-linear evolution, without changing the background cosmology. This description of such a force was introduced in~\cite{Creminelli:2013nua} to illustrate the violation of the single field consistency conditions, which are among the most robust tests of the equivalence principle throughout cosmic history~\cite{Maldacena:2002vr,Creminelli:2004yq,Creminelli:2013mca,Creminelli:2013poa}.

We start from the continuity equation and modified Euler equation for the baryons and dark matter, 
\begin{align}
% \delta_X^{\prime}+\vn \cdot\left[\left(1+\delta_X\right) \v_X\right] & =0 \\
% \v_b^{\,\prime}+\mathcal{H} \v_b+\left(\vec{v}_b \cdot \vn\right) \vec{v}_b & =-\vec{\nabla} \Phi \\
% \vec{v}_{\rm dm}^{\,\prime}+\mathcal{H} \vec{v}_{\rm dm}+\left(\vec{v}_{\rm dm} \cdot \vec{\nabla}\right) \vec{v}_{\rm dm} & =-\vec{\nabla} \Phi-\alpha \vec{\nabla} \varphi  \ ,
\delta_X^{\prime}+\vn \cdot\left[\left(1+\delta_X\right) \v_X\right] & =0 \\
\v_b^{\,\prime}+\mathcal{H} \v_b+\left(\v_b \cdot \vn\right) \v_b & =-\vn \Phi \\
\v_{\rm dm}^{\,\prime}+\mathcal{H} \v_{\rm dm}+\left(\v_{\rm dm} \cdot \vn\right) \v_{\rm dm} & =-\vn \Phi-\alpha \vn \varphi  \ ,
\end{align}
where $' \equiv \partial_\tau$, the derivative with respect to conformal time. Here $\delta_X$ with $X=b, {\rm dm}$ are the baryon and dark matter density contrast respectively, with fluid velocities $\v_X$. The gravitational potential and field $\varphi$ obey the Poisson equations,
\beq\label{eq:Poisson}
\nabla^2 \Phi=4 \pi G \rho_{\mathrm{m}} \delta=4 \pi G \rho_{\mathrm{m}}\left( f_b \delta_b+f_{\rm dm} \delta_{\rm dm}\right) \qquad \nabla^2 \varphi=\alpha  8 \pi G \rho_{\mathrm{dm}} \delta_{\rm dm} \ ,
\eeq
where $f_X =\rho_X / \rho_m$. It is useful to take the divergence of the Euler equations and define the velocity divergence $\theta_X = \vn \cdot \v_X$. Working at linear order in $\delta_X$ and only to order $\alpha^2$, we can write these equations in terms of $\delta_m= \fb \db + \fdm \ddm$ and $\delta_r = \ddm-\db $ as
\bea
\delta'_{m } +\theta_{m} &=& 0 \\
\theta_{m}^{\prime}+\mathcal{H} \theta_{\rm m}+\frac{3}{2} \Omega_m \H^2 (\delta_m +  2\alpha^2 f^2_{\rm dm} \delta_{m}) &=& 0 \label{eq:fluid_dm_fix}\\ 
\delta'_{r } +\theta_{r} &=& 0  \\
\theta_{r}^{\prime}+\mathcal{H}\theta_{r}-\frac{3}{2} \Omega_m \H^2 (2\alpha^2 f_{\rm dm} \delta_{m}) &=& 0  \ .
\eea 
%\joel{Check whether $f_{\rm dm}^2$ should appear in (4.8).}\dg{It is correct as written. The origin of the difference is that $\delta_m \supset \fdm \ddm$ whereas $\delta_r \supset \ddm$ with no additional factor of $\fdm$.}
Using the ansatz, $\delta_m = D_m(\tau) \delta_{m,0}$ and $\delta_r = D_r(\tau) \delta_{m,0}$, we find the equations for the growth functions
\bea
D_m''+\H D_m' - \frac{3}{2} \Omega_m \H^2 (1+ 2 \alpha^2 f_{\rm dm}^2) D_m &=&0  \\
D_r''+\H D_r' -\frac{3}{2} \Omega_m \H^2 ( 2 \alpha^2 f_{\rm dm}^2) D_m &=&0  \ .
\eea
A priori, the ansatz for $\delta_r$ might seem to be too restrictive, but it is sufficiently general for the solution to order $\alpha^2$. 
%Defining the linearized solutions as $\delta^{(1)}_X$ and making the ansatz $\delta_b^{(1)} = D(\tau) \delta_0$ and $\delta_{\rm dm} = b(\tau) \delta_b^{(1)}$ and $\theta_X = - D' \delta_X$, one finds
%\bea
%&& D'' + \H D' + \frac{3}{2} \Omega_m \H^2 D (f_b + f_{\rm dm} b) = 0 \\
%&& b' = 0 \qquad b f_{\rm dm} + f_b (1-b^{-1}) = f_{\rm dm}(1+ 2\alpha^2) \ .
%\eea
%\bea
%\frac{\mathrm{d}^2 D}{\mathrm{~d} \ln a^2}+\left(2-\frac{3}{2} \Omega_{\mathrm{m}}\right) \frac{\mathrm{d} D}{\mathrm{~d} \ln a}-\frac{3}{2} \Omega_{\mathrm{m}}\left(w_b+w_{\rm dm} b\right) D&=&0  \\
%\frac{d}{d\log a } b = 0 \qquad b w_{\rm dm} + w_b (1-b^{-1}) = w_{\rm dm}(1+ 2\alpha^2)
%\eea
During matter domination, where $\Omega_m \approx 1$ and $\H = 2/\tau$, one can find the growth factors (to first order in $\alpha^2$) are
\beq\label{eq:Enhanced_Growth}
 D_m (\tau) = \tau^{2\gamma} = a(\tau)^{\gamma} \qquad  \gamma =1+ \frac{6}{5} f_{\rm dm}^2 \alpha^2 \qquad D_r = 2 \alpha^2 \fdm^2 a(\tau) \ .
\eeq
As expected, the growth function for matter, $D_m(\tau)$, is enhanced by the new long range force. This formula is essentially the same as the growth function with massive neutrinos if we changed the neutrino fraction $f_\nu = \rho_\nu /\rho_m \to - 2 f_{\rm dm}^2 \alpha^2$. 

At non-linear order, the equations of motion become
%so that the equations of motion become
\begin{align}
% \delta_m' + \theta_m' =& - \int \frac{d^3 q}{(2\pi)^3}  \alpha(\q,\k-\q) \theta_m(\vec q) \delta_m(\k-\vec q) \\
% \theta_m^{\prime}+\mathcal{H} \theta_m+\frac{3}{2} \Omega_m \H^2 (1+2\alpha^2 \fdm^2) \delta_m& =- \int \frac{d^3 q}{(2\pi)^3}\beta(\q,\k-\q) \theta_m(\vec q) \theta_m(\k-\vec q) \\
% \delta_r' + \theta_r' = - \int \frac{d^3 q}{(2\pi)^3} &\alpha(\q,\k-\q)\left( \theta_m(\vec q) \delta_r(\k-\vec q) + \theta_r(\vec q) \delta_m(\k-\vec q)\right) \\
% \theta_{\rm r}^{\prime}+\mathcal{H} \theta_r+\frac{3}{2} 2\alpha^2 \fdm^2 \Omega_m \H^2 \delta_m =&-\int \frac{d^3 q}{(2\pi)^3}(\beta(\q,\k-\q)+\beta(\k-\q,\q) ) \theta_{\rm m}(\vec q) \theta_{\rm r}(\k-\vec q)
&\delta_m' + \theta_m' = - \int \frac{d^3 q}{(2\pi)^3}  \alpha(\q,\k-\q) \theta_m(\q) \delta_m(\k-\q) \\
&\theta_m^{\prime}+\mathcal{H} \theta_m+\frac{3}{2} \Omega_m \H^2 (1+2\alpha^2 \fdm^2) \delta_m =- \int \frac{d^3 q}{(2\pi)^3}\beta(\q,\k-\q) \theta_m(\q) \theta_m(\k-\q) \\
&\delta_r' + \theta_r' = - \int \frac{d^3 q}{(2\pi)^3} \alpha(\q,\k-\q)\left( \theta_m(\q) \delta_r(\k-\q) + \theta_r(\q) \delta_m(\k-\q)\right) \\
& \theta_{\rm r}^{\prime}+\mathcal{H} \theta_r+3 \alpha^2 \fdm^2 \Omega_m \H^2 \delta_m =-\int \frac{d^3 q}{(2\pi)^3}(\beta(\q,\k-\q)+\beta(\k-\q,\q) ) \theta_{\rm m}(\q) \theta_{\rm r}(\k-\q)
\end{align}
with
\beq
\alpha(\q_1, \q_2) = \frac{(\q_1+\q_2) \cdot  \q_1}{q_1^2} \qquad \beta(\q_1,\q_2) = \frac{ (\q_1\cdot \q_2)(\q_1 +\q_2)^2 }{2 q_1^2 q_2^2} \ .
\eeq
Notice that the nonlinear term in the evolution of $\delta_m$ is independent of $\alpha$. This is effectively a definition of $\theta_m$ since mass is conserved. In this sense, the origin of the equivalence principle violation appears only directly in the linear equations (and solutions) that feed into the non-linear solutions. However, the long range force also impacts the nonlinear evolution of $\delta_r$ since the baryons and dark matter no longer have the same density contrasts or velocities, even though they are individually conserved.

The galaxy bispectrum offers a novel window into the growth of structure beyond the change to the matter power spectrum measured via CMB lensing. Concretely, the new long range force changes the relative velocities of different components of the universe, due to violations of the equivalence principle. The velocity appears nonlinearly in the continuity equation and this change to the velocities manifests itself at tree-level in the bispectrum.

First, we can calculate the bispectrum of just the matter. To do so, we solve for $\delta_m$ and $\theta_m$ to second order in the initial density fluctuations. We will assume matter domination so that $\Omega_m =1$ and $\fb+\fdm =1$. We will make the  approximation
\bea
\delta_m =\sum_{n} D_m(\tau)^n \delta_m^{(n)} \qquad \theta_m = - {\cal H}\sum_{n} D_m(a)^n \theta_m^{(n)} \ ,
\eea
where $D_m(\tau)$ is the linear growth function. The non-linear terms for the evolution of matter are not affected by the change to the growth, so the iterative solutions to the equations are the same and we find
\beq
\delta^{(2)}_{m} = \int \frac{d^3 q_1 d^3 q_2}{(2\pi)^6}\delta(\k -\q_1-\q_2) F_2(\q_1,\q_2) \delta^{(1)}(\q_1)\delta^{(1)}(\q_2)
\eeq
where $F_2(\q_1,\q_2)$ is given by 
\beq
F_2(\q_1,\q_2) =\frac{5}{7} + \frac{1}{2} \frac{\q_1\cdot \q_2}{q_1 q_2} \left(\frac{q_1}{q_2}
 +\frac{q_2}{q_1} \right) + \frac{2}{7} \frac{(\q_1\cdot \q_2)^2}{q_1^2 q_2^2} \ .
 \eeq
The tree-level matter bispectrum is then
\beq\label{eq:Bmatter}
\begin{aligned}
\langle \delta_{m}(\k_1) \delta_{m}(\k_2) \delta_{m}(\k_3) \rangle =& \bigg( F_2(\k_2,\k_3)P_m(k_2)P_m(k_3)  
+F_2(\k_1,\k_3) P_m(k_1)P_m(k_3) \\
&+ F_2(\k_1,\k_2)P_m(k_1)P_m(k_2) \bigg) (2\pi)^3\delta(\k_1+\k_2+\k_3)  
\end{aligned}
\eeq
In the limit $\k_1 \to 0$, the tree-level matter bispectrum becomes
\beq 
\lim_{\k_1 \to 0} \langle \delta_{m}(\k_1) \delta_{m}(\k_2) \delta_{m}(\k_3) \rangle'\to \left(\frac{27}{14}  + \frac{4}{7} \frac{(\k_1\cdot \k_2)^2}{k_1^2 k_2^2} \right) P_m(k_1) P_m(k_2) \ .
\eeq
Although the equivalence principle is violated, the effect cancels due to the symmetry under exchanging $\k_2$ and $\k_3$. This is also not surprising given that, for matter, $\beta$ only appeared in the growth function and not in the non-linear interactions, leaving a bispectrum that is changed only by the size of $P_m(k_1)$ at a given redshift.

In order to see the violation of the equivalence principle, we need to include the difference between baryon and dark matter over-densities, $\delta_r$. To do so, we need to make an non-linear ansatz for $\delta_r$
\bea
\delta_r = (2\alpha^2 \fdm^2 ) \sum_{n} D_m(\tau)^n \delta_r^{(n)} \qquad \theta_r = - {\cal H}(2\alpha^2 \fdm^2 ) \sum_{n} D_m(a)^n \theta_r^{(n)} \ ,
\eea
This solution is only valid to order $\alpha^2$ so that we can ignore the difference between $D_m(\tau)$ and $D_r(\tau)$ in this ansatz. Substituting into the equation and solving for $\delta_r^{(2)}$, one finds
\bea
\delta^{(2)}_{r} &=& \int \frac{d^3 q_1 d^3 q_2}{(2\pi)^6}\delta(\k -\q_1-\q_2) \tilde F_2(\q_1,\q_2) \delta^{(1)}(\q_1)\delta^{(1)}(\q_2)\\
\tilde F_2(\q_1,\q_2) &=&\frac{2}{5}\beta(\q_1,\q_2) +\frac{3}{10} F_2(\q_1,\q_2)  + \alpha(\q_1,\q_1) \\ 
&=&\frac{17}{14} + \frac{17}{20} \frac{\q_1\cdot \q_2}{q_1 q_2} \left(\frac{q_1}{q_2}
 +\frac{q_2}{q_1} \right) + \frac{17}{35} \frac{(\q_1\cdot \q_2)^2}{q_1^2 q_2^2}  \ . \label{eq:F2noback}
 \eea
Now, because $\tilde F_2 \neq F_2$, there is no cancellation of the $k_1^{-1}$ term in the the squeezed limit of the mixed bispectrum, 
\begin{align}
\lim_{\k_1 \to 0} \langle \delta_{m}(\k_1) \delta_{r}(\k_2) \delta_{m}(\k_3) \rangle' &= (2\alpha^2 \fdm^2) (\tilde F_2(\k_1,\k_2) + F_2(\k_2,\k_1))P_m(k_1) P_{m}(k_2) \\
&= (2\alpha^2 \fdm^2) \frac{7}{10} \frac{\k_1\cdot \k_2}{k_1^2} P_m(k_1) P_{m}(k_2) \ .
\label{eq:mmr_bispectrum}
\end{align}
This is a particularly interesting result because it gives a signal at large distances (small wavenumber, $k_1 \ll k_{2,3}$), away from the nonlinear regime, that is not produced by a conventional gravitational theory at any order in perturbation theory. The challenge is that measuring this signal requires two populations of galaxies that have different biases with respect to baryons and dark matter. Specifically, if we model
\beq
\delta^{A,B}_{g} = b_1^{A,B} \delta_m +b_r^{A,B} \delta_r + \ldots
\eeq
then
\beq\label{eq:multi_B}
\lim_{\k_1 \to 0} \langle \delta^{A}_{g}(\k_1) \delta_g^{A}(\k_2) \delta^{B}_g(\k_3) \rangle \to \frac{7\alpha^2 \fdm^2}{5}\frac{\k_1\cdot \k_2}{2 k_1^2} b_1^A P(k_1) P(k_2)  (b_1^B b_r^A-b_1^A b_r^B)\ .
\eeq
While this is a compelling target, especially because it is enhanced at long distances ($k_1 \to 0$), we see that it will require a somewhat more specialized analysis than even the galaxy bispectrum alone. Specifically, to produce a meaningful constraint or detection, we need to use multiple tracers with different biases and non-zero $b_r$.

\subsection{Backreaction and Expansion History}
\label{subsec:Backreaction}

We have seen that purely at the level of the cosmological density, it is possible to add a long range force that increases clustering via equivalence principle violation. However, long range forces require new light degrees of freedom. Given their coupling to dark matter, it is not a given that we can ignore the energy density or time evolution of the mediator fields on cosmological evolution.

We will use a specific model of a new long-range dark force to illustrate both the simplicity of these models and how they may evade other constraints, following~\cite{Bottaro:2023wkd,Bottaro:2024pcb}. We will start from the model of dark matter, in the form of a scalar $\chi$ and a light scalar force mediator $s$. We will couple the dark matter to this force via 
\beq\label{eq:DM_action}
S\supset \int d^4 x \sqrt{g} \left[\partial_\mu \chi \partial_\mu  \chi - m^2_\chi(s) \chi^2 - \frac{1}{2 G_s}  \partial_\mu s \partial^\mu s +\ldots \right] \ .
\eeq
where $G_s = \beta 4\pi G_{\rm N}$ and $m^2_{\chi}(s) = m^2 (1+2s)$, for simplicity\footnote{We will see that, with this choice, the comparison to the previous section can be made when $\beta \approx 2 \alpha^2$ in the regime of interest.}. Both fields are canonically coupled to the metric. The Standard Model is assumed to couple to this dark sector gravitationally, at least as far as the late universe is concerned. We will assume that the origin of the large density of the dark matter can be achieved without meaningful changes to this model (for example, via a WIMP-like scenario).

The key issue to notice here is that in the non-relativistic limit, $m^2_\chi(s) \chi^2 \approx \rho_{\rm dm}$, so there is a source of $s$ in the Lagrangian at finite density. Therefore, at the level of the homogeneous background, we must consider the time evolution of $s$. Since the number density of dark matter particles redshifts like $a^{-3}$, the energy density will obey
\beq
 \rho'_{\rm dm} + (3 \H  - \tilde m_s s' ) \rho_{\rm dm} = 0 \qquad \tilde m_s = \frac{\partial \log m_\chi(s)}{\partial s} \ .
\eeq
The evolution of dark matter density will therefore be modified if $s$ evolves in time. This is sourced by the non-zero $\rho_{\rm dm}$, via
\beq
s'' + 2 \H s' +4\pi \beta  G_N a^2 \tilde m_s \rho_{\rm dm} = 0
\eeq
Assuming matter domination so that  $8\pi G_N a^2 \rho_m = 3 \H^2 $, and $\tilde m_s \approx$ constant, we find that
\beq
s'' + 2 \H s' + \frac{3}{2} \H^2 \beta \tilde m_s \fdm =0 \to s - s_{\rm eq}= - 2 \beta \tilde m_s \fdm \log \tau/\tau_{\rm eq}  \ .
\eeq
where we used $\H = 2/\tau$ since we are working at linear order in $\beta$ and $s = {\cal O}(\beta)$. At linear order, we can now use $s' = - \H \beta \tilde m_s \fdm$, to find
\beq\label{eq:DM_scaling}
\rho_\chi= a^{-3 - \epsilon}  \qquad \epsilon \equiv  \beta \tilde m_s^2 \fdm\ .
\eeq
 Of immediate relevance is that the dark matter redshifts faster than would be expected in a $\Lambda$CDM cosmology. For example, if $\epsilon = 5\times 10^{-3}$, then $\Omega_{\rm dm}^{\rm BAO} \approx 0.97 \Omega_{\rm dm}^{\rm CMB}$. The impact on the expansion rate at linear order in $\beta$ is
\beq\label{eq:Modified_Expansion}
H^2 \propto  a^{-3}\left(1- \fdm \epsilon \log (a/a_{\rm eq}) \right) \approx a^{-3- \fdm \epsilon } \ .
\eeq
While the second equality is understood to be applied at linear order, the exponential will simplify the derivation of a number of equations.

The $s$ field itself carries energy density, mainly given by its kinetic energy $\sim \dot{s}^2$. The potential energy or energy of interaction of $s$ with the $\chi$ field has already been included in the $\chi$ energy density above.  From our solutions for the $s$ background, it can be seen that this kinetic energy $\sim \epsilon \rho_m$ and redshifts mainly like matter.  As a result, the $s$ background and fluctuations will behave like matter to order $\epsilon$, as shown in detail in~\cite{Bottaro:2023wkd}.

We see that in this model, $\beta \neq 0$ will appear to generate both kinds of negative neutrino mass. The results of the previous subsection tells us to expect $\mnuclustering < 0$ while backreaction should generate an effective $\mnubao < 0$. However, to determine the precise values, we will need to calculate the change to the growth including both the long range force and non-standard redshifting of the dark matter. There are two key changes that are relevant. First, the momentum density $\rho_\chi \v_\chi$ redshifts like $a^{-4}$ so that $\theta' +\H \theta \to \theta' +\H(1-\epsilon) \theta$. Second, because of the change to the redshifting of the dark matter, $\H^2 \propto a^{-1-\fdm \epsilon}$ so that $a(\tau) \propto \tau^{2-\fdm \epsilon}$ and $\H = (2-\fdm \epsilon)/\tau$. Putting this together, the linear evolution is described by
\bea
\delta'_{\rm dm } +\theta_{\rm dm} &=& 0 \\
\theta_{\rm dm}^{\prime}+\mathcal{H}(1-\epsilon) \theta_{\rm dm}+\frac{3}{2} \Omega_m \H^2 (\delta_m +  \beta f_{\rm dm} \tilde m_s \delta_{\rm dm}) &=& 0 \ .
\eea 
We can combine this with the evolution of the baryons to arrive at the evolution of $\delta_m =f_b \delta_b + f_{\rm dm} \delta_{\rm dm}$ and $\delta_r = \delta_{\rm dm} -\delta_b$. The equations we get are
\bea
\delta'_{m } +\theta_{m} &=& 0 \\
\theta_{m}^{\prime}+\mathcal{H}(1-f_{\rm dm}\epsilon) \theta_{\rm m}+\frac{3}{2} \Omega_m \H^2 (\delta_m +  f_{\rm dm} \epsilon \delta_{m}) &=& 0 \label{eq:fluid_dm_back} \ ,
\eea 
where we used $\beta \delta_{\rm dm} = \beta f_{\rm dm} \delta_m$ at linear order in $\beta$. Comparing Equation~(\ref{eq:fluid_dm_back}) and~(\ref{eq:fluid_dm_fix}) we see that the models with and without backreaction are related by $\epsilon =2\alpha^2 f_{\rm dm}$, or $\beta \tilde m_s^2 = 2\alpha^2$. Solving these equations for the growth function $\delta_m \propto D(\tau)$, we get
\beq
D(\tau) = \tau^\gamma \qquad \gamma = 2 + \frac{2}{5} f_{\rm dm}\epsilon \ .
\eeq
Finally, using $\H^2 \propto a^2 \rho_m$, we have
\beq
a(\tau) = c \tau^{2 -2 \fdm \epsilon} \to D(a) = a(\tau) \left(1 + \frac{6}{5} \fdm \epsilon \log a/a_{\rm eq} \right) \ .
\eeq
 We can compare this to the case without a change to the background, using $2\alpha^2 f_{\rm dm} =  \epsilon$, so that 
\beq
D_{\rm fixed-background}(a) = a(\tau) \left(1 + \frac{3}{5} \fdm \epsilon \log a/a_{\rm eq} \right) \ .
\eeq
In this sense, we see that the growth of structure in this model is enhanced by the evolution of the background.

The change to the background evolution also slightly changes the solutions to the perturbative equations. For example, $\tilde F_2$ is derived in~\cite{Bottaro:2023wkd}, where they find
\beq\label{eq:F2back}
\tilde F_{2,{\rm backreaction}}(\q_1,\q_2)  
=\frac{59}{30} + \frac{17}{12} \frac{\q_1\cdot \q_2}{q_1 q_2} \left(\frac{q_1}{q_2}
 +\frac{q_2}{q_1} \right) + \frac{13}{15} \frac{(\q_1\cdot \q_2)^2}{q_1^2 q_2^2}  \ .
 \eeq
This result can be derived using the same methods as in the previous section, with the appropriate changes to the equation for the time evolution. The key signal in the multi-tracer bispectrum is the same with and without back-reaction, but the exact relationship between $\epsilon$ and  signal amplitude changes slightly due to the different expansion history.

\subsection{Observational Status}
\label{subsec:Dark_Force_Observations}

Cosmological and astrophysical observations have a unique potential to probe dark sector physics. On these scales, the gravitational influence of the dark matter has been measured at high significance, allowing for more precise determinations of the microphysics that influences the distribution of dark matter in space and evolution in time. However, the nature of the observables on (sub-) galactic and cosmological scales is sufficiently distinct that we will discuss each separately.

\subsubsection*{Current Cosmological Measurement}

Long range dark forces have two distinct physical effects in linear cosmology, as we demonstrated above. First, they change the clustering of matter by changing the growth rate for the dark matter during matter domination:
\beq
D(a) = a(\tau)\left(1+ \frac{6}{5} f_{\rm dm} \epsilon \log a\right) \ .
\eeq
This is equivalent to a negative clustering mass
\bea
&&f_\nu^{\rm clustering} = f_\nu- 2 f_{\rm dm} \epsilon \\
% \to && \mnuclustering = - 60 \, {\rm meV} \left( \left(\frac{f_{\rm dm}}{0.84} \right)^2 \frac{\beta}{3 \times 10^{-3}} - 1\right) \ ,
\to && \mnuclustering =  60 \, {\rm meV} - 60 \, {\rm meV} \left(\frac{f_{\rm dm}}{0.84} \right)^2 \frac{\beta}{3 \times 10^{-3}}  \ ,
\label{eq:Dark_Force_mnuclustering}
\eea
where the first term is present because we defined $\sum m_\nu = 60$ meV in $\Lambda$CDM ($\beta=0$). This is a relevant point of comparison. The measurement of $ \mnuclustering  =  -160 \pm 90$~meV corresponds to $\beta = \left( 1.1 \pm 0.45\right) \times 10^{-2}$. This is much larger than the measured value of $\beta = (4 \pm 2)\times 10^{-3}$. Furthermore, if we eliminate the change to the background from the calculation of the growth, the corresponding value of $\beta$ is larger by a factor of two. 

Importantly, back-reaction on the force mediator changes the mass of the dark matter particles over cosmic history. This causes the energy density in dark matter to decrease faster than $a^{-3}$, scaling instead as
\beq
\rho_{\rm dm} \propto a^{-3 - f_{\rm dm} \epsilon}
\eeq
Therefore, between redshift of  $z=1100$ and DESI observations at $z\approx 1$, the change to $\Omega_m$ from the $\Lambda$CDM value is 
\beq
\Omega_m = \Omega_m^{(\Lambda \mathrm{CDM})}  \left(1 - 0.02 \times \frac{f_{\rm dm} \epsilon}{3 \times 10^{-3}} \right)
\eeq
This again will behave like negative neutrino mass, now through the effect on the BAO via the change to the expansion rate, such that  
\bea
&&f^{\rm BAO}_\nu =f_\nu  - 0.013 \times \left(\frac{f_{\rm dm}}{0.84} \right)^2 \frac{\beta}{3 \times 10^{-3}}\\
\to && \mnubao = 60\, {\rm meV} - 190 \, {\rm meV} \left(\frac{f_{\rm dm}}{0.84} \right)^2 \frac{\beta}{3 \times 10^{-3}} 
\label{eq:Dark_Force_mnubao}
\eea
We see that for a given $\beta$, the effect on $\Omega_m$ is about 3 times larger than the effect on the growth of structure. Moreover, if we take the measurement $\beta = (4 \pm 2) \times 10^{-3}$ using DESI DR1 in~\cite{Bottaro:2024pcb} and apply this estimate we would have $\sum  m^{\rm BAO}_{\nu} = -190 \pm 130$. This is consistent with the direct measurement of $\sum m^{\rm BAO}_{\nu} = -193 \pm 83$ meV~\cite{Lynch:2025ine} using similar data.

%%%%%%%%%%%%%%
%% mnu Dark Force
%%%%%%%%%%%%%%
%%%%%%%%%%%%
\begin{figure}[t!]
    \centering
    \includegraphics[width=0.95\textwidth]{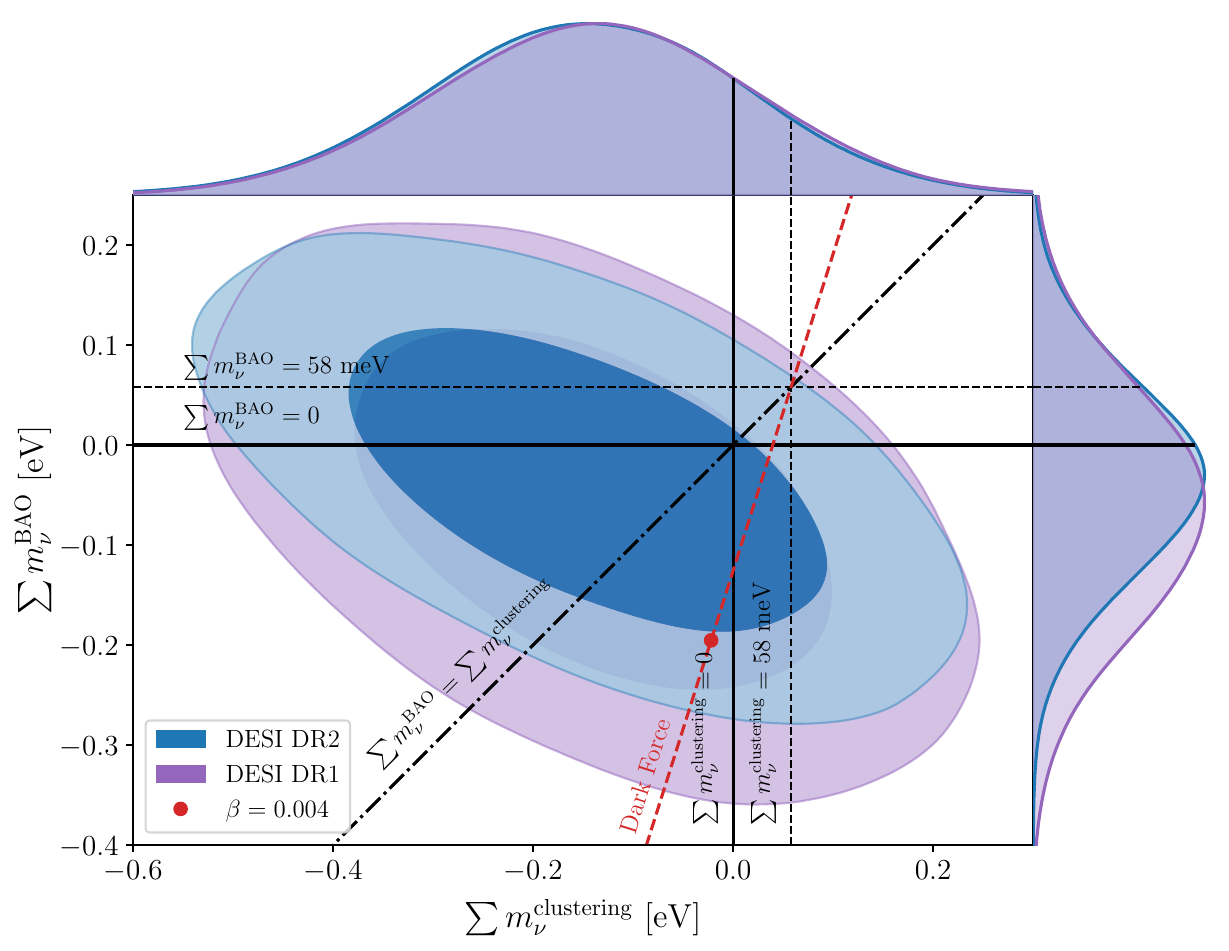}
    \caption{Predictions of dark force model translated to $\mnubao$ and $\mnuclustering$, compared to constraints on these parameters derived from CMB+DESI BAO.  We show constraints for both DESI DR1 and DESI DR2, and plot the best-fit point of the parameter $\beta$ found in Ref.~\cite{Bottaro:2024pcb}, derived there using DESI DR1 BAO.  For the purpose of plotting these predictions, we assume a model that contains neutrinos with $\mnu=58$~meV such that the line passes through the point $\mnuclustering=\mnubao=58$~meV when $\beta=0$.}
    \label{fig:mnu_dark_force}
\end{figure}
%%%%%%%%%%%%%

As we have emphasized from the beginning, we can understand these results better if we work in the two-dimensional clustering-BAO plane. In Figure~\ref{fig:mnu_dark_force}, we show the line predicted by the long range force model in the $\mnuclustering$ and $\mnubao$ plane and plot the point $\beta = 4 \times 10^{-2}$ along the line. We see that this is quite consistent with our measurement when restricted to this line.  Additionally, it is possible to anticipate from our constraints on $\mnuclustering$ and $\mnubao$ that an analysis of the long-range force model with DESI DR2 data would provide a constraint of approximately $\beta=0.0025 \pm 0.0015$. We can also understand why the uncertainty in the measurement of $\mnubao$ in~\cite{Lynch:2025ine} is much smaller than our estimate using $\beta$: if we restrict to the line $\mnuclustering = 58$~meV, we are restricting to a corner of the 68\% confidence region and therefore have a stronger constraint than the $\beta$-line which is closer to the center of the allowed region. Although clustering has a smaller effect on the overall signal, the impact of clustering meaningfully changes the uncertainties derived from current data for this reason.

\subsubsection*{Future Tests with the Bispectrum}

The key property of these models is the violation of the equivalence principle for the dark matter. This can lead to violations of the single-field consistency conditions in the bispectrum that are potentially measurable in galaxy surveys. We would therefore like to assess if this specific aspect\footnote{The changes to the growth of structure will also impact the bispectrum and therefore will contribute to the Fisher matrix for $\beta$. However, the unique feature of this model is the equivalence principle violation, which sets it apart from other contributions to the bispectrum.} of the bispectrum can be observed with DESI. 

The potential for observing equivalence principle violation from the bispectrum was first explored in~\cite{Creminelli:2013nua}. They found a 1 $({\rm Gpc}/h)^3$ survey with $k_{\rm max}=0.1 \, h \, {\rm Mpc}^{-1}$ would have a sensitivity of $\sigma(\beta) = 2\times 10^{-3}$, where we used $\beta = 2\alpha^2$. This question was revisited more recently in~\cite{Bottaro:2023wkd}, where it was found that $\sigma(\beta)\approx 5 \times 10^{-3}$ with Euclid and $k_{\rm max} = 0.11 \, h \, {\rm Mpc}^{-1}$. This forecast was calculated using the 1d Fisher matrix with only the squeezed modes where $5 k_1 < k_2,k_3$. The full bispectrum contains more information relevant to $\beta$ (for 1d Fisher) but they find this is subdominant to the CMB + BAO constraints when marginalizing over cosmological parameters.

Although the bispectrum provides a weaker constraint on $\beta$ than CMB + BAO, the fact that CMB + DESI favors $\beta \neq 0$ makes such a measurement compelling nonetheless. Concretely, the squeezed bispectrum signal is completely independent of the CMB, and therefore a measurement via the bispectrum with similar sensitivity could provide an non-trivial check on the non-zero measurement of $\beta$ from the expansion history. 

\begin{figure}[t!]
    \centering
    \includegraphics[width=\textwidth]{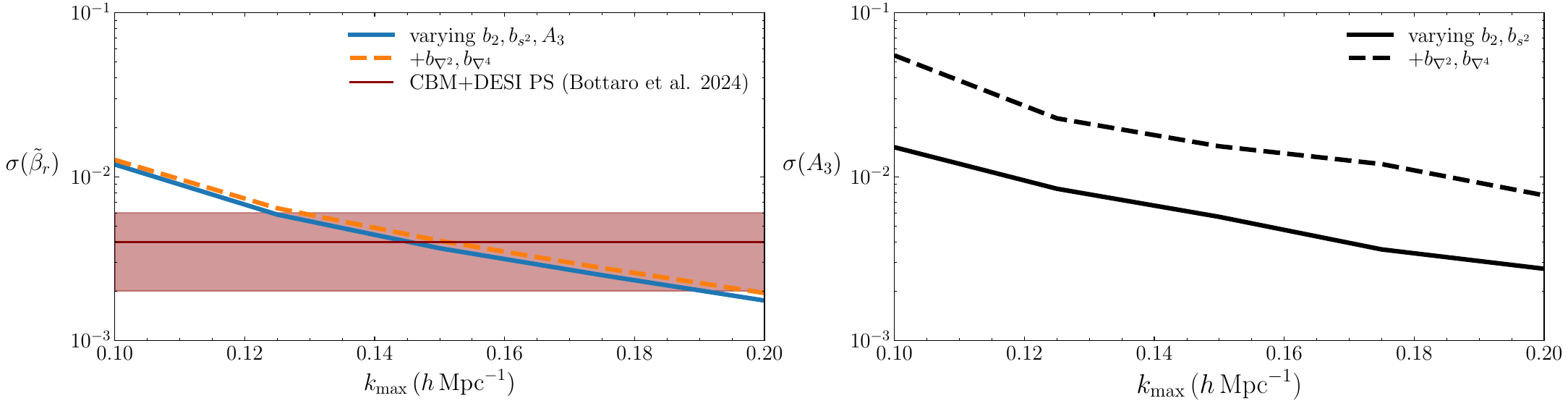}
    \caption{{\it Left:} Forecasts for $\sigma(\tilde \beta_r)$, where $\tilde \beta_r = \beta \fdm^2 (b^B_1 b^A_r -b_1^A b_r^B)/(\bar b_1^2)$, for the DESI bispectrum. The CMB+PS result from~\cite{Bottaro:2024pcb}, $\beta = 0.004\pm 0.002$, is shown in red assuming $\tilde \beta_r = \beta$. The precise relationship between $\beta$ and $\tilde \beta_r$ depends the possible values of $b_1 b_r$ when splitting the DESI samples which are unknown. {\it Right:} Forecasts for $\sigma(A_3)$, the amplitude of the matter bispectrum, with DESI. Since $A_3$ is degenerate with small scale physics, comparing the left and right panels demonstrates a useful comparison of the behavior of the forecasts when the signal is and is not protected from astrophysical degeneracies by the equivalence principle.}
    \label{fig:bispectrum}
\end{figure}

We forecast the sensitivity of the DESI bispectrum to equivalence principle violation as follows: we define the bias expansion to second order as
\beq
\delta_g(\x) = (b_1+b_{\nabla^2} \nabla^2 + b_{\nabla^4} \nabla^4) \delta_m + b_r \delta_r +b_2 \delta_m^2 + b_{s^2} \nabla_i \nabla_j \Phi \nabla^i \nabla^j \Phi \ .
\eeq
We additionally allow the amplitude of the matter bispectrum, shown in Eq.~(\ref{eq:Bmatter}), to vary as a parameter $A_3$. The parameter $A_3$ captures in the information in $\beta$ associated with changes to the growth rate (up to the $z$-dependence). We have introduced gradient bias terms $b_{\nabla^2}$ and $b_{\nabla^4}$ in order to ensure we are getting most of the constraining power from the squeezed modes and not from the detailed shape in the equilateral limit. Rather than forecasting $\beta$ directly, we use $\tilde \beta \equiv \beta \fdm^2 (b_1^B b_r^A- b_1^A b_r^B)/b_1^2$ so that 
\beq
\begin{aligned}
\frac{\partial}{\partial \tilde \beta} B_g^{AAB}(k_1,k_2,k_3) =  b_1^3 \bigg( & F_2(\k_2,\k_3)P_m(k_2)P_m(k_3)  
+\tilde F_2(\k_1,\k_3) P_m(k_1)P_m(k_3) \\
&+ F_2(\k_1,\k_2)P_m(k_1)P_m(k_2) \bigg) \ ,
\label{eq:AAB_bispectrum}
\end{aligned} 
\eeq
where $B^{AAB}$ is the multi-tracer bispectrum defined\footnote{There are additional contributions to $B^{AAB}$ that we are neglecting because they do not contribute to the equivalence violating behavior when $k_1 \ll k_2,k_3$. Figure~\ref{fig:bispectrum} confirms that the forecasts for $\tilde \beta$ are insensitive to how we model the bispectrum away from this limit.} by Equation~(\ref{eq:multi_B}). From here we can forecast $\tilde \beta$ using either $\tilde F_2$ from Eq.~(\ref{eq:F2noback}) without backreaction or $\tilde F_2$ from Eq.~(\ref{eq:F2back}) with backreaction. The difference between these two can be absorbed into the definition of $\tilde \beta$, but we use the result without backreaction as our baseline forecast. We will compare forecasts with and without these gradient terms to illustrate the sensitivity of the forecasts to short distance scales. Additional details can be found in Appendix~\ref{app:bispectrum}.

The results of our forecast show that the bispectrum sensitivity of DESI to $\beta$ is potentially within reach of the CMB + BAO measurement, for sufficiently large $k_{\rm max}$ (taken to be $z$-independent). Figure~\ref{fig:bispectrum} shows the sensitivity of the equivalence principle violating bispectrum to $\beta$ as a function of $k_{\rm max}$. The red line and band show the current best fit and 68\% confidence interval from Planck + DESI DR1 BAO, $\beta = 0.004\pm 0.002$. The solid lines is marginalized over $b_2$ and $b_{s^2}$ while the dashed line includes $b_{\nabla^2}$ and $b_{\nabla^4}$ as well. We see from the contrast between the left and right panel that the $\tilde b_r$ forecast is less sensitive to marginalizing over the short distance physics, as we would expect given the long distance nature of the signal.

%%%%%%%%%%%%
%%%%%%%%%%%%%

The forecasts suggest that a bispsectrum measurement could provide useful information about $\beta$. However, to be competitive with measures from CMB + BAO, we require a higher value of $k_{\rm max}$ than was used by~\cite{Bottaro:2023wkd} in their otherwise similar forecast. It is important that we have not included any contributions from the $\beta$-dependence of the bispectrum itself, meaning that all the signal-to-noise arises from the violation of the single-field consistency conditions. This information should be robust to nonlinear physics on smaller scales and therefore more amenable to a larger value of $k_{\rm max}$.

The most significant limitation of the equivalence principle violating signal is the requirement of measuring multiple tracers with different values of $b_1 b_r$. While there is some preliminary work on values of $b_r$ theoretically~\cite{Chen:2019cfu,Barreira:2019qdl}, to determine a meaningful upper limit on $\beta$ would require some way to determine $b_r$ reliably for each sample. Nevertheless, detection of a signal does not require that we know the true values of $b_r$, only that $b_1^A b_r^A \neq b_1^B b_r^B$. Given the unusual hints of such a signal, performing such a multi-tracer bispectrum analysis could be very informative, despite these limitations.

\subsubsection*{Astrophysics}

Inside collapsed objects such as galaxies or galaxy clusters, dark matter plays a dominant role in the dynamics of objects and the trajectory (and redshifts) of light. Careful measurements of these phenomena can, in principle, provide important constraints on, or even a signal of, new forces that impact the distribution of dark matter and its influence on other tracers. Some of the best current constraints on these forces come from the observations of the Bullet Cluster \cite{Bogorad:2023wzn} and tidal streams of the Sagittarius dwarf galaxy \cite{Kesden:2006zb, Kesden:2006vz, Keselman:2009oaz}. In both cases, the dynamics of baryons compared to dark matter provides important window into the dark sector.

In the case of the Bullet Cluster, a new force on dark matter would cause the two clusters to be attracted together, thus raising the impact velocity in the collision.  Observations of the shock front of the gas in the cluster constrain this impact velocity and thus constrain the strength of a new force acting on dark matter \cite{Bogorad:2023wzn}. This is one of the strongest constraints on such a new force.  However at long ranges for this new force the constraint asymptotes to $\beta \gtrsim 1$.  Thus it does not constrain forces with infinite range and small $\beta$ which is our region of interest here. In this sense, measurements of the Bullet Cluster are consistent with the observation of $\beta = (4\pm2)\times 10^{-3}$ from cosmology, and will be challenging to improve with additional observations of cluster dynamics.

Observations of the tidal tails of the Sagittarius dwarf galaxy can test the equivalence principle on dark matter versus baryons \cite{Kesden:2006zb, Kesden:2006vz, Keselman:2009oaz}.
If dark matter couples to a new, attractive force that does not couple to baryons, this will change the tidal tails of satellite galaxies undergoing disruption.  More stars will be found in the trailing tidal tail than in the leading tail.
Observations of roughly equal tails in the Sgr dwarf galaxy have thus been used to constrain the strength of a new force coupling only to dark matter to about 10\% of gravity \cite{Kesden:2006zb}, thus setting the limit $\beta \lesssim 0.1$, at infinite range\footnote{Interestingly, however it does not constrain forces with large $\beta$, and in fact $\beta \approx 1$ is not constrained \cite{Keselman:2009oaz}.  It is thus complementary to the Bullet Cluster constraint which constrains larger $\beta$. The exact upper limit of the Sagittarius constraint is not known.}.
However, these observations of and predictions for the shape of the Sgr dwarf have not been made precisely enough to probe the region of interest here with $\beta \lesssim 10^{-2}$.  For a plot comparing these astrophysical limits from the bullet cluster and the Sgr dwarf and the cosmological limits see for example Figure 1 of \cite{Bogorad:2023wzn}.

Note that the cosmological limits on new forces (at the longest ranges) have  been calculated in the region $\beta < 1$ and the expansions break down at large $\beta$, thus the limits do not cover the region $\beta \gtrsim 1$ \cite{Archidiacono:2022iuu, Bottaro:2023wkd}.  The astrophysical constraints arising from the Bullet Cluster only cover the region $\beta > 1$.  Thus the astrophysical and cosmological limits are quite complementary.
%Cosmology appears sensitive to weaker forces (lower $\beta$) than can be probed with these astrophysical systems. 
Yet, it is noteworthy that the physical origin of the astrophysical constraints share similar behavior to the signal of long range forces in cosmology.  We have already seen that there is a qualitative change in the origin of the cosmological constraint when we only treat the system as if it were a new long-range force on dark matter. The current bounds require that we describe the backreaction on the force mediator (the scalar field $s$ in our notation).  And the resulting dynamics of this new light field changes the cosmological expansion in a way that is the dominant effect on current data.  Similarly, in astrophysical systems such as the Bullet Cluster, in large parts of  parameter space the dynamics of the force mediator must be considered and the equation of motion for the mediator must be solved as well, just as we have seen in cosmology.  However, in astrophysical systems such as the Bullet Cluster or ultra-faint dwarf galaxies, the region where the dynamics of the mediator matters is generally at larger $\beta \gg 1$, not surprisingly since the dark matter densities in these astrophysical systems tend to be lower than the relevant densities in cosmology (e.g.~at last scattering).

Ultimately though, the region of parameter space of interest here for these cosmological observations, at low $\beta$ and infinite range for the new force, is not currently constrainable by astrophysical observations. It appears difficult to significantly strengthen these astrophysical constraints, and thus cosmology appears to be the best way to probe new forces with very long range and small $\beta$.

\section{Expanding the Possibilities}\label{sec:expanding}

We have seen in the previous section that there are a wide range of models that can be mapped to the $\mnubao$-$\mnuclustering$ plane. This parameterization shows the potential for many ideas to resolve the current tension with the experimental lower limit on $\sum m_\nu$ of 58~meV. In this section, we will discuss other ideas that might achieve the same qualitative changes to cosmic observables through different microphysics.

\subsection{Altering the BAO}
\label{subsec:AlteringBAO}
\subsubsection*{Curvature}

The observation that $\mnubao <0$ is a good fit to CMB + BAO tells us that the measurements of energy densities of the universe during matter domination at $z=1100$ and $z= {\cal O}(1)$ appear to be inconsistent with $\Lambda$CDM. Introducing spatial curvature is a natural way to try accommodate this difference (see also~\cite{Chen:2025mlf}).

As far as the background evolution is concerned, for $1100> z> 1$, we have 
\beq
H^2 = H_0^2 (\Omega_m a^{-3} +\Omega_K a^{-2}) \ .
\eeq
Therefore, if our effective $\Omega_m(z=1100)^{\Lambda{\rm CDM}} = \Omega+ 1100 \, \Omega_K$, then $f_\nu^{\rm BAO} \approx -  1100 \Omega_K/\Omega_m$. Therefore, we expect marginalizing over $\Omega_K$ should have some approximate degeneracy with $\mnubao$.

%%%%%%%%%%%%
\begin{figure}[t!]
    \centering
    \includegraphics[width=0.95\textwidth]{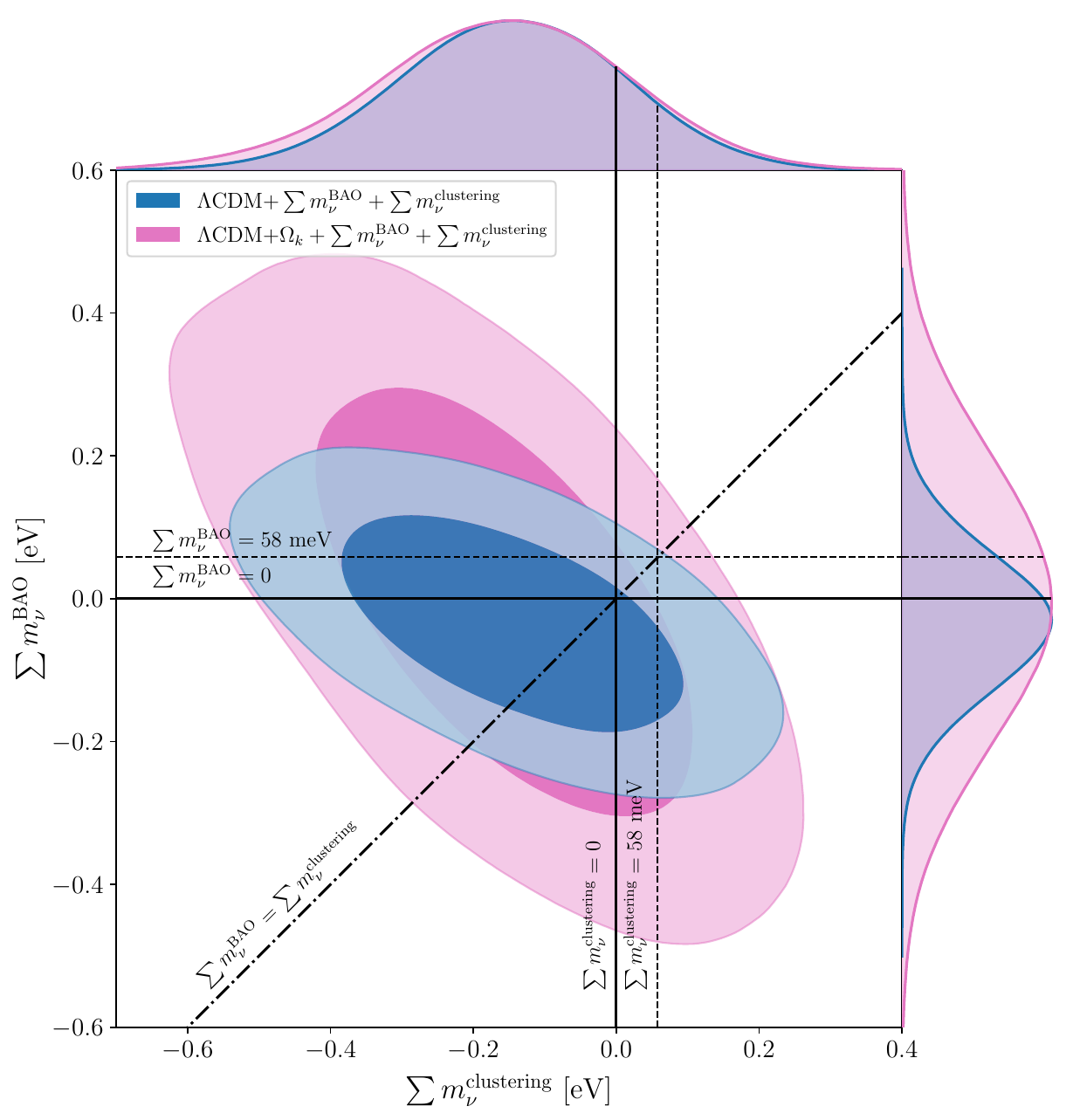}
    \caption{Same as Figure~\ref{fig:mnu_BAO_clustering}, including constraints for the $\Lambda$CDM+$\Omega_K$+$\mnubao$+$\mnuclustering$ cosmology.  }
    \label{fig:mnu_omk}
\end{figure}
%%%%%%%%%%%%%

In Figure~\ref{fig:mnu_omk}, we show the impact of allowing $\Omega_K$ to vary when measuring $\mnuclustering$ and $\mnubao$. As expected, the uncertainty on $\mnubao$ increases significantly; we find $\mnubao=0\pm 200$~meV, an uncertainty twice as large as in $\Lambda$CDM$+\mnubao+\mnuclustering$ cosmology. In contrast, marginalizing over $\Omega_K$ has no meaningful impact on $\mnuclustering$, which remains peaked at negative values, giving $\mnuclustering=-160\pm 180$~meV.  We note that we find the curvature parameter to be consistent with zero, $\Omega_K = 0.0005\pm 0.0023$.

Curvature illustrates a common element of trying to address $\sum m_\nu$ by changing the expansion history. Since $\mnubao$ is defined by expansion, it is generally sensitive to the model and can be shifted toward positive values fairly generically. However, these models do not improve the clustering measurement or, in this case, even appreciably change the uncertainty. In this regard, the large amplitude of the CMB lensing power spectrum appears to be fairly robust to the changes to the expansion history.

\subsubsection*{Decaying Dark Matter}

A natural strategy for achieving $\mnubao < 0$ is to have the inferred value of $\Omega_m$ decrease between $z =1100$ and $z=1$. A physically reasonable way one could achieve this is by having a fraction of the dark matter decay. This specific idea was proposed in Ref.~\cite{Lynch:2025ine} as a physical model for $\mnubao$. It has also been pursued in other specific models, such as in Ref.~\cite{DESI:2025ffm}. The details of how the dark matter decays and its other interactions leads to different predictions in detail, but the essence of how they address the tension between DESI BAO and the CMB is the same.

One of the challenges with decaying dark matter observed by~\cite{Lynch:2025ine} is that reducing the physical amount of matter in the universe, rather than just the matter contributing to expansion, is that it reduces the lensing amplitude as well. This means the lensing power spectrum will be suppressed, rather than enhanced. In this sense, having a fraction of the matter $f_{\rm decay}$ decay will look like $\mnubao \propto - f_{\rm decay}$ and $\mnuclustering \propto +f_{\rm decay}$. In principle, with the correct slope of this line in the $\mnubao$-$\mnuclustering$ plane, such a model could be consistent with observations. However, detailed analyses such as \cite{McCarthy:2022gok,Lynch:2025ine} do not find that the models they studied reduced the tension.

There are two straightforward extensions of this model that could allow for better agreement with data. This first is to further change the low redshift expansion history~\cite{DESI:2025ffm}. Alternatively, the fraction of dark matter that decays (or annihilates) could be coupled to a long range force along the lines of our model in Section~\ref{sec:darkforce}. We have already seen that this alters both the clustering and expansion, while introducing a decay would allow for additional freedom in this plane.

\subsubsection*{Negative Energy}

A negative contribution to the energy density of the universe could mimic $\mnu < 0$. Simply adding a fluid with an arbitrary equation of state would then include $\mnubao < 0$ but also dynamical dark energy, as discussed in Section~\ref{subsec:constraints}. Dynamical dark energy has been investigated from many perspectives given the suggestion that DESI BAO prefers such models. However, given complete freedom in the equation of state, the data does seem to prefer phantom dark energy, which would violate the null energy condition (NEC).

On theoretical grounds, violating the NEC raises a number of concerns. Fluctuations of these models are often unstable, although this conclusion depends on details of the equation of state and interactions. In addition, quantum field theory in flat space is known to obey the averaged null energy condition, which would seem to strongly constrain the possibility of violating the NEC with conventional degrees of freedom.

However, it is possible to find sources of NEC violation associated with non-dynamical objects or boundaries. One such example is orientifold planes in string theory, whose source of negative energy has long been used in string compactification to stabilize extra dimensions~\cite{Verlinde:1999fy,Giddings:2001yu}. More recently, interest in time-like boundaries in our four-dimensional universe has also raised the possibility of controlled NEC violation relevant for the expansion of the universe. A specific example with a uniform density of spherical boundaries was recently introduced and analyzed in~\cite{Philcox:2025faf}. The main phenomenological feature is to introduce a negative source of energy density that redshifts as $a^{-1}$ without additional fluctuations, as the boundaries are non-dynamical. The difference between an energy scaling as $a^{-1}$ from boundaries and from $\Omega_K$, which scales as $a^{-2}$ does not appear to be significant over the range of redshifts probed by DESI. The analysis of~\cite{Philcox:2025faf} finds impact on cosmological parameters, including neutrino mass, is similar in both cases. In this regard, a collection of spherical boundaries is likely to improve the $\mnubao$ bounds without meaningfully impacting the allowed range of $\mnuclustering$ as we found for $\Omega_K$.

\subsection{Altering the CMB}
\label{subsec:AlteringCMB}
\subsubsection*{Biasing the Optical Depth}

\begin{figure}[t!]
    \centering
    \includegraphics[width=0.95\textwidth]{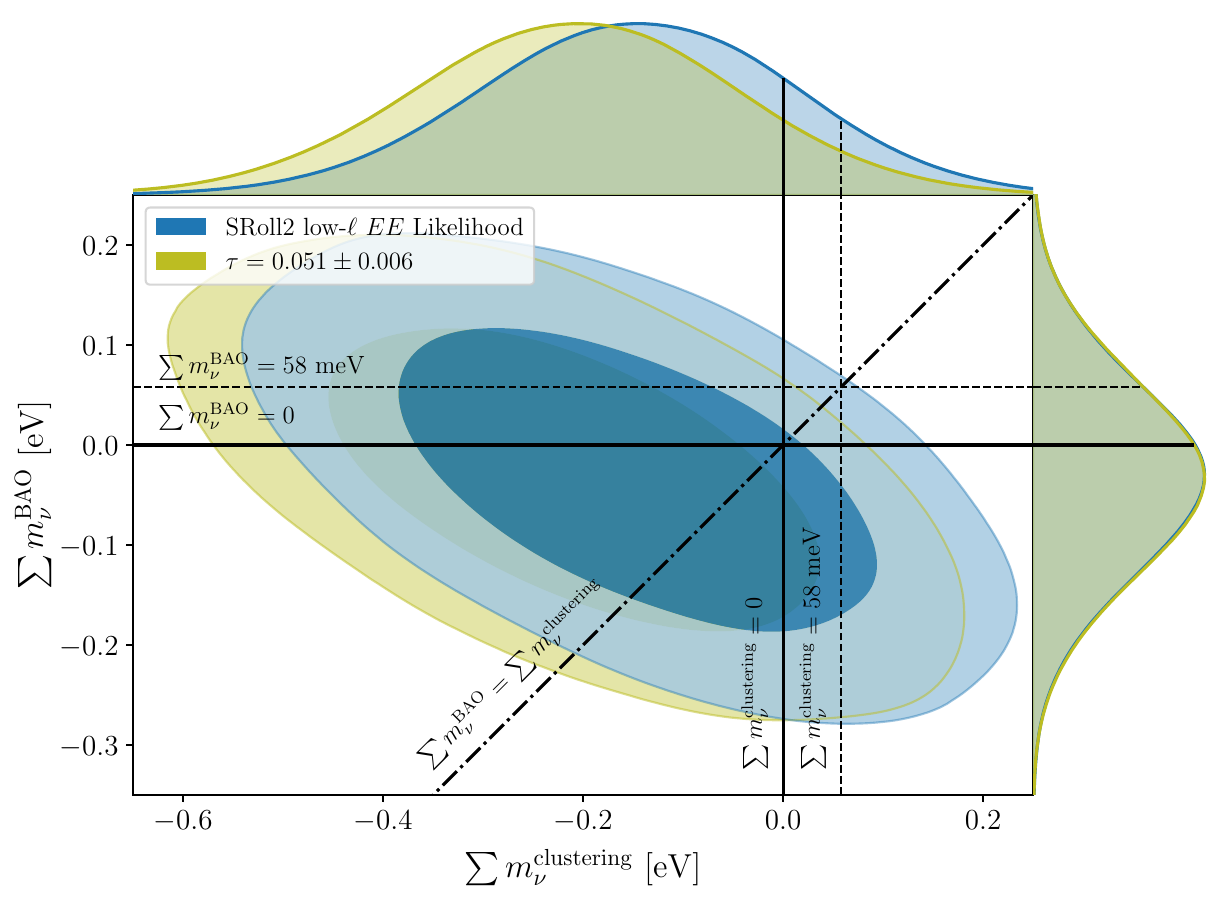}
    \caption{Same as Figure~\ref{fig:mnu_BAO_clustering}, comparing constraints derived from using the \texttt{SRoll2} analysis of low-$\ell$ $EE$ likelihood~\cite{Pagano:2019tci} to those derived by instead using a prior on the optical depth $\tau = 0.051\pm0.006$~\cite{Planck:2020olo} as in the SPT-3G D1 analysis~\cite{SPT-3G:2025bzu}.  The case with the $\tau$ prior favors more negative values of $\mnuclustering$ while $\mnubao$ is unaffected.}
    \label{fig:mnu_tauprior}
\end{figure}

It has long been known that the uncertainty of the optical depth is a central challenge to a cosmological neutrino mass measurement~\cite{Allison:2015qca}.  Neutrino mass constraints are sensitive to the treatment of the low-$\ell$ CMB polarization and the prior on the optical depth that is imposed.  The recent analysis of SPT-3G data~\cite{SPT-3G:2025bzu} utilized a prior on the optical depth of $\tau=0.051\pm0.006$~\cite{Planck:2020olo} rather than the \texttt{SRoll2} low-$\ell$ $EE$ likelihood that was employed in the ACT DR6 analysis~\cite{ACT:2025fju,ACT:2025tim} and that we have employed throughout this work (the \texttt{SRoll2} analysis favors $\tau=0.059\pm0.006$ in $\Lambda$CDM cosmology~\cite{Pagano:2019tci}).  In Figure~\ref{fig:mnu_tauprior}, we show constraints on the neutrino mass parameters imposing the same $\tau$ prior as used in the SPT-3G analysis.  As can be seen, the lower value of $\tau$ imposed by the prior favors lower values of $\mnuclustering$ while $\mnubao$ is unaffected.  This choice of prior on the optical depth also at least partially accounts for tighter constraints on the physical neutrino mass found in the SPT-3G analysis, as can be seen reflected in how the two-dimensional contours intersect with the $\mnuclustering=\mnubao$ line in Figure~\ref{fig:mnu_tauprior}.

%%%%%%%%%%%%
%%%%%%%%%%%%%

A biased measurement of $\tau$ was raised as a possible explanation for the negative neutrino mass signal in~\cite{Craig:2024tky}. It was further shown that an optical depth of $\tau =0.09 \pm 0.012$ is preferred by just the high-$\ell$ data~\cite{Sailer:2025lxj,Jhaveri:2025neg} (see also~\cite{Giare:2023ejv,Allali:2025wwi}). Not only is this value consistent with the WMAP central value, it would bring both the clustering and expansion history into agreement with $\Lambda$CDM cosmology including $\mnu=58$~meV for CMB+BAO (the tension with uncalibrated supernovae remains~\cite{Jhaveri:2025neg}). A large physical value of the optical depth has implications for the ionization history that can be tested using the Lyman-$\alpha$ forest (Gunn-Peterson)~\cite{Fan:2005es,McGreer:2014qwa,Becker:2014oga}, CMB secondaries~\cite{Ferraro:2018izc,Yu:2018tem,Alvarez:2020gvl}, and direct observations of high redshift objects~\cite{Robertson:2013bq,Robertson:2015uda,Munoz:2024fas}. At present, this large value of $\tau$ is in $2\sigma$ tension with measurements of the patchy kinematic Sunyaev-Zeldovich effect from SPT when accounting for constraints on the end of reionization from Lyman-$\alpha$ forest observations~\cite{Cain:2025usc}.

New physics that affects low-$\ell$ polarization directly could bias the measurement of the optical depth. At high-$\ell$, we measure $e^{-2\tau} A_s$ with very high precision, but at low-$\ell$ we are sensitive to $A_s$ (or $\tau$) directly. In $\Lambda$CDM, the signal of a larger value of $\tau$ is enhanced power, particularly in $EE$, for $\ell < 30$. If new physics were to suppress the polarization signal on these scales, it could accommodate a larger value of $\tau$ and therefore a larger value of $A_s$.
For such a model to work, we would need to change the $EE$ power spectrum on scales $\ell < 30$, while leaving higher-$\ell$ data unchanged. One such example was recently proposed in Ref.~\cite{Namikawa:2025doa} using an ultra-light axion ($\phi$) with the usual coupling to photons, $g_{a \gamma\gamma} a \tilde F F$.  If the axion evolves significantly during reionization, the polarization rotation caused by the axion field can wash out the low-$\ell$ polarization generated around reionization, similarly to how early-time axion oscillation can wash out polarization generated during recombination~\cite{Fedderke:2019ajk}.  This axion evolution will generically lead to (isotropic) rotation of the polarization sourced at recombination, which leads to constraints on this scenario from limits on the $TB$ and $EB$ power spectra~\cite{Komatsu:2022nvu}.  However, if the rotation angle is nearly an integer multiple of 180~degrees, the impact on the observed polarization sourced at recombination is small (and the rotation angle can be made to account for recent hints of non-zero isotropic polarization rotation~\cite{Minami:2020odp,Diego-Palazuelos:2022dsq,Eskilt:2022wav,Eskilt:2022cff}). This mechanism thereby allows for suppressed low-$\ell$ polarization, allowing for the true optical depth $\tau$ to be larger than that inferred from the CMB when assuming $\Lambda$CDM cosmology.

Of course, this axion model requires a few orders of magnitude tuning to set the overall polarization rotation angle from recombination to today close to an integer multiple of 180 degrees.
However there are modifications to the minimal axion model that can still bias the measurement of optical depth with reduced or eliminated tuning, at the cost of a more complicated model.
For example, the axion could start near zero at recombination and then be driven to a larger value by energy transfer from another field before reionization. In this case the amplitude of oscillation at reionization can naturally be much larger than the value of the axion field at recombination.  This already reduces the tuning needed: as we will see below, the value of the axion field today just needs to be about 10\% tuned to be near enough to zero to satisfy the bound on isotropic polarization rotation.

In more detail, if the polarization angle oscillates with amplitude $\sim 90^\circ$ during reionization  (and a period somewhat shorter than the length of reionization), it could wash out the polarization produced at that time by a factor  $\sim 2$ \cite{Fedderke:2019ajk}  which would allow the large values of $\tau$ suggested by the high-$\ell$ CMB data~\cite{Sailer:2025lxj,Jhaveri:2025neg}.  If the energy in the axion is primarily in the zero mode (or nonrelativistic modes) then the axion field redshifts $\propto a^{-3/2}$ and hence the amplitude of the oscillation of the polarization rotation angle does as well.  Thus if we start with a $90^\circ$ oscillation amplitude at reionization ($z = {\cal O}(10)$) the amplitude of the oscillation of the angle today will be around $3^\circ$.
The period of the oscillation is set by the (inverse) axion mass, $m_a^{-1}$,
%$m_\phi^{-1}$, 
and has a few constraints.  It must be shorter than about tens of millions of years in order to oscillate sufficiently rapidly at reionization, and longer than about 30 years for the energy density in the axion to be below the dark matter density at reionization\footnote{The energy density in the axion is roughly $m_a^2 a^2$. In order to wash out the signal at reionization, the  polarization rotation angle we need is $g_{a \gamma\gamma} a \approx \pi/2$. Imposing that the maximum coupling strength allowed by CAST is $g_{a \gamma\gamma} \lesssim 7 \times 10^{-11} \, \text{GeV}^{-1}$, one finds the axion period must be longer than about 30 years to achieve a 90 degree rotation angle at reionization.}.  
Since the period is longer than 30 years, the axion field has an effectively fixed value in CMB measurements today and this gives a static polarization rotation angle somewhere in the 3 degree range.  But it must be less than about 0.3 degrees to satisfy the isotropic polarization rotation constraints, which is about a 10\% tuning.  In this case we would expect that the isotropic polarization rotation will be measured to be as large as possible, near the current hints of a non-zero value \cite{Minami:2020odp,Diego-Palazuelos:2022dsq,Eskilt:2022wav,Eskilt:2022cff}.

Even this 10\% tuning could possibly be eliminated by giving the axion a new friction mechanism that damps it sufficiently between reionization and today.  For example thermal friction (see e.g.~\cite{Berghaus:2019whh}) with the axion coupling to a dark gauge sector could work, though the axion would need a coupling strength to that new gauge sector several orders of magnitude stronger than to the photon in order to have sufficient friction.
In this case, the isotropic birefringence from the axion may be too small to be observable. Interestingly, the transfer of energy from the axion into the dark sector could behave like decaying dark matter, which has been shown in Ref.~\cite{Lynch:2025ine} to mitigate the preference for negative neutrino mass that comes from measurements of the expansion history.  In the language we use here, the axion construction could address the preference for negative $\mnuclustering$ via polarization washout at reionization, and it could address the preference for negative $\mnubao$ by transferring energy from matter to radiation.

There are multiple mechanisms that could dump energy into the axion to push the axion to large values before reionization.  For example, if it is coupled to another field which stores the energy density and then starts oscillating at the relevant time, then it will drive the axion up its potential (either directly or through parametric resonance).  Or possibly a new contribution to the axion potential with a different minimum could turn on around the time of reionization (though would need to turn off afterwards).  Either way the physical effect would be to washout the polarization produced during reionization. 
Furthermore, the evolution of the axion between reionization and today gives a model-independent prediction that various astronomical sources would exhibit differing amounts of polarization rotation based on the the source redshift.  This could potentially be observed~\cite{Galaverni:2014gca} with polarization measurements of ultraviolet sources~\cite{Cimatti:1993yc}, radio sources~\cite{Leahy:1997wj}, and gamma-ray bursts~\cite{Gotz:2013dwa}. 
Of course it would be interesting to explore a real model of this which is important for testing the actual viability of this scenario.  We leave this for future work.

One could in principle also avoid the isotropic birefringence bounds by having the axion period be short enough today that any CMB measurement happens over many periods and sees a very small average value.  Unfortunately, this possibility requires far too much energy density in the axion, as described above.  This energy density constraint could be avoided if either the axion mass $m_a$ or coupling $g_{a \gamma\gamma}$ changes between reionization and today.
Such models could could still generate a time-dependent cosmic birefringence of the CMB. Searches for this signal with periods from 12 hours to 100 days have been conducted by BICEP/Keck~\cite{BICEPKeck:2020hhe,BICEPKeck:2021sbt}, SPT~\cite{SPT-3G:2022ods}, and POLARBEAR~\cite{POLARBEAR:2023ric} and exclude oscillations today with amplitude $\gtrsim 0.5$ degrees.
It would be interesting to extend these searches to a larger range of frequencies.

\subsubsection*{Adding More Lenses}

A universe with more structure will generally have a larger lensing amplitude. However, as we have seen, achieving more structure through low redshift dynamics can give rise to a variety of other effects that can also change the expansion history at a significant level. If we would like to directly impact lensing while having minimal changes to expansion, it may be easier to directly change the clustering in the initial conditions, thus removing the issue of backreaction. In particular, we need only change the initial conditions to provide additional sources of lensing.

One possible strategy is to introduce a sub-component of the dark matter that contains an isocurvature perturbation that has a larger density contrast than ordinary dark matter. Naively, it would seem this possibility is already excluded by cosmological constraints. We need to increase the lensing amplitude by approximately 5\%, while bounds on isocurvature from Planck~\cite{Planck:2015fie,Planck:2018jri} and ACT~\cite{ACT:2025tim} exclude more than a 3\% matter-radiation isocurvature component.

%%%%%%%%%%%%
\begin{figure}[t!]
    \centering
    \includegraphics[width=0.85\textwidth]{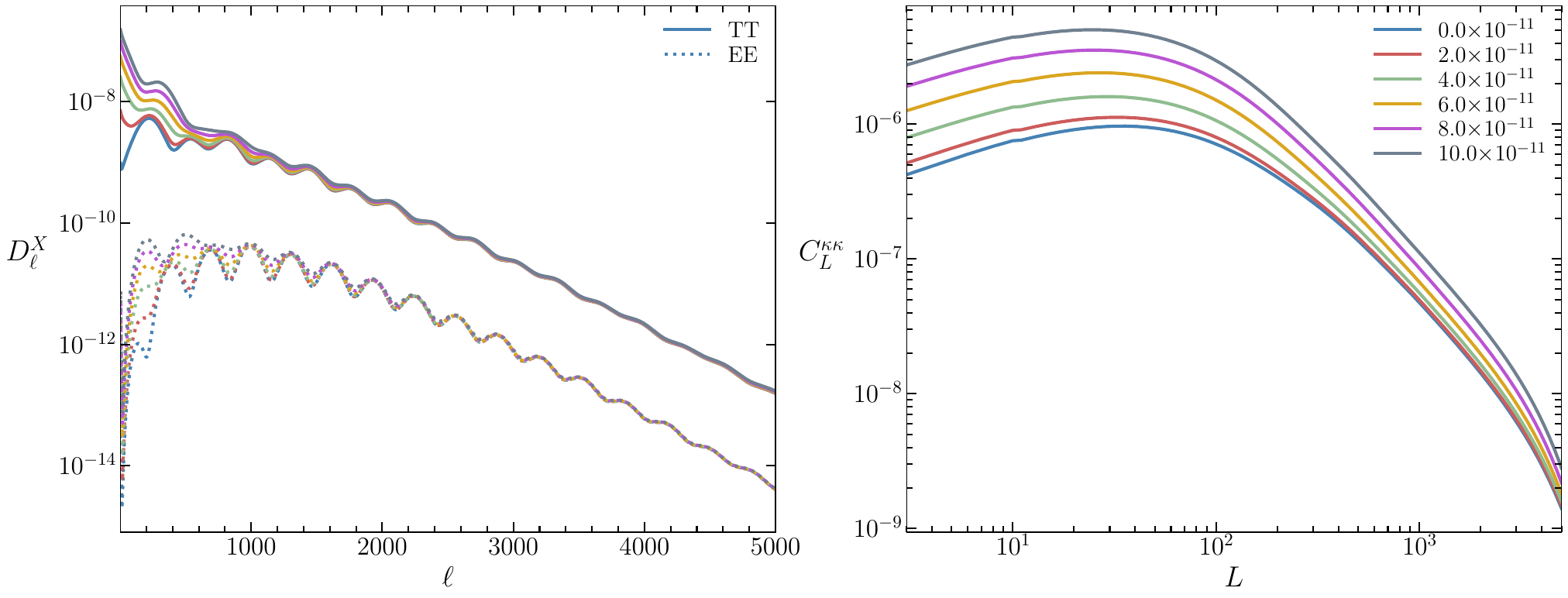}
    \caption{Comparison of the effect of scale invariant isocurvature perturbations on primary CMB (left) and lensing power spectrum (right). The amplitude of the uncorrelated lensing amplitude, $A_{\rm iso}$ is show in the legend of the right panel. We see that the primary CMB is only affected at $\ell <1000$ where matter domination is most relevant. In contrast, the effect on CMB lensing is present for $L \sim 100-1000$ which is entirely radiation dominated.}
    \label{fig:iso}
\end{figure}
%%%%%%%%%%%%%

However, it is important that the physical modes that are best measured by CMB lensing, corresponding to angular scales with multipoles $L \in [50,300]$, involve a much smaller range of scales than those measured in the primary CMB, involving multipoles $\ell \in [2,4000]$. In addition, because the lensing modes correspond to density fluctuations that are closer to us than the CMB, the physical length scales are shorter than those contributing to CMB temperature and polarization modes at the same angular scale. In practice, the lensing modes responsible for the constraints on neutrino mass ($\mnuclustering$) correspond to density fluctuations that entered the horizon during the radiation era. Uncorrelated dark matter isocurvature modes  are much less constrained in this regime because the matter fluctuations are a subleading component of the universe. In this sense, the most stringent constraints on isocurvature arise from lower $\ell$ modes in the CMB and are less relevant to the lensing signal.

This can be seen explicitly in Figure~\ref{fig:iso}. Introducing uncorrelated isocurvature in the dark matter, the impact on the CMB is limited to $\ell < 1000$, where the modes that entered the horizon during matter domination reside. On the other hand, since we are directly increasing the dark matter fluctuations, the lensing amplitude is increased on all scales. A scale-dependent isocurvature contribution that impacts $L \sim 50-300$ at $z=1$ would correspond to modes at $\ell \sim 200-1200$ in the CMB and therefore avoid\footnote{Technically speaking, because cosmic variance scales as $1/(2\ell +1)$, the higher-$\ell$ modes are measured with more precision. It is straightforward to check that the signal of isocurvature vanishes faster than cosmic variance for $\ell >200$ and therefore the constraints are dominated by lower-$\ell$.} the most constraining parts of the primary CMB. 

Isocurvature has many observable effects on both the CMB~\cite{White:1994sx,Sievers:2002tq} and large scale structure~\cite{Efstathiou:1986pba,Crotty:2003rz}. Most signals are present in the linearized fluctuations, making it straightforward to include in existing analyses~\cite{Planck:2018jri}. Deeper maps of the CMB~\cite{CMB-S4:2016ple,SimonsObservatory:2018koc,NASAPICO:2019thw} and the distribution of galaxies~\cite{EuclidTheoryWorkingGroup:2012gxx,Chung:2023syw} are both expected to improve the sensitivity to a wide range of isocurvature models by an order of magnitude or more. Strongly scale dependent isocurvature signals like those discussed here can be easily tested with current and future data (see for example the recent analysis from ACT~\cite{ACT:2025tim} that shows improved constraints on small-scale isocurvature power).

\section{Conclusions}\label{sec:conclusions}

Cosmological measurements combining the early and late universe are extremely sensitive to the overall mass scale of neutrinos~\cite{Dvorkin:2019jgs}. Unfortunately, this sensitivity is limited entirely by degeneracies with other cosmological parameters including the optical depth and overall matter density. As a result, the current discrepancy between cosmological neutrino mass measurements and neutrino oscillation experiments is hard to resolve without an independent measurements of these parameters.

Fortunately, we have a wide range of cosmological measurements that encode the neutrino mass through multiple distinct signals. Naturally, it is important to know if all or just some of these distinct effects show the same unusual preference for negative neutrino mass. We found that there are really two distinct effects that could account for the observations: the low value of $\Omega_m$ in the measurement of expansion from DESI and/or a high value of the CMB lensing amplitude. In principle, an explanation of one of these two discrepancies with new physics, or systematics, can resolve the other tension.

It is still unknown if the preference for negative neutrino mass is due to systematic effects associated with the surveys or is a signal of new physics. In the absence of independent and improved measurements of the cosmological parameters in $\Lambda$CDM, it is useful to explore beyond the Standard Model explanations for these discrepancies. These models can point to additional measurements that exclude or point to potential resolutions. In addition, these models give us new appreciation for the origin of these signals that can help us better understand effects that could be important. The central feature of the broad classes of models discussed in this paper is that they rarely lead to predictions that are equivalent to a shift in the $\Lambda$CDM parameters.

One particularity interesting observable for understanding the nature of the current tension is the lensing amplitude as measured by different estimators. For true gravitational lensing, these effects all come together with the same amplitude. However, models that modulate the CMB that mimic lensing~\cite{Yadav:2009eb}, or biases in the lensing reconstruction from foregrounds~\cite{Su:2011ff}, behave in a unique way in temperature- and polarization-based reconstruction. A consistent bias in all of these measurements would favor a physical change to the matter in the universe, or a bias in another parameter like the optical depth~\cite{Sailer:2025lxj,Jhaveri:2025neg}.

An important take-away from this work is that there is value in expanding the parameter space of neutrino masses, even if one only cares about the physical regime. Due to the degeneracies, data from current surveys are unlikely to meaningfully change the posterior for physical neutrino masses. As a result, we cannot depend on statistics alone to pinpoint the origin of the tension with the observed neutrino oscillations. Fortunately, as we have shown in this work, expanding the physical signal in terms of its unique influences provides new pathways to understanding cosmological data, both in the physical neutrino mass regime and in extensions of the Standard Model.

\paragraph{Acknowledgments}
We are grateful to Eugene Chen,
Kyle Dawson, Simone Ferraro, Adrienne Erickcek, Raphael Flauger, Jiashu Han, Colin Hill, Austin Joyce, Lloyd Knox, Marilena Loverde, Vincent Lee, Joshua Perez, Surjeet Rajendran, Eva Silverstein, Cynthia Trendafilova, Ben Wallisch, Risa Wechsler, and Zach Weiner for helpful discussions. 
PWG acknowledges support by NSF Grant No.~PHY-2310429, Simons Investigator Award No.~824870,  the Gordon and Betty Moore Foundation Grant No.~GBMF7946, and the John Templeton Foundation Award No.~63595.
DG is supported by the US~Department of Energy under grant~\mbox{DE-SC0009919}.
JM is supported by the US~Department of Energy under Grant~\mbox{DE-SC0010129}, by NASA through Grant~\mbox{80NSSC24K0665}, and by NSF through grant \mbox{AST-2510926}. 
Computational resources for this research were provided by SMU’s Center for Research Computing.  
We acknowledge the use of \texttt{CAMB}~\cite{Lewis:1999bs}, \texttt{CLASS}~\cite{Blas:2011rf}, %\texttt{CLASS-PT}~\cite{Chudaykin:2020aoj}, 
%\texttt{FAST-PT}~\cite{McEwen:2016fjn},
\texttt{CosmoRec}~\cite{Chluba:2010ca,Chluba:2010fy}
\texttt{cobaya}~\cite{Torrado:2020dgo},
\texttt{GetDist}~\cite{Lewis:2019xzd},
\texttt{IPython}~\cite{Perez:2007ipy}, %\texttt{MontePython}~\cite{Audren:2012wb, Brinckmann:2018cvx}, 
and the Python packages \texttt{Matplotlib}~\cite{Hunter:2007mat}, \texttt{NumPy}~\cite{Harris:2020xlr}, and~\texttt{SciPy}~\cite{Virtanen:2019joe}.

\newpage
% -----------------------------------------------------------------------------------------------------------------------------------------
\appendix

%%%%%%%%%%%%%%%%%

\section{Fisher Matrices}\label{app:fisher}

We would like to make quantitative predictions for our models and understand if the parameters relevant to the current signals could be tested with additional data. The purpose of this Appendix is to provide the details for those forecasts throughout the paper.

\subsection{CMB+BAO Forecasts}
\label{app:DE_forecasts}

Here we present forecasts for neutrino mass parameters in cosmological models with dynamical dark energy, described by parameters $w_0$ and $w_a$.  Dynamical dark energy impacts the expansion history and the clustering of matter, affecting the same observables that are primarily used to constrain neutrino mass.  We therefore expect weaker constraints on neutrino mass in $w_0w_a$ cosmology as compared to models where dark energy is treated as a cosmological constant.  In Table~\ref{tab:w0wa_Forecasts}, we present constraints on $\mnu$, $\mnuclustering$, and $\mnubao$ in models with dynamical dark energy using the same configurations as for the $\Lambda$CDM forecasts presented in Table~\ref{tab:Forecasts}.  As expected, the constraints for cosmology involving dynamical dark energy are weaker across the board.  In the $w_0w_a$ models, the high redshift BAO measurements provided by Spec-S5~\cite{Spec-S5:2025uom} provide a non-trivial improvement on the forecasted constraints, since those measurements better constrain the impact of dynamical dark energy at $z>2$.

%%%%%%%%%%%%%%%%%%
% Forecast table w/physical mnu
%%%%%%%%%%%%%%%%%%%%%
\begin{table}
    \centering
    \scriptsize
    \begin{tabular}{c|c||>{\cellcolor{gray!15}}ccc|>{\cellcolor{gray!15}}ccc}
         \multicolumn{2}{c||}{ }   & $\mnu$ & $\mnuclustering$ & $\mnubao$ & $\mnu$ & $\mnuclustering$ & $\mnubao$ \\
         \hline
         $\Delta_T$~[$\mu$K-arcmin] & $\sigma(\tau)$  & \multicolumn{3}{c|}{+DESI 5yr} & \multicolumn{3}{c}{+DESI 5yr+Spec-S5} \\
        \hline \hline
        10 & 0.006  & 85 & 92 & 134 & 74 & 88 & 109 \\
        6 & 0.006  & 82 & 87 & 125 & 71 & 84 & 103 \\
        1 & 0.006  & 74 & 68 & 108 & 66 & 67 & 91 \\
        \hline
        10 & 0.002  & 78 & 80 & 134 & 66 & 74 & 109 \\
        6 & 0.002  & 75 & 75 & 124 & 64 & 70 & 103 \\
        1 & 0.002  & 67 & 59 & 108 & 59 & 57 & 91 \\
    \end{tabular}
    \caption{Forecasts for $1\sigma$ errors on $\mnu$ in units of meV for $w_0w_a$CDM+$\mnu$ cosmology shown in gray and on $\mnuclustering$ and $\mnubao$ in units of meV for $w_0w_a$CDM+$\mnubao$+$\mnuclustering$ cosmology.  For the forecasts shown here, we assume that CMB surveys are modeled by white noise spectra at $\ell\geq30$ with a $1.4$~arcmin beam covering $f_\mathrm{sky}=0.5$. }
    \label{tab:w0wa_Forecasts}
\end{table}
%%%%%%%%%%%%%%%%%%%%%

\subsection{CMB Non-Gaussianity Forecasts}\label{app:CMB_for}

In the main text, we focus on the bias of the lensing reconstruction map, $\hat \phi$, due to the presence of a modulating field $\Phi$. Of course, our primary interest is the impact on $\mnuclustering$ through the change to the lensing amplitude. If we assume the lensing and modulating fields are uncorrected, then total reconstruction power spectrum is simply
\beq
P(L) = \langle \hat \phi(\L) \hat \phi(-\L)\rangle' = A_{\rm lens} P_{\rm lens}(L) + A_\phi P_{\phi,\Phi} (L) \ .
\eeq
where $P_{\phi,\Phi}(L)$ is the power spectrum of the field after reconstruction. At this point, we could still vary both $A_{\rm lens}$ and $A_\phi$ to isolate the contribution from each term. We can additionally define a unique estimator for each modulating field $\hat \Phi$ and reconstruct the individual power spectra.

Given several modulating fields, $\Phi_i$, each with its own reconstruction $\hat \Phi_i$, we can define a set of power spectra $A_i P_i(L)$ where $A_i$ is the amplitude. The Fisher matrix for the simultaneous measurement of all the $A_i$ is given in the usual way by
\beq
F_{ij} = \int \frac{d^2 L}{(2\pi)^2} \frac{P_i(L) P_j(L)}{(\bar A_i P_i(L) +N_i(L)) (\bar A_j P_j(L)+ N_j(L))}
\eeq
where $\bar A_k$ are the fiducial values for the amplitudes and $N_k(L)$ are the reconstruction noise curves given by Equation~(\ref{eq:noisecurve}),
\beq
% N_\Phi(L) = \left( \int \frac{d^2\ell}{(2\pi)^2} \frac{|f^{(\Phi)}(\vl,\L-\vl)|^2}{2 \tilde C^{TT}(\ell) \tilde C^{TT}(\L-\ell)  }\right)^{-1} \ ,
% N_\Phi(L) = \left( \int \frac{d^2\ell}{(2\pi)^2} \frac{|f^{(\Phi)}(\vl,\L-\vl)|^2}{2 \tilde C^{TT}_\ell \tilde C^{TT}_{|\L-\vl|}  }\right)^{-1} \ ,
N_\Phi(L) = \left( \int \frac{d^2\ell}{(2\pi)^2} \frac{|f^{(\Phi)}(\vl,\L-\vl)|^2}{2 C^{TT}_\ell C^{TT}_{|\L-\vl|}  }\right)^{-1} \ ,
\eeq
Here 
% $C^{TT}(\ell)= C^{TT}(\ell) + N^{TT}(\ell)$
$C^{TT}_\ell= \tilde C^{TT}_\ell + N^{TT}_\ell$
is the observed temperature power spectrum including noise. The $N^{TT}_\ell$ noise curves can be calculated in a straightforward way for Planck or any future CMB survey following~\cite{Wu:2014hta,CMB-S4:2016ple} to determine the primary CMB noise curves to use in (\ref{eq:noisecurve}). With this, the reconstruction noise and Fisher matrix are determined by replacing the integral with a sum over $L$.

%%%%%%%%%%%%
\begin{figure}[t!]
    \centering
    \includegraphics[width=0.5\textwidth]{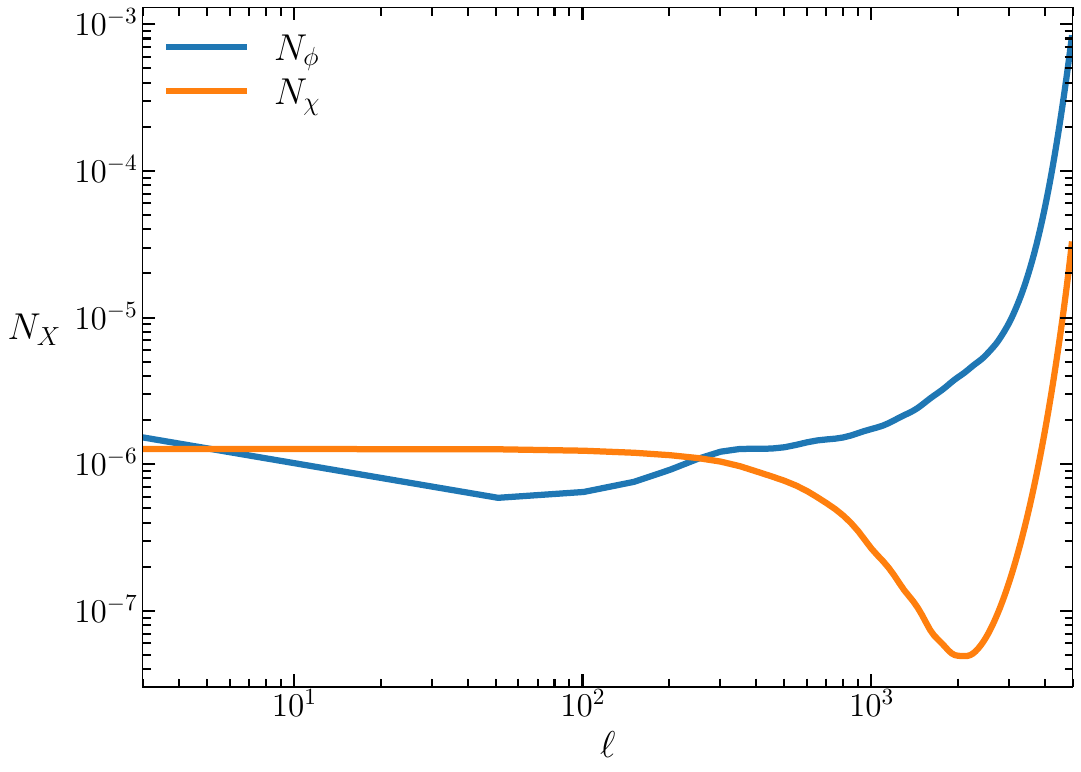}
    \caption{Comparison of reconstruction noise curves CMB lensing, $N_\phi$, and a scalar modulation, $N_\chi$, using the $TT$ estimators. The primary CMB is assumed to be Planck-like. }
    \label{fig:rnoise}
\end{figure}
%%%%%%%%%%%%%

For a field like $\sigma$ that biases $\hat \phi$ without any additional suppression, the Fisher matrix determines how hard it is to distinguish two different power spectra. Following~\cite{Babich:2004gb}, we define a cosine
\beq
\cos(A_i,A_j) = \frac{F_{ij}}{\sqrt{F_{ii}F_{jj}}} \ .
\eeq
For example, if $P_\sigma(L) \propto L^{-2}$, something we would associate with a scale-invariant spectrum, then $\cos(A_{\rm lens},A_\sigma) \approx 0.1$ for Planck noise levels.

It is also useful to check that this formalism is consistent with direct measurements of $\fnl$ or $\taunl$ that are also parameterizations of modulating fields, via the bispectrum or trispectrum. The reconstruction noise curves for the lensing estimator $\hat \phi$ and for a scalar modulator $\hat \chi$ are shown in Figure~\ref{fig:rnoise}. For case of $\chi$, the amplitude of the power spectrum $A_\chi$ is related to $\taunl$ by $A_\chi = A_s \taunl$. For a power spectrum that scales as $\ell^{-2}$, we would expect the largest signal at low $\ell$. Using $N_\chi(\ell =10) \approx 10^{-6}$ and $A_s\ \sim 10^{-9}$, we would expect the constraint on $\taunl$ to be $\sigma(\tau_{\rm NL}) \approx 10^3$. Using this noise curve to calculate the the Fisher matrix, we find $\sigma(\taunl) = 830$. This is in good agreement with the bound from Planck of $\taunl < 1700$ at 95\% confidence~\cite{Marzouk:2022utf}, derived from the amplitude of the $\hat \chi$ reconstruction. Directly constraining the trispectrum gives a somewhat better constraint of $\taunl< 1500$~\cite{Philcox:2025wts}.

\subsection{Large Scale Structure Bispectrum Forecasts}\label{app:bispectrum}

The large scale structure bispectrum is a potentially powerful tool for cosmology, although still in its infancy in terms of analyses of real data. For example, there is currently no measurement of the bispectrum using DESI observations. The equivalence principle violating signal we are looking for requires a multi-messenger bispectrum analysis and is therefore an additional level removed from current techniques. Given the uncertainties in what can be achieved with realistic analyses, we will give a somewhat simplified forecasts to illustrate the potential, acknowledging that it is difficult to assess how realistic the forecasts will be in practice.

Our forecasting will largely follow the approach described in~\cite{Smith:2006ud,Baumann:2021ykm}. Specifically, we will treat each redshift bin a three-dimensional box with volume $V(z)$ and a number density of galaxies $\bar n(z)$ with biases $b_1(z)$. With this information, the contribution to the Fisher matrix for each redshift is given by 
\beq
\begin{aligned}
F_{i j}(z)=&\frac{V(z)}{6} \int \frac{\mathrm{~d}^3 k_1}{(2 \pi)^3} \frac{1}{k_1} \frac{3!}{(2 \pi)^2}    \int_{k_1 / 2}^{k_1} k_2 \mathrm{~d} k_2 \int_{k_1-k_2}^{k_2} k_3 \mathrm{~d} k_3  \\
&\times \frac{B_i\left(k_1, k_2, k_3;z\right) B_j\left(k_1, k_2, k_3;z\right)}{\left(P\left(k_1;z\right)+N^2\right)\left(P\left(k_2;z\right)+N^2\right)\left(P\left(k_3;z\right)+N^2\right)} \ ,
\end{aligned}
\eeq
where $N(z)= 1/\bar n(z)$  and $B_i = \partial_{\theta_i} B$ with $\theta_i \in \{b_2,A_3,b_{\nabla^2}, b_{\nabla^4}, b_{s^2},\beta \}$. We then define the full Fisher matrix as
\beq
F_{ij} = \sum_{z_k} F_{ij}(z_k) \ .
\eeq
Here we define $\partial_{A_3} B  = b_1^3 B_{m}$ where $B_m$ is the tree-level matter bispectrum in Equation~(\ref{eq:Bmatter}). Our forecasts for the full DESI survey use the survey definitions from~\cite{DESI:2016fyo,Baumann:2017gkg} and are found in the form used for our forecasts in Appendix~A.2 of~\cite{Green:2023uyz}. 

There are several additional details of the forecast that require some explanation. We are holding $b_1$ fixed because it will generally be measured to very high accuracy with the galaxy power spectrum. In addition, we are varying $A_3$ which is generated with $b_1$ in a bispectrum-only analysis. We have not included parameters like $b_3$ and other higher order biases, as these are better constrained by higher point statistics~\cite{Baumann:2021ykm}. We instead include higher derivative biases $b_{\nabla^2}, b_{\nabla^4}$ in order to marginalize over short distance physics specifically. These higher derivative biases are also determined by the power-spectrum so our baseline forecast holds them fixed. Since we are only calculating the bispectrum to linear order in our parameters, the fiducial values of the nonlinear bias parameters $\beta$ and $A_3$ do not impact the results.

The forecasts for $\beta$ are written in terms of the parameter
\beq
\tilde \beta  = \beta \fdm^2 (b_1^A b_r^B - b_1^B b_r^A)/b_1^2 \ .
\eeq
With this definition the signal of interest appears in the bispectrum as 
\beq
\begin{aligned}
B_{\tilde \beta} = & b_1^3 \bigg( F_2(\k_2,\k_3)P_m(k_2)P_m(k_3)  
+\tilde F_2(\k_1,\k_3) P_m(k_1)P_m(k_3) \\
&+ F_2(\k_1,\k_2)P_m(k_1)P_m(k_2) \bigg)  \ .
\end{aligned} 
\eeq
This forecast is approximate, as we have absorbed all of the multi-tracer information into $\tilde \beta$ without determining the samples or changes to the shot noise from splitting the sample. While this is important for a realistic forecast it is less important than our lack of knowledge of $b_r^{A,B}$ for any of the samples in DESI.

\section{Review of Lensing in the Flat Sky Limit}\label{app:lensing}

Polarization of the CMB is a subtle topic that is not always intuitive, particularly on the full sky. Fortunately, a lot of the physics relevant to the CMB is understandable in the flat sky limit where many aspects of the problem become more familiar. In this appendix, we will review the construction and properties of the polarization $E$ and $B$ modes in the flat sky description, following~\cite{Zaldarriaga2000b,Lewis:2006fu}.

To define the flat sky coordinate system, we assume that light travels in the $\hz$-direction with a classical electric field defined by
\beq
% \vec E = {\rm Re}\left[ \left(E_x(\n)\hat x + E_y(\n) \hat y \right) e^{i k(z-c t)} \right] \ .
{\bm E} = {\rm Re}\left[ \left(E_x(\n)\hx + E_y(\n) \hy \right) e^{i k(z-c t)} \right] \ .
\eeq
Importantly, these are both function of $\n = (x,y)$, the coordinates on the sky. We can then define a polarization tensor as $P_{ab} = E_a E_b$ where $a=1,2$ are the $x$ and $y$ components of the field.  In these coordinates, it will be useful to define complex unit vectors
\beq
% \vec e_{\pm} = \vec e_x \pm i \vec e_y \ .
\he_{\pm} = \he_x \pm i \he_y \ .
\eeq
It will be important that 
% $\vec e_+ \cdot \vec e_- =2$ and $\vec e_{\pm} \cdot \vec e_{\pm} = 0$.
$\he_+ \cdot \he_- =2$ and $\he_{\pm} \cdot \he_{\pm} = 0$.
We can then use these unit vectors to relate the polarization tensor and the Stokes' $Q$ and $U$ parameters.
\beq
P(\n)= e^a_+ e^b_+ P_{ab}  = (E_x + i E_y)^2 = Q(\n) + i U(\n) \ .
\eeq
Again, we highlight that $P(\n)$, $Q(\n)$, and $U(\n)$ are fields in two-dimensions.

Naturally, one would like to express $P(\n)$ in terms of curl-like ($P_B$) and divergence-like ($P_E$) fields 
\beq
P_{ab} = \nabla_{\langle a} \nabla_{b\rangle} P_E + \epsilon_{\phantom{c} [ a}^c \nabla_{b]} \nabla_c P_B
\eeq
where $\epsilon$ in the fully anti-symmetric tensor and $\langle a, b\rangle$ means symmetric-traceless component and $[a, b]$ means the anti-symmetric component.

In practice, we write this in terms of the Fourier transform $\n \to \vl$ and harmonics 
\beq
P_{a b}=\frac{1}{\sqrt{2}} \int \frac{d^2 \ell}{(2\pi)^2}\left[E_\vl Q_{ \vl a b}^G+B_\vl Q_{\vl a b}^C\right]
\eeq
where 
\beq
Q_{\vl a b}^G=N_\ell \nabla_{\langle a} \nabla_{b \rangle } Q_\vl \quad Q_{\vl a b}^C=N_\ell \, \epsilon_{\phantom{c} [a}^c \ \nabla_{b]} \nabla_{c} Q_\vl
\eeq
and $Q_\vl = e^{i \vl \cdot \n}$. We will take the normalization, $N_\ell = \sqrt{2} /\ell^2$ so that we have uniform weighting in $\ell$. Using $\epsilon^{ab} = -i e_-^{[a} e_+^{b]}$, one can check that 
\beq
% P = \int \frac{d^2 \ell}{(2\pi)^2} \frac{N_\ell}{\sqrt{2}}\left(E_k+i B_k\right) \mathbf{e}_{+}^a \mathbf{e}_{+}^b \nabla_a \nabla_b Q_\ell \ .
% P = \int \frac{d^2 \ell}{(2\pi)^2} \frac{N_\ell}{\sqrt{2}}\left(E_k+i B_k\right) e_{+}^a e_{+}^b \nabla_a \nabla_b Q_\vl \ .
P = \int \frac{d^2 \ell}{(2\pi)^2} \frac{N_\ell}{\sqrt{2}}\left(E_{\l}+i B_{\l}\right) e_{+}^a e_{+}^b \nabla_a \nabla_b Q_\vl \ .
\eeq
Using $e_-^a e_+^b  \nabla_a \nabla_b = \nabla^2$, and $\ell^{-2} \to -\nabla^{-2}$, we can write
\beq
\nabla^2 (E+ i B) =- e_-^a e_-^b \nabla_a  \nabla_b (Q+i U) \ .
\eeq
Specifically, for the B modes, this means
\beq
\nabla^2 B = \left(\partial_y^2 -\partial_x^2\right) U + 2 \partial_x \partial_y Q \ .
\eeq
These expressions can be used to derive the behavior under modulation of the scalar metric fluctuations described in the main text.

When the polarization tensor is modulated directly, it is useful to write $E$ and $B$ in terms of the observed polarization field. Using the flat sky mode function 
\beq
\begin{aligned}
% \frac{N_\ell}{\sqrt{2}} \mathbf{e}_{+}^a \mathbf{e}_{+}^b \nabla_a \nabla_b Q_\ell & =\frac{1}{\ell^2} \mathbf{e}_{+}^a \mathbf{e}_{+}^b \nabla_a \nabla_b e^{i \vl \cdot \x} \\
% %& =\frac{1}{\ell^2} \left(\partial_x+i \partial_y\right)^2 e^{i  \vl \cdot \x} \\
% & =-\left(\cos \phi_\vl+i \sin \phi_\vl\right)^2 e^{i  \vl \cdot \x} \\
% & =-e^{2 i \phi_\vl} e^{i i \vl \cdot \x}
\frac{N_\ell}{\sqrt{2}} e_{+}^a e_{+}^b \nabla_a \nabla_b Q_\vl & =\frac{1}{\ell^2} e_{+}^a e_{+}^b \nabla_a \nabla_b e^{i \vl \cdot \n} \\
%& =\frac{1}{\ell^2} \left(\partial_x+i \partial_y\right)^2 e^{i  \vl \cdot \x} \\
& =-\left(\cos \phi_\vl+i \sin \phi_\vl\right)^2 e^{i  \vl \cdot \n} \\
& =-e^{2 i \phi_\vl} e^{i \vl \cdot \n}
\end{aligned}
\eeq
where we have define the vector in polar coordinates $\vl = \ell e^{i \phi_\vl}$. Using the Fourier transform, one arrives at the useful expression
\beq
E(\vl)+i B(\vl)=-\int d^2 x P e^{-2 i \phi_\vl} e^{-i \vl \cdot \x} \ .
\eeq
The impact on $E$ and $B$ for modulations of $P$ are most easily evaluated in this form.

\clearpage
\phantomsection
\addcontentsline{toc}{section}{References}
\bibliographystyle{utphys}
\bibliography{Refs}

\end{document}